\documentclass[12pt]{article}
\usepackage[dvips]{graphicx}
\usepackage{epsfig}
\usepackage{latexsym,amsmath,amsfonts,amssymb}
\usepackage[latin1]{inputenc}
\usepackage[american]{babel}
\usepackage[dvips]{graphicx}
\usepackage{bbm}
\usepackage{color}
\usepackage[unicode]{hyperref}
\usepackage{epstopdf}
\usepackage{float}
\usepackage{caption}
\usepackage{subcaption}
\def\im{Invent. Math.}

\def\a{\alpha}
\def\b{\beta}
\def\c{\gamma}
\def\d{\delta}
\def\eps{\epsilon}           
\def\f{\phi}               
\def\vf{\varphi}  \def\tvf{\tilde{\varphi}}
\def\vp{\varphi}
\def\g{\gamma}
\def\h{\eta}
\def\j{\psi}
\def\k{\kappa}                    
\def\l{\lambda}
\def\m{\mu}
\def\n{\nu}
\def\o{\omega}  \def\w{\omega}
\def\p{\pi}
\def\q{\theta}  \def\th{\theta}                  
\def\r{\rho}                                     
\def\s{\sigma}                                   
\def\t{\tau}
\def\u{\upsilon}
\def\x{\xi}
\def\z{\zeta}
\def\pt{\tilde{\varphi}}
\def\tt{\tilde{\theta}}
\def\lab{\label}
\def\6{\partial}
\def\wg{\wedge}
\def\bpsi{\bar{\psi}}
\def\bt{\bar{\theta}}
\def\bvf{\bar{\varphi}}

\DeclareMathOperator{\tr}{tr}

\newcommand{\be}{\begin{equation}}
\newcommand{\ee}{\end{equation}}
\newcommand{\beq}{\begin{equation}}
\newcommand{\eeq}{\end{equation}}
\newcommand{\bea}{\begin{eqnarray}}
\newcommand{\eea}{\end{eqnarray}}
\newcommand{\nn}{\nonumber}

\newcommand{\ba}{\begin{eqnarray}}
\newcommand{\ea}{\end{eqnarray}}

\newcommand{\beqs}{\begin{eqnarray}}
\newcommand{\eeqs}{\end{eqnarray}}
\newcommand{\bal}{\begin{aligned}}
\newcommand{\eal}{\end{aligned}}

\begin{document}
\baselineskip=15.5pt
\pagestyle{plain}
\setcounter{page}{1}


\def\del{{\partial}}
\def\vev#1{\left\langle #1 \right\rangle}
\def\cn{{\cal N}}
\def\co{{\cal O}}
\def\IC{{\mathbb C}}
\def\IR{{\mathbb R}}
\def\IZ{{\mathbb Z}}
\def\RP{{\bf RP}}
\def\CP{{\bf CP}}
\def\Poincare{{Poincar\'e }}
\def\tr{{\rm tr}}
\def\tp{{\tilde \Phi}}

\def\TL{\hfil$\displaystyle{##}$}
\def\TR{$\displaystyle{{}##}$\hfil}
\def\TC{\hfil$\displaystyle{##}$\hfil}
\def\TT{\hbox{##}}
\def\HLINE{\noalign{\vskip1\jot}\hline\noalign{\vskip1\jot}}
\def\seqalign#1#2{\vcenter{\openup1\jot
   \halign{\strut #1\cr #2 \cr}}}
\def\lbldef#1#2{\expandafter\gdef\csname #1\endcsname {#2}}
\def\eqn#1#2{\lbldef{#1}{(\ref{#1})}%
\begin{equation} #2 \label{#1} \end{equation}}
\def\eqalign#1{\vcenter{\openup1\jot

     \halign{\strut\span\TL & \span\TR\cr #1 \cr

    }}}
\def\eno#1{(\ref{#1})}
\def\href#1#2{#2}
\def\half{\frac{1}{2}}


\def\ads{{\it AdS}}
\def\adsp{{\it AdS}$_{p+2}$}
\def\cft{{\it CFT}}

\newcommand{\ber}{\begin{eqnarray}}
\newcommand{\eer}{\end{eqnarray}}
\newcommand{\beqar}{\begin{eqnarray}}
\newcommand{\cN}{{\cal N}}
\newcommand{\cO}{{\cal O}}
\newcommand{\cA}{{\cal A}}
\newcommand{\cT}{{\cal T}}
\newcommand{\cF}{{\cal F}}
\newcommand{\cC}{{\cal C}}
\newcommand{\cR}{{\cal R}}
\newcommand{\cW}{{\cal W}}
\newcommand{\eeqar}{\end{eqnarray}}
\newcommand{\tht}{\thteta}
\newcommand{\lm}{\lambda}\newcommand{\Lm}{\Lambda}


\newcommand{\nonu}{\nonumber}
\newcommand{\oh}{\displaystyle{\frac{1}{2}}}
\newcommand{\dsl}
   {\kern.06em\hbox{\raise.15ex\hbox{$/$}\kern-.56em\hbox{$\partial$}}}
\newcommand{\id}{i\!\!\not\!\partial}
\newcommand{\as}{\not\!\! A}
\newcommand{\ps}{\not\! p}
\newcommand{\ks}{\not\! k}
\newcommand{\D}{{\cal{D}}}
\newcommand{\dv}{d^2x}
\newcommand{\Z}{{\cal Z}}
\newcommand{\N}{{\cal N}}
\newcommand{\Dsl}{\not\!\! D}
\newcommand{\Bsl}{\not\!\! B}
\newcommand{\Psl}{\not\!\! P}
\newcommand{\eeqarr}{\end{eqnarray}}
\newcommand{\ZZ}{{\rm \kern 0.275em Z \kern -0.92em Z}\;}



\def\del{{\delta^{\hbox{\sevenrm B}}}} \def\ex{{\hbox{\rm e}}}
\def\azb{A_{\bar z}} \def\az{A_z} \def\bzb{B_{\bar z}} \def\bz{B_z}
\def\czb{C_{\bar z}} \def\cz{C_z} \def\dzb{D_{\bar z}} \def\dz{D_z}
\def\im{{\hbox{\rm Im}}} \def\mod{{\hbox{\rm mod}}} \def\tr{{\hbox{\rm Tr}}}
\def\ch{{\hbox{\rm ch}}} \def\imp{{\hbox{\sevenrm Im}}}
\def\trp{{\hbox{\sevenrm Tr}}} \def\vol{{\hbox{\rm Vol}}}
\def\rl{\Lambda_{\hbox{\sevenrm R}}} \def\wl{\Lambda_{\hbox{\sevenrm W}}}
\def\fc{{\cal F}_{k+\cox}} \def\vev{vacuum expectation value}
\def\nodiv{\mid{\hbox{\hskip-7.8pt/}}}
\def\ie{{\em i.e.}}
\def\ie{\hbox{\it i.e.}}
\def\CC{{\mathchoice
{\rm C\mkern-8mu\vrule height1.45ex depth-.05ex
width.05em\mkern9mu\kern-.05em}
{\rm C\mkern-8mu\vrule height1.45ex depth-.05ex
width.05em\mkern9mu\kern-.05em}
{\rm C\mkern-8mu\vrule height1ex depth-.07ex
width.035em\mkern9mu\kern-.035em}
{\rm C\mkern-8mu\vrule height.65ex depth-.1ex
width.025em\mkern8mu\kern-.025em}}}

\def\RR{{\rm I\kern-1.6pt {\rm R}}}
\def\NN{{\rm I\!N}}
\def\ZZ{{\rm Z}\kern-3.8pt {\rm Z} \kern2pt}
\def\IB{\relax{\rm I\kern-.18em B}}
\def\ID{\relax{\rm I\kern-.18em D}}
\def\II{\relax{\rm I\kern-.18em I}}
\def\IP{\relax{\rm I\kern-.18em P}}
\newcommand{\CS}{{\scriptstyle {\rm CS}}}
\newcommand{\CSs}{{\scriptscriptstyle {\rm CS}}}
\newcommand{\rc}{\nonumber\\}
\newcommand{\bear}{\begin{eqnarray}}
\newcommand{\eear}{\end{eqnarray}}

\newcommand{\LL}{{\cal L}}

\def\mani{{\cal M}}
\def\calo{{\cal O}}
\def\calb{{\cal B}}
\def\calw{{\cal W}}
\def\calz{{\cal Z}}
\def\cald{{\cal D}}
\def\calc{{\cal C}}
\def\to{\rightarrow}
\def\ele{{\hbox{\sevenrm L}}}
\def\ere{{\hbox{\sevenrm R}}}
\def\zb{{\bar z}}
\def\wb{{\bar w}}
\def\nodiv{\mid{\hbox{\hskip-7.8pt/}}}
\def\menos{\hbox{\hskip-2.9pt}}
\def\dr{\dot R_}
\def\drr{\dot r_}
\def\ds{\dot s_}
\def\da{\dot A_}
\def\dga{\dot \gamma_}
\def\ga{\gamma_}
\def\dal{\dot\alpha_}
\def\al{\alpha_}
\def\cl{{closed}}
\def\cls{{closing}}
\def\vev{vacuum expectation value}
\def\tr{{\rm Tr}}
\def\to{\rightarrow}
\def\too{\longrightarrow}


\def\a{\alpha}
\def\b{\beta}
\def\c{\gamma}
\def\d{\delta}
\def\e{\epsilon}           
\def\F{\Phi}
\def\f{\phi}               
\def\vf{\varphi}  \def\tvf{\tilde{\varphi}}
\def\vp{\varphi}
\def\g{\gamma}
\def\h{\eta}
\def\j{\psi}
\def\k{\kappa}                    
\def\l{\lambda}
\def\m{\mu}
\def\n{\nu}
\def\o{\omega}  \def\w{\omega}
\def\q{\theta}  \def\th{\theta}                  
\def\r{\rho}                                     
\def\s{\sigma}                                   
\def\t{\tau}
\def\u{\upsilon}
\def\x{\xi}
\def\X{\Xi}
\def\z{\zeta}
\def\pt{\tilde{\varphi}}
\def\tt{\tilde{\theta}}
\def\lab{\label}
\def\6{\partial}
\def\wg{\wedge}
\def\atanh{{\rm arctanh}}
\def\bpsi{\bar{\psi}}
\def\bt{\bar{\theta}}
\def\bvf{\bar{\varphi}}

%



\newfont{\namefont}{cmr10}
\newfont{\addfont}{cmti7 scaled 1440}
\newfont{\boldmathfont}{cmbx10}
\newfont{\headfontb}{cmbx10 scaled 1728}






\newcommand{\re}{\,\mathbb{R}\mbox{e}\,}
\newcommand{\hyph}[1]{$#1$\nobreakdash-\hspace{0pt}}
\providecommand{\abs}[1]{\lvert#1\rvert}
\newcommand{\Nugual}[1]{$\mathcal{N}= #1 $}
\newcommand{\sub}[2]{#1_\text{#2}}
\newcommand{\partfrac}[2]{\frac{\partial #1}{\partial #2}}
\newcommand{\bsp}[1]{\begin{equation} \begin{split} #1 \end{split} \end{equation}}
\newcommand{\calF}{\mathcal{F}}
\newcommand{\calO}{\mathcal{O}}
\newcommand{\calM}{\mathcal{M}}
\newcommand{\calV}{\mathcal{V}}
\newcommand{\bbZ}{\mathbb{Z}}
\newcommand{\bbC}{\mathbb{C}}
\newcommand{\cK}{{\cal K}}
\newcommand{\ep}{{\epsilon}}
\newcommand{\ul}{{u_{\Lambda}}}
\newcommand{\rhol}{{\rho_{\Lambda}}}
\newcommand{\rhoT}{{\rho_{T}}}
\def\rhoz{{\rho_{0}}}
\newcommand{\RD}{{R_{D_p}}}


\newcommand{\Thq}{\Theta\left(\r-\r_q\right)}
\newcommand{\Dq}{\d\left(\r-\r_q\right)}
\newcommand{\kten}{\kappa^2_{\left(10\right)}}
\newcommand{\pbi}[1]{\imath^*\left(#1\right)}
\newcommand{\ho}{\hat{\omega}}
\newcommand{\tth}{\tilde{\th}}
\newcommand{\tf}{\tilde{\f}}
\newcommand{\tj}{\tilde{\j}}
\newcommand{\tw}{\tilde{\omega}}
\newcommand{\tz}{\tilde{z}}
\newcommand{\prj}[2]{(\partial_r{#1})(\partial_{\j}{#2})-(\partial_r{#2})(\partial_{\j}{#1})}
\def\atanh{{\rm arctanh}}
\def\sech{{\rm sech}}
\def\csch{{\rm csch}}
\allowdisplaybreaks[1]

\def\red{\textcolor[rgb]{0.98,0.00,0.00}}
\numberwithin{equation}{section}
\newcommand{\Tr}{\mbox{Tr}}    

%

\renewcommand{\theequation}{{\rm\thesection.\arabic{equation}}}
\begin{titlepage}
\vspace{0.1in}

\begin{center}
\Large \bf  Confinement, Phase Transitions and non-Locality in the Entanglement Entropy
\end{center}
\vskip 0.2truein
\begin{center}
Uri Kol$^{a,}$\footnote{urikol@post.tau.il},
Carlos N\'u\~nez$^{b,}$\footnote{c.nunez@swansea.ac.uk. Also at CP3-Origins, SDU Odense (Denmark).},
Daniel Schofield$^{b,}$\footnote{pyschofield@swansea.ac.uk}, Jacob Sonnenschein$^{a,}\footnote{cobi@post.tau.ac.il}$\\  and Michael Warschawski$^{b,}\footnote{pymw@swansea.ac.uk}$
\vskip 0.2truein

\vskip 4mm

{\it $a$: School of Physics and Astronomy,\\

The Raymond and Beverly Sackler Faculty of Exact Sciences,\\ Tel Aviv University, Ramat Aviv 69978, Israel}

\vspace{0.2in}
{\it $b$: Department of Physics, Swansea University\\
 Singleton Park, Swansea SA2 8PP, United Kingdom.}

\vskip 5mm

\vspace{0.2in}
\end{center}
\vspace{0.2in}
\centerline{{\bf Abstract}}
In this paper we study the conjectural  relation
between confinement in a quantum field theory and the presence of a phase transition
in its corresponding  entanglement entropy.  We  determine
the sufficient conditions for the latter and compare to
the conditions for having a confining Wilson line.
We demonstrate the relation in several examples.
Superficially, it may seem that
certain  confining field theories with a non-local high energy behavior,
like the dual of D5 branes wrapping a two-cycle, do not admit the corresponding
phase transition. However, upon closer inspection we find that,
through the introduction
of a regulating
UV-cutoff, new eight-surface configurations appear,
that satisfy the correct concavity condition and
recover the phase transition in the entanglement entropy.
We show that a local-UV-completion to the confining
non-local theories has a similar effect to that of
the aforementioned cutoff.

\smallskip
\end{titlepage}
\setcounter{footnote}{0}

\tableofcontents

\setcounter{footnote}{0}
\renewcommand{\theequation}{{\rm\thesection.\arabic{equation}}}

 \newpage
\section{Introduction}
The holographic gauge-strings duality \cite{Maldacena:1997re}
provides an effective and controllable
toolkit to study the dynamics of non-perturbative phenomena.
An important  such tool  is   the Entanglement Entropy (EE).
This quantity defined originally  in  quantum mechanical systems
has found a wide-range  of  applications in different branches of Physics.
We refer the reader to the papers \cite{various} for a review of  these
applications and formalism.

The holographic prescription to calculate the EE was proposed
 by
Ryu and Takayanagi \cite{Ryu:2006bv},\cite{Ryu:2006ef}. For a $d$ dimensional conformal  field theory dual to an $AdS_{d+2}$ background, the holographic EE is given by minimizing the $d$ dimensional area  in $AdS_{d+1}$ whose boundary coincides with the boundary of the region that defines the EE.\footnote { For a  review  and
 for a partial list of references to  follow up papers see \cite{various}}

 Klebanov, Kutasov and Murugan (KKM)
\cite{Klebanov:2007ws} generalized the prescription of \cite{Ryu:2006bv} to non-conformal field theories. In particular they found that certain backgrounds, which are
 holographically dual to confining systems, admit
a  first order phase transition upon varying
the width of a strip that sets the entangled regions.
In this paper we further explore various aspects of the holographic EE of confining systems.

Another well known observable that in holography 
involves minimizing an area is the Wilson line (WL). We show that the functional forms of the length of strip associated with EE   and of a Wilson line are similar and further so are the EE and the energy of the WL as a function of the length. We discuss the similarities and the differences of these forms  for systems that admit confinement.  In an analogous manner to the determination of sufficient conditions for an area law WL \cite{Kinar:1998vq}, we find the sufficient condition for a first order phase transition of the holographic EE.
We apply these conditions to several examples including the $AdS_5\times S^5$, D$p$ branes compactified on $S^1$,
the hard and soft wall models and the Klebanov-Strassler 
model \cite{Klebanov:2000hb}.

A special feature  of the holographic EE  as a  diagnostic tool  for
confinement is the fact that  it relates  not only the IR behavior of the geometry, but also
has implications on the UV behavior of the background
\footnote{Here and below,
we follow the `common parlance' according to which the large
and small radial position in the string background are
associated with the UV and IR of the dual QFT}.  We  will make clear what sort of connection
we expect between confinement and the UV-behavior of the QFT.

Making these points and connections explicit will
bring us to study the calculation of the EE in non-local QFTs.
We deal with this complicated problem  by
using holographic duals based on D5 or higher D$p$ branes ($p>5$).
Then, performing the usual calculation we  find that in spite of
the background in question having the IR-geometry suitable to be
dual to a confining QFT \cite{Kinar:1998vq},
the phase transition in the EE is absent. We will observe that
this is an effect of the UV non-locality of the QFT. We will propose
a way to fix this situation, by introducing a hard UV cutoff and
observing that new configurations appear that would not only
recover the phase transition argued in \cite{Klebanov:2007ws}, but also
solve an stability problem of the configurations
that miss the phase transition.
Finally, since the UV-cutoff may look like a `bad
fix' for the problem, we will show
(with two examples) how a suitable UV-completion to give a local theory,
plays a similar role to the UV-cutoff, at least for
the purposes of the transition.

This suggest that the EE is not only a quantity
useful to diagnose confinement, but
also to determine if the QFT in question is local in the far UV.
Let us explain in a bit more detail some ideas that
motivated this paper.

\subsection{General Idea of This Paper}
Consider a theory like QCD: it was argued that this theory has
a Hagedorn density of states. For a mass $M$,
the number of states $N(M)$  is
\bea
N(M)\sim \left(\frac{M}{T_H}\right)^{2b}e^{+\frac{M}{T_H}}.\nonumber
\eea
Here $T_H$ is some energy scale and $b$ a number.
The Partition Function will roughly be,
\bea
Z\sim \int DM \,M^{2b}e^{+\frac{M}{T_H} - \beta M}.\nonumber
\eea
We see that for a high enough temperature, this Partition Function
is divergent and not well-defined.

In the case of QCD, this is not an effect one would
actually measure, because hadrons
have a natural width and at energies high enough, they
would start decaying into each other, there would also be
pair creation, etc.
But in a `truly confining' QFT,
like for example Yang-Mills or minimally SUSY Yang-Mills, we would
have --- if we take the large $N_c$ limit --- infinite hadrons (glueballs)
that will be very narrow and they will present the
Hagedorn behaviour above (although generically, the deconfinement occurs before the Hagedorn
transition).
Just like it happens in a typical theory of strings.

The authors of \cite{Klebanov:2007ws}, have argued that
something similar happens to the
EE of a confining theory. An intuitive reasoning lead them to write
the EE as
\bea
S\sim \int DM \,M^{2b} e^{+\beta_H M- 2ML_{EE}},\nonumber
\eea
where they used that for scalar non-interacting degrees of freedom ---
our `large $N_c$'
glueballs for example --- the EE goes like $e^{- 2ML_{EE}}$, where $L_{EE}$
is the separation between the entangled regions.
So it was argued in \cite{Klebanov:2007ws}, that `truly confining'
QFTs should present a phase transition in the EE, when $2L>\beta_H$.
This phase transition is {\it phenomenologically} similar to the
confinement-deconfinement one. So, it is expected that
for a given $L_{EE,crit}$ the EE behaves
in a `constant' manner --- that is $N^0$ --- for  $L>L_{EE,crit}$,
 or grows very
rapidly, like $N^2$, for
for
$L<L_{crit}$
respectively. We will see this in various examples below.

One may argue that in some QFTs the reasoning above
might fail, if for example the density of states grows differently
or if the behavior of the EE for many scalars changes
and hence the argument breaks down. We will see that a way of having some
analytic control
over the problem is to first introduce a cutoff at high energy,
find the phase transition exploring
regions close to the cutoff and then take the cutoff to infinity.
These limits need not commute. A similar situation but
in a different context was encountered in \cite{Faedo:2014naa} and
we will find this to be the case
in theories with non-local UV-behavior.
Similarly
we will observe that a local and
honest UV-completion of non-local theories plays the
same role as the regularisation with a cutoff.

The material of this paper is organised as follows:
In Section \ref{relationsWEE}, we will provide
a criteria (valid beyond the examples presented in this paper) to decide if
the EE
can present a phase transition. The criteria will be carefully illustrated
using confining models based on D$p$ branes compactified on circles.
Specifically, the understanding of
the case of D5 branes on $S^1$,
where the phase transition is curiously absent,
will be the subject of Section \ref{newmaterial}.
We will argue that a naive holographic calculation of the EE misses
some key-configurations. We also
motivate these configurations by introducing the
UV-cutoff already mentioned. In Section \ref{sectionabsence},
we make explicit the analysis above, on a four dimensional model of
confining QFT, but with a non-local UV --- the need for a cutoff appears
here also.
In Section \ref{recoverbb},
we will see that a suitable UV-completed and local QFT version of the
confining model above,
has an EE that behaves
like the one in the cutoff-QFT does. This emphasises the point that the
cutoff is capturing
`real Physics'.

The paper is of high technical content, as reflected in various appendixes,
where interesting material was relegated to ease the reading
of the main body of this work.

\section{Entanglement Entropy and Wilson Loops as Probes of Confinement}\label{relationsWEE}

\subsection{Review - Entanglement Entropy in Confining Backgrounds}

We start with a brief summary of the results of Klebanov, Kutasov and
Murugan (KKM) \cite{Klebanov:2007ws}, who studied the Entanglement Entropy (EE) in gravity duals of confining large $N_c$ gauge theories.
KKM have generalized the Ryu-Takayanagi conjecture to non-conformal theories, and suggested that the EE in these cases is given by
\begin{equation}\label{conjecture}
  S=\frac{1}{G_{N}^{(10)}}\int_{\gamma} d^8 \sigma e^{-2\phi} \sqrt{G_{ind}^{(8)}}
\end{equation}
where $G_{N}^{(10)}$ is the $10$-dimensional Newton constant and $G_{ind}^{(8)}$ is the induced string frame metric on $\gamma$.
The EE is obtained by minimizing the action \eqref{conjecture} over all surfaces that approach the boundary of the entangling surface.
KKM have considered, as the entangling surface, a strip of length $L$. In this case, they found that there are two local minima of the action \eqref{conjecture} for a given $L$.
The first is a disconnected surface, which consists of two cigars which are separated by a distance $L$. The second is a connected surface, in which the two cigars are connected by a tube whose width depends on $L$.

The gravitational background in the string frame is of the form
\begin{equation}\label{background}
  ds^2 = \alpha (\rho) \left[  \beta(\rho) d\rho^2 +dx^{\mu} dx_{\mu}  \right]  +g_{ij} d\theta^i d\theta^j
\end{equation}
where $x^{\mu}$ $\left(\mu=0,1,\dots,d \right)$ parameterize $\mathbb{R}^{d+1}$, $\rho$ is the holographic radial coordinate
\begin{equation}
  \rhol<\rho<\infty
\end{equation}
($\rhol$ can be zero in some cases) and $\theta^{i}$ $\left(  i=d+2,\dots,9 \right)$ are the $8-d$ internal directions. There is also
a dilaton field that we denote with $\phi$. Some
RR and NS fluxes complete the background, but they
will not be relevant to our analysis.
The volume of the internal manifold
(described by the $\vec{\theta}$ coordinates)
is $V_{int}=\int d\vec{\theta}\sqrt{\det[g_{ij}]}$. We also define the quantity
\begin{equation}\label{H}
  H(\rho) = e^{-4\phi} V_{int}^2 \alpha ^d
\end{equation}
The functions $H(\rho)$ and $\beta(\rho)$ will play an essential role in the following and all along the rest of this paper.
We will mention few important properties of these functions.
KKM have argued that in confining backgrounds $H(\rho)$ is typically a monotonically \emph{increasing} function while $\beta(\rho)$ is typically a monotonically \emph{decreasing} function.
Since $H(\rho)$ includes a factor of the volume of the internal manifold, it typically shrinks to zero size at $\rho=\rhol$, in agreement with the vanishing of the central charge at zero
energies. On the other hand, $\beta(\rho)$ is less restricted and it can either diverge or approach a finite value at $\rho=\rhol$.

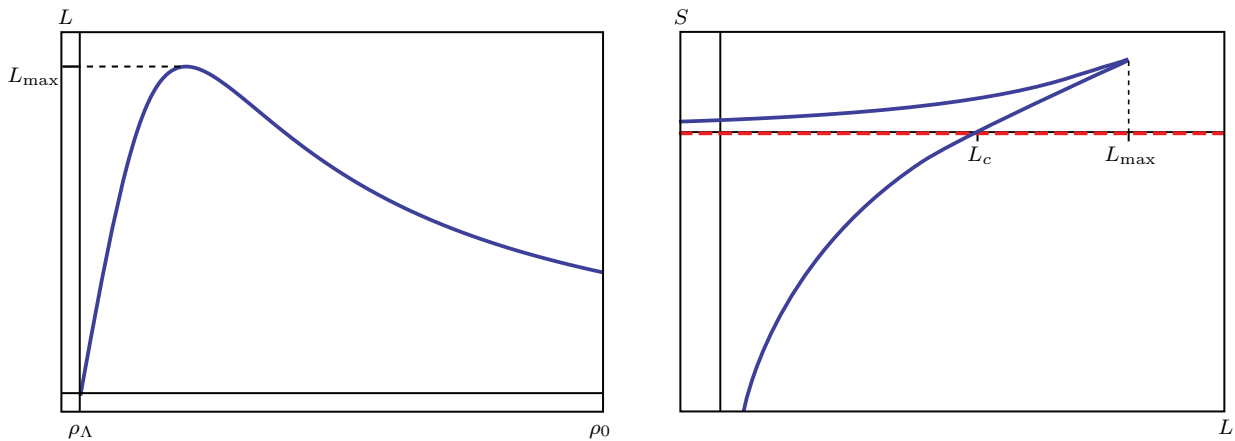
\begin{figure}[ht]
\begin{center}
\begin{picture}(220,170)
\put(-110,10){\includegraphics[height=5.15cm]{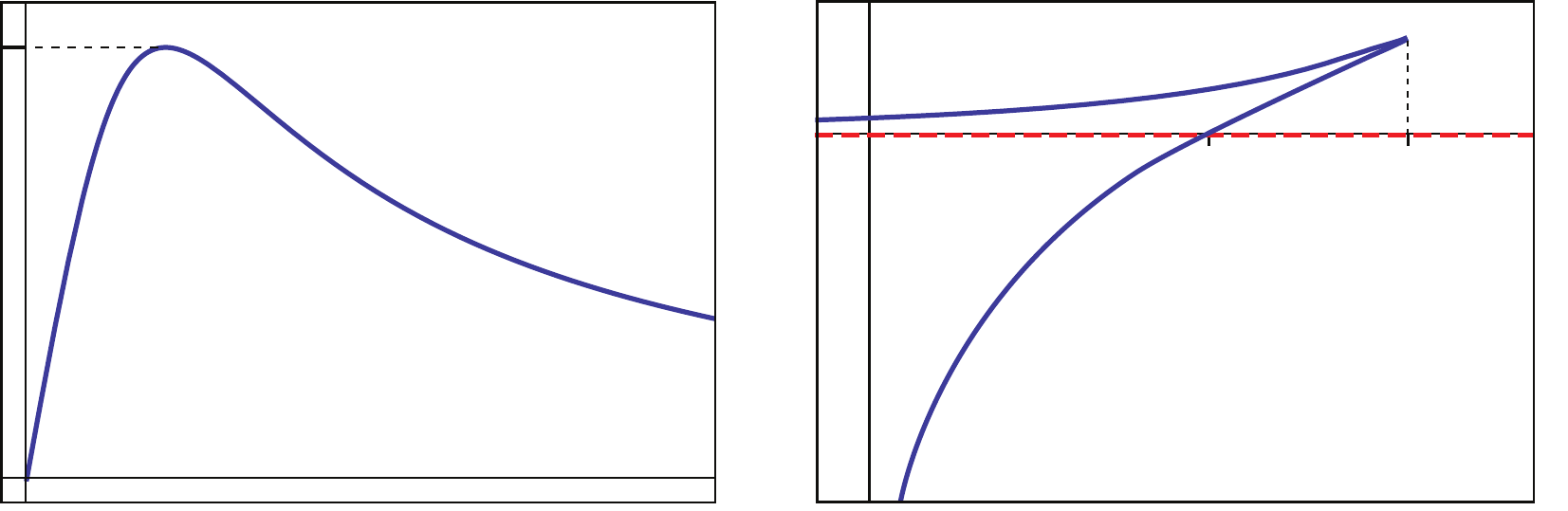}}
\put(-111,159){{\scriptsize{$L$}}}
\put(-107,4){{\scriptsize{$\rhol$}}}
\put(90,4){{\scriptsize{$\rho_{0}$}}}
\put(-130,137){{\scriptsize{$L_{\text{max}}$}}}
\put(122,159){{\scriptsize{$S$}}}
\put(328,4){{\scriptsize{$L$}}}
\put(233,107){{\scriptsize{$L_{c}$}}}
\put(285,107){{\scriptsize{$L_{\text{max}}$}}}
\end{picture}
\caption{The phase diagram for the entanglement entropy in confining theories. On the left, the length of the connected solution as a function of the minimal radial position in the bulk $L(\rhoz)$, which is a non-monotonic function in confining theories. On the right, the entanglement entropy of the strip as a function of its length. The solid blue line represent the connected solution while the dashed red line is the disconnected solution. At the point $L=L_c$ there is a first order phase transition between the two solutions. This type of first-order phase transition behavior is called
the ``butterfly
shape'' in the bibliography.}
\label{KKM}
\end{center}
\end{figure}

Denoting the minimal value of $\rho$ along the connected surface in the bulk by $\rhoz$, its EE is given by
\begin{equation}\label{EEstripC}
  S_{C}(\rhoz) = \frac{V_{d-1}}{2G_N^{(10)}}  \int _{\rhoz}^{\infty} d\rho \sqrt{\frac{\beta(\rho)H(\rho)}{1-\frac{H(\rhoz)}{H(\rho)}}}
\end{equation}
The length of the line segment for the connected solution as a function of $\rhoz$ is
\begin{equation}\label{length}
  L (\rhoz) = 2   \int_{\rhoz}^{\infty} d\rho \sqrt {   \frac{  \beta(\rho) }{   \frac{  H(\rho)  }{  H(\rhoz)  }  -1 }   }.
\end{equation}
On the other hand, the EE of the disconnected solution does not depend on $\rhoz$ and is given by
\begin{equation}\label{EEstripD}
  S_{D}(\rhoz) = \frac{V_{d-1}}{2G_N^{(10)}}  \int _{\rhol}^{\infty} d\rho \sqrt{\beta(\rho)H(\rho)}
\end{equation}
The EE is in general UV divergent, but the difference between the EE of the connected and disconnected phases is finite
\begin{equation}\label{difS}
  \frac{2G_N^{(10)}}{V_{d-1}}S(\rhoz)\equiv \frac{2G_N^{(10)}}{V_{d-1}}(S_C-S_D) =
    \int _{\rhoz}^{\infty} d\rho   \sqrt{\frac{\beta(\rho)H(\rho)}{1-\frac{H(\rhoz)}{H(\rho)}}}
  -  \int _{\rhol}^{\infty} d\rho \sqrt{\beta(\rho)H(\rho)}
\end{equation}
Depending on the value of $L$ (or alternatively of $\rhoz$) $S$ would either be positive or negative. In the first case the true solution will be
a disconnected surface, while in the later case the connected solution will be the true one. The phase transition between the two solutions is a characteristic of confining theories, and is described in Figure \ref{KKM}.

The connected solution exists only in a finite range of lengths $0<L<L_{\text{max}}$.
In this range there are two possible values for the connected solution, corresponding to the two branches in Figure \ref{KKM}.
The upper branch is an unstable solution.
This double valuedness, which is called
the ``butterfly shape'', corresponds to the double valuedness in the graph of $L(\rhoz)$.
As a result of the double valuedness, there is a first order phase transition at the point $L=L_{c}$ between the connected and the disconnected solutions.
For this reason KKM have argued that 
a signal for a phase transition, and therefore also for confinement, is the 
non-monotonicity of the function $L(\rhoz)$.
Indeed, as we will show later in different examples, every peak in
$L(\rhoz)$ corresponds to a possible
phase transition in the entanglement entropy $S(L)$.


\subsection{Review - Wilson Loops in Confining Backgrounds}

In this subsection we will review the results of \cite{Kinar:1998vq} for the rectangular Wilson loop in confining backgrounds. We define the
following function
\begin{equation}\label{g}
  g(\rho) \equiv  \alpha\sqrt{\beta}.
\end{equation}
The authors of \cite{Kinar:1998vq} then found that the regularized energy of the Wilson loop is given by\footnote{In the notations of \cite{Kinar:1998vq} $\alpha=f$.}
\begin{equation}\label{energyWL}
  E_{WL} (\rhoz) = 2 \int _{\rhoz}^{\infty} \frac{g(\rho)}{\alpha(\rho)} \sqrt{\frac{\alpha^2(\rho)}{1-\frac{\alpha^2(\rhoz)}{\alpha^2(\rho)}}}-2\int _{\rhol}^{\infty}d\rho g(\rho),
\end{equation}
where $\rhoz$ is the minimal radial position of the dual string in the bulk.
The first term in the equation above is the bare energy of the Wilson loop while the second term is the energy of two straight string segments (stretched from the horizon to the boundary) which is subtracted from the bare expression in order to regularize its divergence.
The length of the Wilson loop as a function of $\rhoz$ is given by
\begin{equation}\label{lengthWL}
  L_{WL} (\rhoz) = 2 \int _{\rhoz}^{\infty} \frac{g(\rho)}{\alpha(\rho)} \frac{1}{\sqrt{\frac{\alpha^2(\rho)}{\alpha^2(\rhoz)}-1}}
\end{equation}

The authors of \cite{Kinar:1998vq}
found that the background admits linear confinement if one of the following conditions is satisfied
\begin{equation}\label{WLconditions}
\begin{aligned}1.\quad & \alpha\left(\rho\right)\:\text{has a minimum}\\
2.\quad & g\left(\rho\right)\:\text{diverges}
\end{aligned}
\end{equation}
and the corresponding string tension is $\alpha(\rhol)\neq0$, where $\rhol$ is either the point where $\alpha$ minimizes or where $g$ diverges.
In other words, if one of the two conditions above is satisfied, the potential at long distances will behave linearly in $L$
\begin{equation}
  E_{WL}\left(L_{WL}\right) = \alpha(\rhol) L_{WL}+\dots
\end{equation}
where $``\dots"$ stand for subleading corrections in $\frac{1}{L_{WL}}$.

\begin{figure}[ht]
\begin{center}
\begin{picture}(220,170)
\put(-110,10){\includegraphics[height=5.15cm]{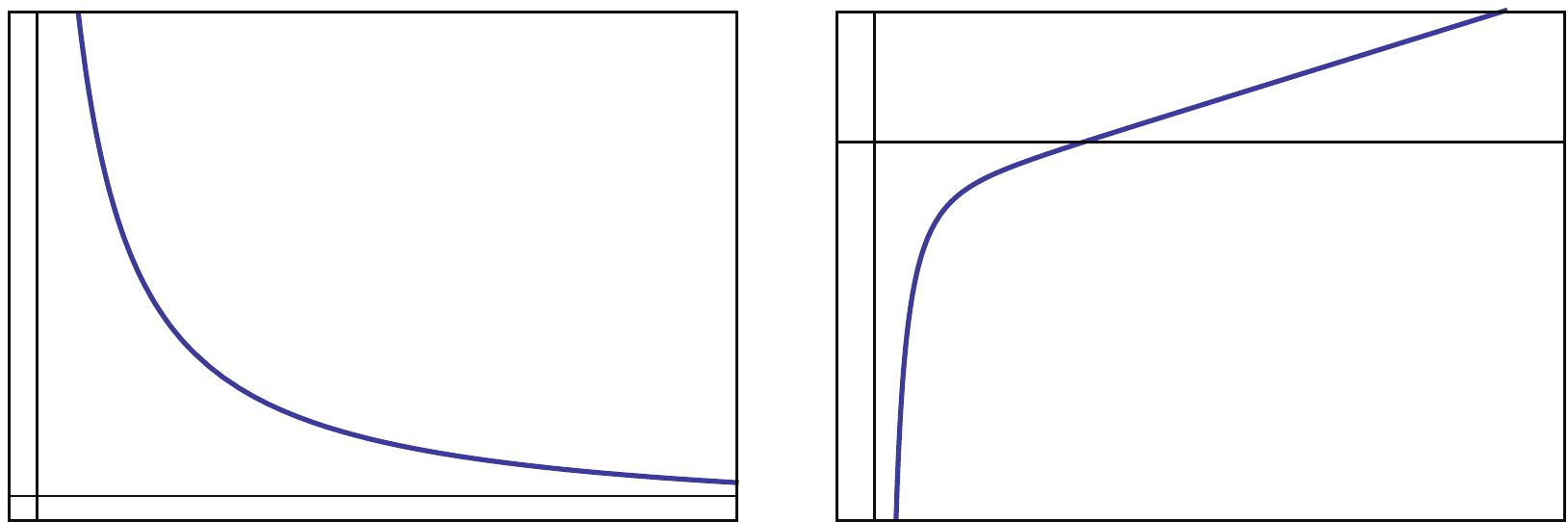}}
\put(-110,155){{\scriptsize{$L$}}}
\put(-107,2){{\scriptsize{$\rhol$}}}
\put(90,2){{\scriptsize{$\rho_{0}$}}}
\put(122,155){{\scriptsize{$E$}}}
\put(325,2){{\scriptsize{$L$}}}
\end{picture}
\caption{On the left, the length of the Wilson loop as a function of the minimal radial position in the bulk, which is a monotonically decreasing function. On the right, the energy of the Wilson loop as a function of its length. At long distances the energy is linear in length, which correspond to linear confinement.}
\label{WL}
\end{center}
\end{figure}
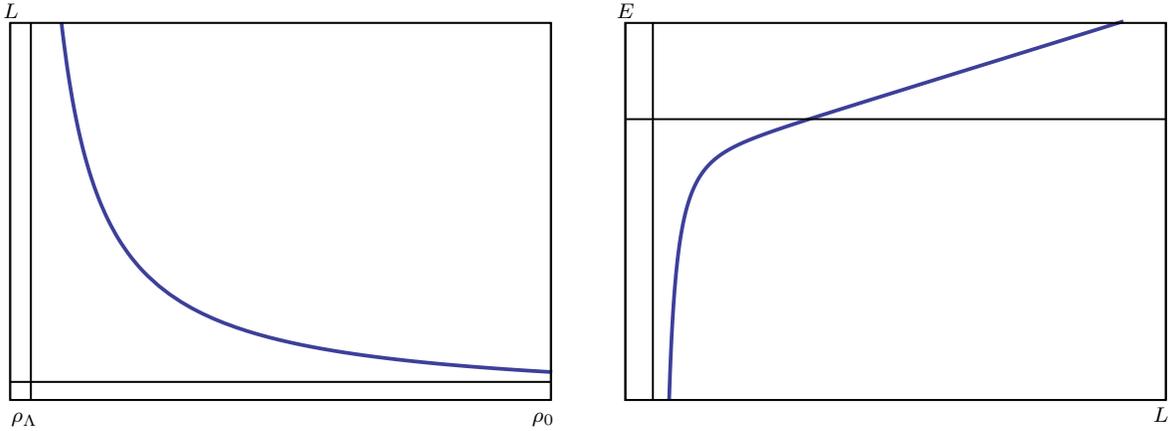

The authors of \cite{Kinar:1998vq} proved that in confining backgrounds $L_{WL}(\rhoz)$ is a monotonically decreasing function. This behavior corresponds to the monotonicity of $E_{WL}\left(L_{WL}\right)$, which always increase with the length.
This features, as well as the linearity of $E_{WL}\left(L_{WL}\right)$ at long distances, is described in figure \ref{WL}.

\subsection{Similarities and Differences}

So far we have reviewed well-known results for the entanglement entropy and the Wilson loop in confining backgrounds.
\emph{A priori}, these two quantities are not related to each other, and indeed, they show very different behaviors - the first presents a phase transition in the form of a ``butterfly" shape, while the later is monotonic.
However, the fact that both are probes of confinement suggests that there maybe some deeper relation between them.
Moreover, as we will discuss below, the functional form of both quantities is very similar.

The length of the entangling strip \eqref{length} and the length of the Wilson loop \eqref{lengthWL} can both be written in the form
\begin{equation}\label{ell}
  L(\rhoz) = 2\int_{\rhoz}^{\infty} d\rho \sqrt{\frac{\beta(\rho)}{\frac{M(\rho)}{M(\rhoz)}-1}}
\end{equation}
where $M(\rho)$ is different for the two cases and is given by
\begin{eqnarray}
   M_{EE}(\rho) &=& H(\rho) \\
   M_{WL}(\rho) &=& \alpha^2(\rho)
\end{eqnarray}
The entropy of the strip \eqref{difS} and the energy of the Wilson loop \eqref{energyWL} can also be written in a similar way as
\begin{equation}\label{S}
  \left(
  \begin{array}{c}
     \frac{4G_N^{(10)}}{V_{d-1}}S \\
      E_{WL}
  \end{array}
  \right)
   = 2\int _{\rhoz}^{\infty} d\rho   \sqrt{\frac{\beta(\rho)M(\rho)}{1-\frac{M(\rhoz)}{M(\rho)}}}
  -  2\int _{\rhol}^{\infty} d\rho \sqrt{\beta(\rho)M(\rho)}
\end{equation}
The expression \eqref{S} can also be written in the following form
\begin{eqnarray}\label{Sl}
  \left(
  \begin{array}{c}
     \frac{4G_N^{(10)}}{V_{d-1}}S \\
      E_{WL}
  \end{array}
  \right)
   &=& \sqrt{M(\rhoz)} L(\rhoz) -2 K(\rhoz)\\
   K(\rhoz) &\equiv&
   \int _{\rhol}^{\infty} d\rho \sqrt{\beta(\rho)M(\rho)} -   \int _{\rhoz}^{\infty} d\rho   \sqrt{\beta(\rho)\left(M(\rho)-M(\rhoz)\right)}
\end{eqnarray}
The form \eqref{Sl} emphasizes the linear nature of $E_{WL}$ at long distances.
In both cases, $M(\rho)$ is a monotonically increasing function.
Therefore, the functional form of the EE and the Wilson loop is very similar.
Of course, these mathematical similarities in equations \eqref{ell} and \eqref{S}
are based on the fact that both observables are
solutions to a minimization problem.

The question arises as to what is the difference between them.
We claim that the qualitative
difference between these two observables (at least in the case of confining QFTs)
is due to the behaviour of the function $M(\rho)$ close to $\rho=\rhol$.
For both cases $M(\rho)$ is a monotonically increasing function, but the behavior close to $\rho=\rhol$ is different.
For the entanglement entropy $M(\rho)=H(\rho)$
shrinks to zero close to $\rhol$, since $H(\rho)$
includes a factor of the internal volume --- see \eqref{H} ---
which always goes to zero at the end of the geometry\footnote{The vanishing of the internal volume is in agreement with the vanishing
of the central charge at zero energies. This is characteristic of confining field theories.}.
On the other hand, for the Wilson loop in confining backgrounds $M(\rhol)=\alpha^2(\rhol)\neq0$ since this quantity
is related to the confining-string tension.
Therefore $M(\rhol)$ behaves very differently for the two observables, when calculated
in a  generic confining background. This is the source for the qualitative difference
between these two quantities.

To be concrete, let us focus on D$p$ branes compactified on a circle.
These backgrounds are dual  to confining field
theories in $p$ space-time dimension.
The background
metric and dilaton are a generalisation of
those written by
Witten, as a dual to a Yang-Mills-like
four dimensional QFT \cite{Itzhaki:1998dd,Witten:1998zw} (with $\alpha'=g_s=1$),
\bea
& & \label{DbraneMetricx} ds^2 = \left(\frac{\rho}{R}\right)^{\frac{7-p}{2}}
            \left[         \left(\frac{R}{\rho}\right)^{7-p}
\frac{d\rho^2}{h(\rho)} +  dx_{1,p-1}^2   \right]
        +  h(\rho)      \left( \frac{\rho}{R}\right)^{\frac{7-p}{2}}
d\varphi_{c}^2
       + \left( \frac{\rho}{R}  \right)^{\frac{p-3}{2}}
R^2 d\Omega_{8-p}^2,\\
& &    h(\rho) = 1-\left(\frac{\rhol}{\rho}\right)^{7-p},
\;\;  e^{-4\phi} = \left(  \frac{\rho}{R} \right) ^{(3-p)(7-p)},\;\;
\alpha^2(\rho) = \left( \frac{\rho}{R} \right)^{7-p},\;\;
g^2(\rho) =\frac{1}{1-\left(\frac{\rhol}{\rho}\right)^{7-p}}.\nonumber
\eea
which implies that,
\begin{eqnarray}
\label{MEE}
M_{EE}(\rho) = (S_{7-p}\times2\pi R_c)^2 R^{7-p} \rho ^{9-p} h(\rho), \;\;
M_{WL}(\rho) = \left(\frac{\rho}{R}\right)^{7-p}.
\end{eqnarray}
$R_c$ is the radius of the compact cycle (see for example \cite{Kol:2010fq}),
\begin{equation}
  R_c = \frac{2}{7-p}\left( \frac{R}{\rhol} \right)^{\frac{7-p}{2}} \rhol
\end{equation}
and $S_{n-1}=\frac{2\pi^{\frac{n}{2}}}{\Gamma\left(\frac{n}{2}\right)}$ is the surface area of the $n$-sphere.
Therefore $M_{WL}(\rhol)\sim \rhol^{7-p}\neq0$, while $M_{EE}(\rhol)=0$.
This difference of behavior in the function $M(\rho)$,
creates a "butterfly" shape (and a phase transition) in $S(L)$,
whereas it gives place to a monotonic behavior and a long-separation
linear law in $E_{WL}(L_{WL})$.


\subsection{Sufficient Conditions for Phase Transitions}
The conditions a background must satisfy so that the Wilson loop shows a
confining behaviour \eqref{WLconditions}, were derived in \cite{Kinar:1998vq}.
The difference in behavior, of the Wilson loop and the Entanglement Entropy as
probes of the phenomenon of confinement, leads to the following question:
\begin{itemize}
\item{{\it Question:} What are the conditions on the background to present a phase transition in the EE?}
\end{itemize}
Using a similar logic to that of \cite{Kinar:1998vq},
in the following we will derive sufficient conditions on the
background to present a phase transition in the Entanglement Entropy.
More explicitly, we will derive the conditions on
the background such that $L(\rho_0)$ (the length of the entangled
strip as a function of the minimal radial position)
will increase for $\rhoz$ close to $\rhol$ (the IR of the dual QFT)
and decrease for asymptotically large value of $\rhoz$ (the UV of the dual QFT).
Hence, the quantity $L$ will present (at least) a maximum and the required double valuedness needed for a
phase transition.

We start by deriving the conditions on the background
such that $L(\rhoz)$ is an increasing function close to $\rhoz=\rhol$.
Let us assume that the functions $H(\rho)$ and $\beta(\rho)$ have the following expansions around $\rhoz=\rhol$
\begin{eqnarray}
  \label{HIR} H(\rho) = h_r (\rho-\rhol)^{r} + \mathcal{O}(\rho-\rhol)^{r+1},\;\;\;
\beta(\rho) &=& \beta_t   (\rho-\rhol)^{-t}  + \mathcal{O}(\rho-\rhol)^{-t+1}
\end{eqnarray}
with $r,t>0$. Then,
the integrand corresponding to $L$ in \eqref{length}
is divergent close to $\rho=\rhol$, and therefore the
integral gets most of its contribution from this region.
That allows us to approximate the integrand using eq.\eqref{HIR},
\begin{eqnarray}
\label{sasaza}& &   \lim_{\rhoz\rightarrow\rhol}L(\rhoz) = 2\sqrt{\beta_t} \int_{\rhoz}^{\infty }
d\rho (\rho-\rhol)^{-\frac{t}{2}}\left[ \left(\frac{\rho-\rhol}{\rhoz-\rhol}\right)^{r}-1  \right]^{-\frac{1}{2}}\\
& &= (\rhoz-\rhol)^{1-\frac{t}{2}}  \frac{2\sqrt{\beta_t}}{r} \int_1^{\infty} dz  (z-1)^{-\frac{1}{2}}z^{-\frac{t+2r-2}{2r}} =
(\rhoz-\rhol)^{1-\frac{t}{2}}   \frac{2\sqrt{\pi\beta_t}\Gamma\left(\frac{t+r-2}{2r}\right)}{r\Gamma\left(\frac{t+2r-2}{2r}\right)}.\nonumber
\end{eqnarray}
We have changed variables to $z\equiv\left(\frac{\rho-\rhol}{\rhoz-\rhol}\right)^r$.
We find that $L(\rhoz)$ is monotonically increasing when
\begin{equation}\label{t}
t<2
\end{equation}
which means that $\beta(\rho)$ should not diverge faster
than $\frac{1}{(\rho-\rhol)^2}$ close to $\rhol$.

Next, we derive the conditions on the background such that $L(\rhoz)$
is a decreasing function at asymptotically large value of $\rho$.
Close to the boundary $\rho=\infty$ we can expand
\begin{eqnarray}\label{boundaryexpansion}
  H(\rho) = h_k \rho^k + \mathcal{O}(\rho^{k-1}) \;\;\;\; \beta(\rho) = \beta_j \rho^{-j} + \mathcal{O}(\rho^{-j+1}).
\label{mamana}\end{eqnarray}
Plugging these expansions in \eqref{length} we find,
\begin{equation}
L (\rhoz) = 2\sqrt{\beta_j} \int _{\rhoz}^{\infty} d\rho \rho^{-\frac{j}{2}} \left[  \left( \frac{\rho}{\rhoz} \right)^{k} -1  \right] ^{-\frac{1}{2}}.
\end{equation}
Changing variables to $z\equiv \frac{\rho}{\rhoz}$ we find the asymptotic behavior of $L(\rhoz)$ near the boundary
\begin{eqnarray}
L(\rhoz) = 2\sqrt{\beta_j} \rhoz^{1-\frac{j}{2}} \int _{1}^{\infty} dz  \frac{z^{-\frac{j}{2}}}{\sqrt{z^{k} -1}} =
\frac{2\sqrt{\pi \beta_j}\Gamma\left(\frac{k+j-2}{2k}\right)}{k\Gamma\left(\frac{2k+j-2}{2k}\right)}   \rhoz^{1-\frac{j}{2}} .
\end{eqnarray}
We see that for
\begin{equation}\label{j}
j>2
\end{equation}
the length $L$ will go to zero as $\rhoz\to\infty$.
There is a maximum somewhere in the middle, hence a double
valuedness for $\rho_0(L)$ and the possibility of a
phase transition in the quantity $S(\rho_0[L])$.
On the other hand, for $j\leq2$, the quantity $L$
will either diverge, or saturate at a constant value, in the same limit
and in this case we do not expect to have a phase transition.

Let us present some examples
of absence or presence of phase transitions in the EE, in agreement with the criteria of this section.
\subsection{Examples of the Criteria for Phase Transitions}
As anticipated, we will study here different non-confining and confining
models to illustrate our criteria above. $AdS_5\times S^5$, D$p$
branes compactified on a circle, hard and soft walls and the
Klebanov-Strassler model \cite{Klebanov:2000hb}
will serve as confirmation of our treatment.
\subsubsection{$AdS_5\times S^5$}
\begin{figure}[h]
\begin{center}
\begin{picture}(220,160)
\put(-125,0){\includegraphics[height=5.35cm]{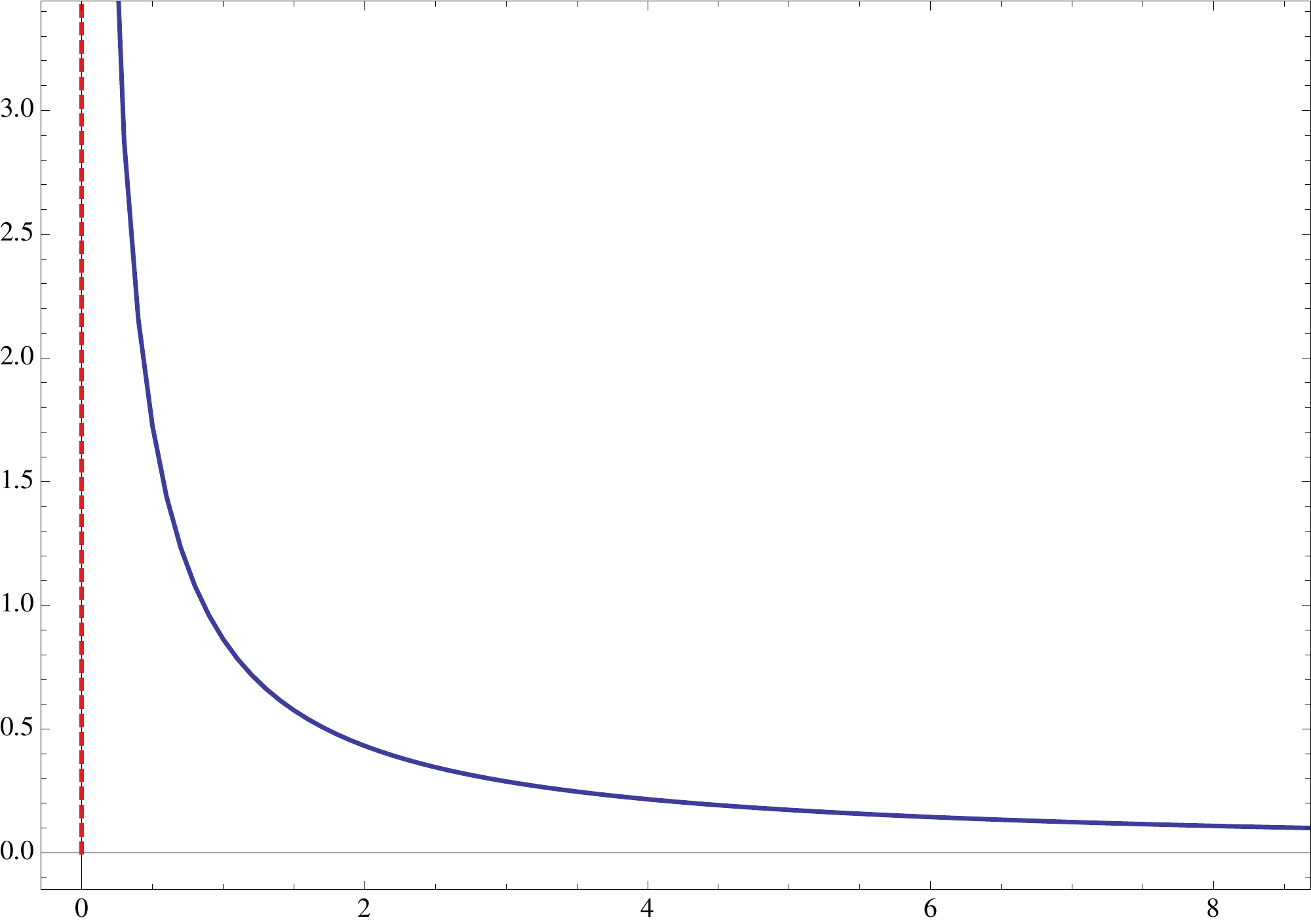}}
\put(-119,155){{\scriptsize{$L$}}}
\put(85,-3){{\scriptsize{$\r_{0}$}}}
\put(120,0){\includegraphics[height=5.35cm]{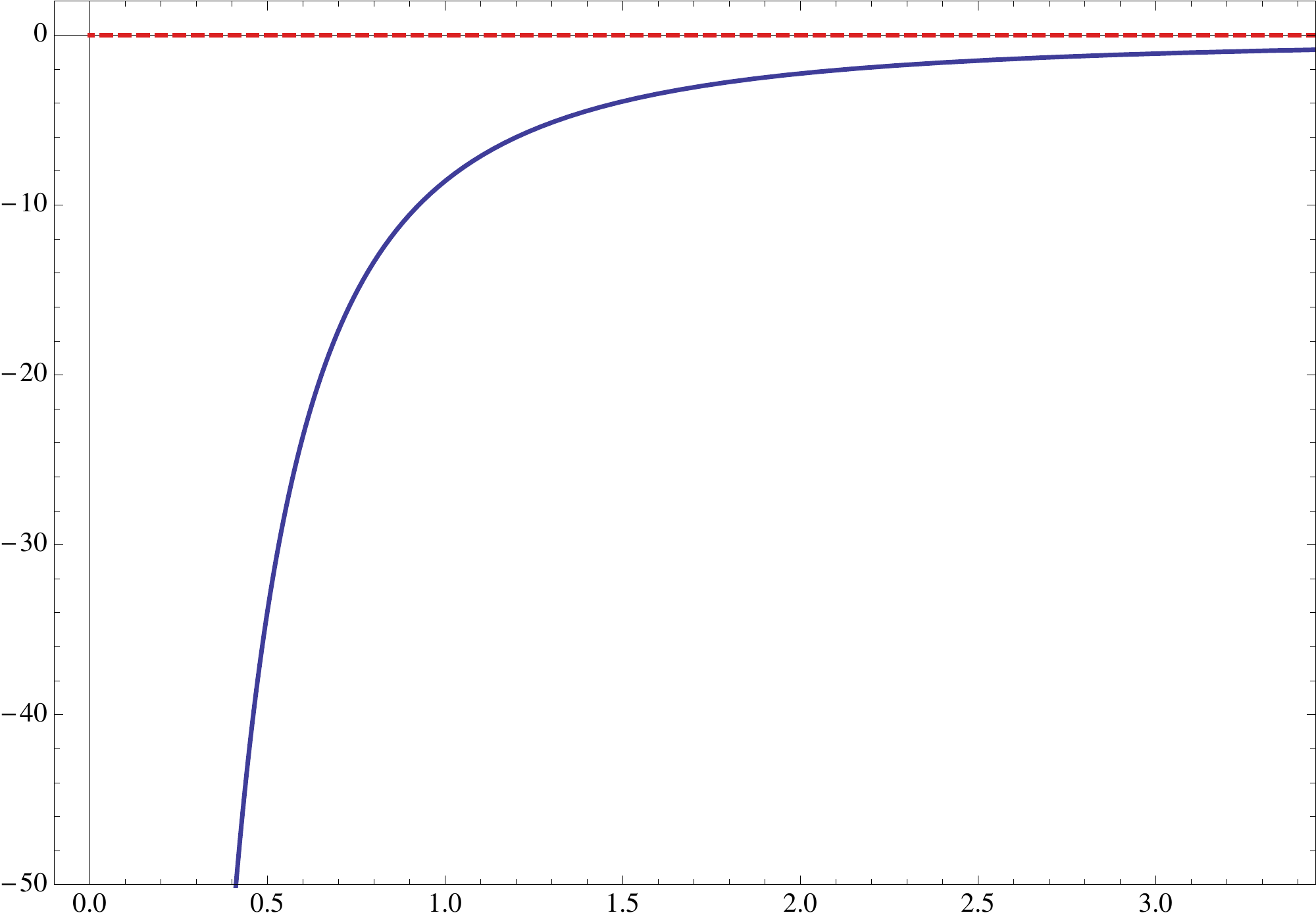}}
\put(128,155){{\scriptsize{$S$}}}
\put(335,-3){{\scriptsize{$L$}}}
\end{picture}
\caption{The case of $AdS_5\times S^5$ --- Here we plot $L(\rho_0)$ and $S(L)$.}
\label{FAdS}
\end{center}
\end{figure}
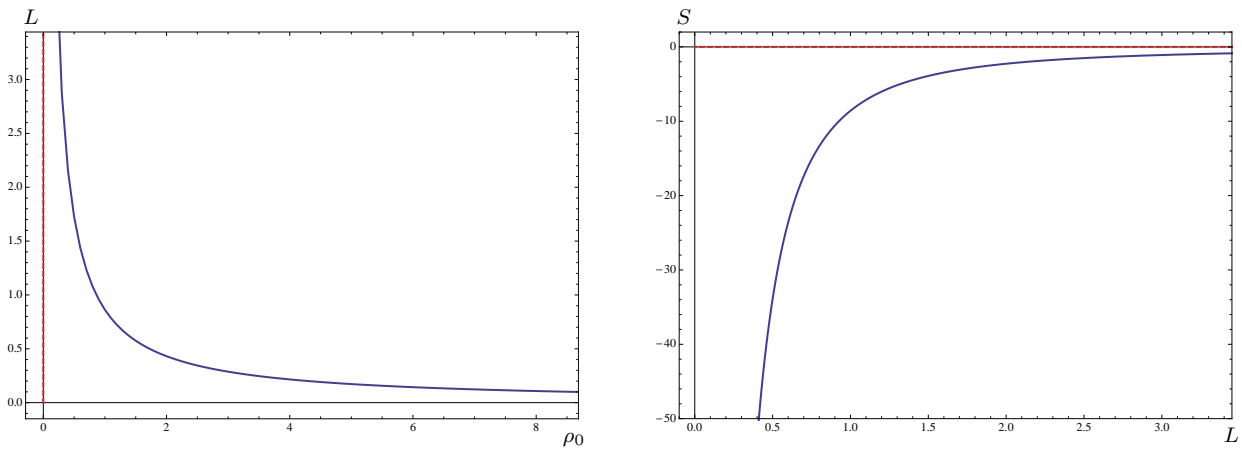

As a first example we discuss the
EE of $N=4$ Super-Yang-Mills, to demonstrate that it does not present a
phase transition.
The metric and other
relevant functions for the case of $AdS_5\times S^5$ are given by
\begin{equation}\label{metricAdS}
  ds^2_{AdS} = \frac{R^2}{\rho^2} d\rho^2  + \frac{\rho^2}{R^2} dx_{1,3}^2 + R^2 d\Omega_5^2,\;\;\;  \beta(\rho) = \frac{R^4}{\rho^4},\;\;\\
  H(\rho) = \left(\frac{8\pi^2}{3}\right)^2 R^4 \rho ^6.
\end{equation}
Notice that in this case $t=4$ and the condition \eqref{t} is not attained.
The length of the strip $L(\r_0)$
as well as the $S$ can be exactly computed and are plotted in Figure \ref{FAdS}.
We see that $L(\rhoz)$ is a monotonically decreasing
function which diverges at the origin and goes to zero at the boundary.
The EE shows two possible phases: in the first, it monotonically grows with $L$
while in the second (the disconnected phase) it is constant.
The second phase is not favoured,
since the EE in this phase is always larger than the EE in the `connected' phase.
Therefore there is no phase transition, as appropriate for a conformal field theory.
Incidentally, also notice that the concavity of the $S(L)$ is such that the
correct
condition
\cite{Bachas:1985xs},\cite{Brandhuber:1999jr},
\beq
\frac{d^2 S}{dL^2}<0,
\label{concavityxx}
\eeq
is achieved.

\subsubsection{D$p$ brane on a Circle}
Next we consider the background generated by D$p$ brane compactified on a circle described in eq.(\ref{DbraneMetricx}).
The functions $\alpha(\rho)$ and $\beta(\rho)$ are,
\begin{eqnarray}
  \alpha(\rho) =  \left(\frac{\rho}{R}\right)^{\frac{7-p}{2}}, \;\;\;
  \beta(\rho) =\frac{1}{1-\left(\frac{\rhol}{\rho}\right)^{7-p}}\left(\frac{R}{\rho}\right)^{7-p}
\end{eqnarray}
Close to the horizon $\rho=\rhol$, we can expand
\begin{equation}
  \beta(\rho)= \frac{\rhol}{7-p}\left( \frac{R}{\rhol}  \right)^{7-p}  \frac{1}{\rho-\rhol}+\dots
\end{equation}
where $"\dots"$ stands for subleading (finite) terms.

Comparing with \eqref{HIR} we find that in this case $t=1$.
This means that for this background, and for any value of $p$, the condition \eqref{t} will be satisfied and $L(\rhoz)$ will always go to zero at the horizon.
Close to the boundary $\beta(\rhoz) \sim \rhoz^{-(7-p)}$ and therefore in
this case, comparing with eq.(\ref{mamana}), we have $j=7-p$.
Hence, for $p\leq4$, the condition \eqref{j} will be satisfied and $L(\rhoz)$ will decrease to zero close to the boundary.
However, for $p=5$, $L(\rhoz)$ will saturate to a finite value and for $p\geq6$ it will increase towards the boundary.

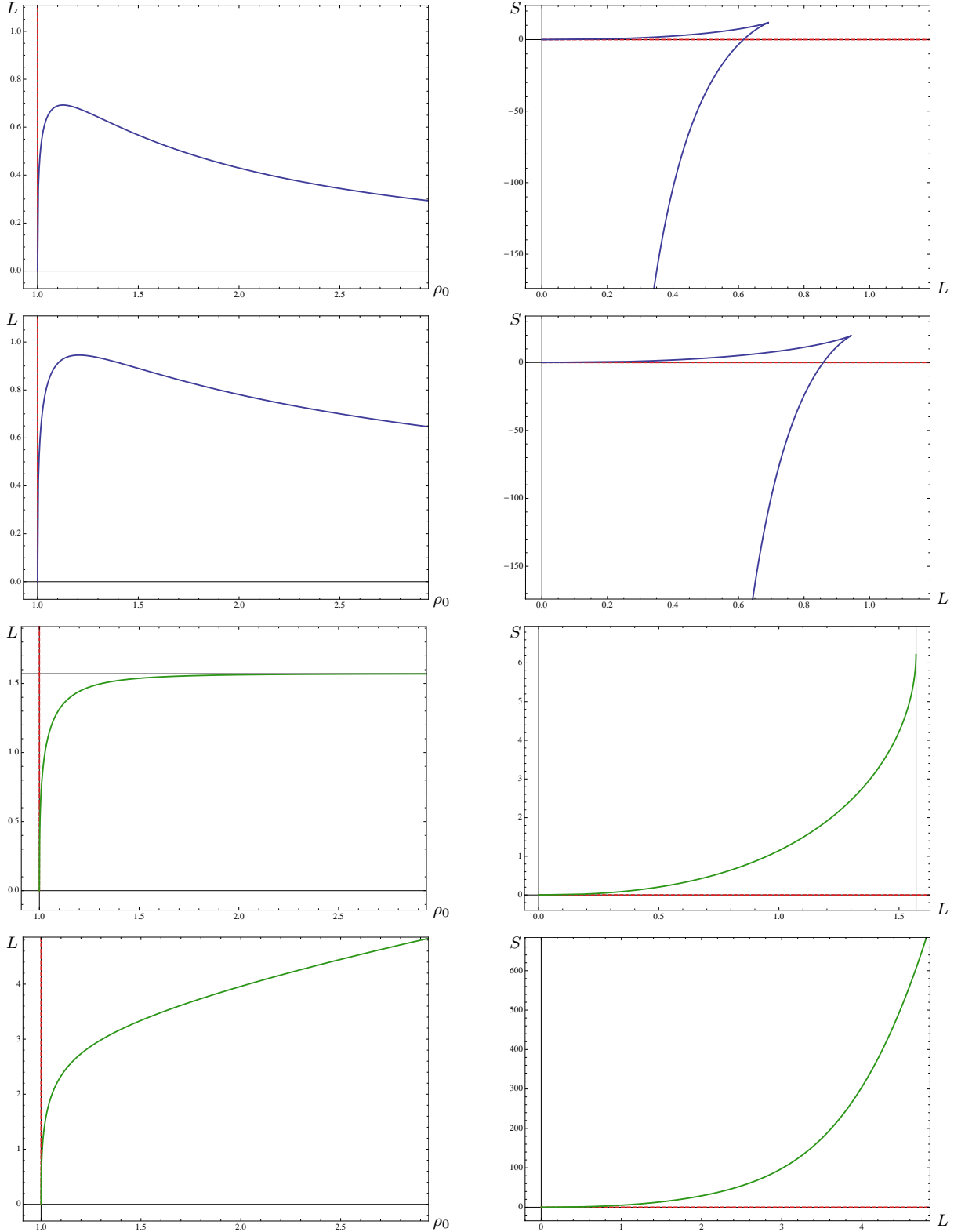
\begin{figure}
\begin{center}
\begin{picture}(220,560)
\put(-115,450){\includegraphics[height=5cm]{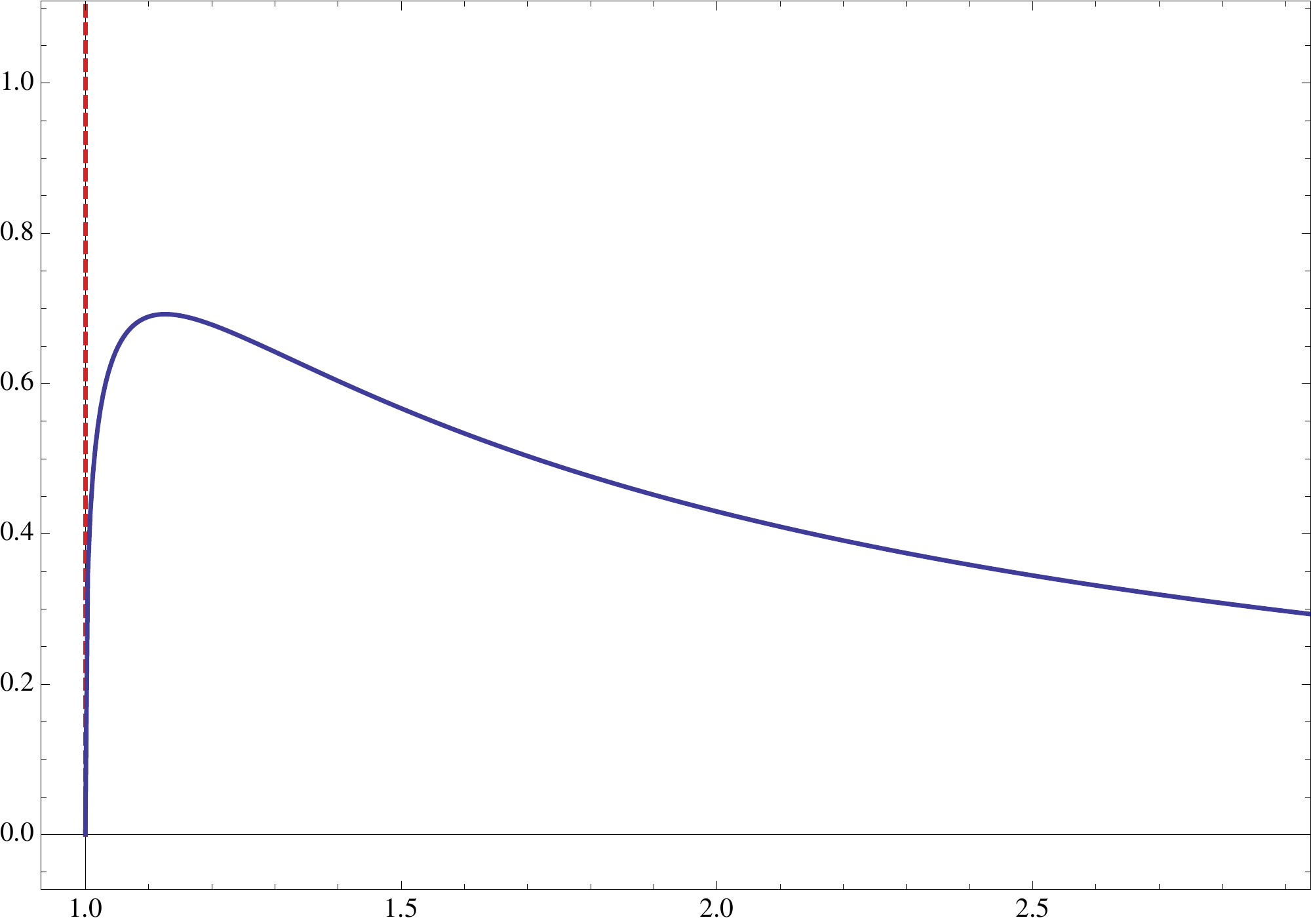}}
\put(-117,588){{\scriptsize{$L$}}}
\put(89,453){{\scriptsize{$\r_{0}$}}}
\put(123,450){\includegraphics[height=5cm]{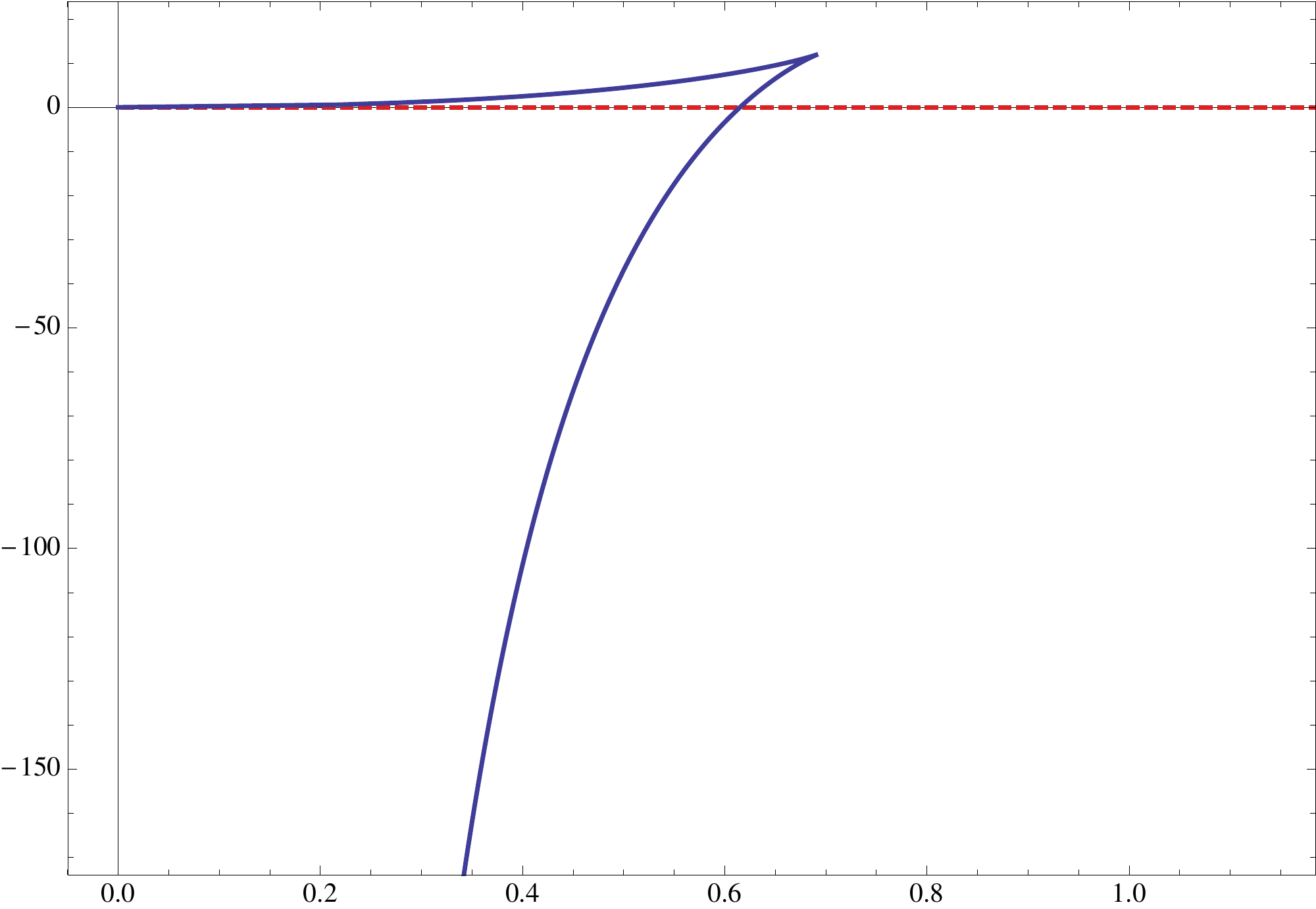}}
\put(126,588){{\scriptsize{$S$}}}
\put(332,453){{\scriptsize{$L$}}}
\put(-115,300){\includegraphics[height=5cm]{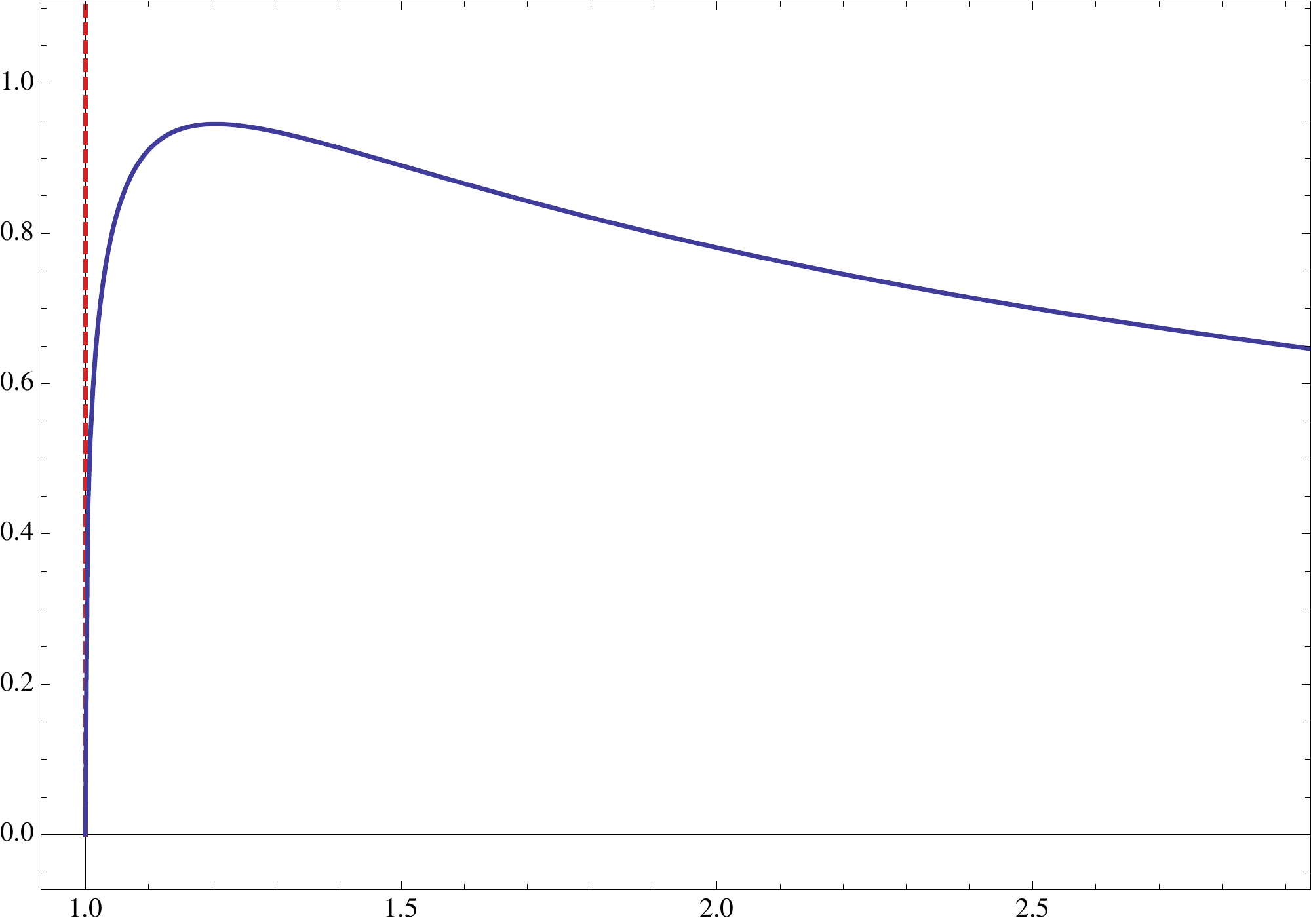}}
\put(-117,438){{\scriptsize{$L$}}}
\put(89,303){{\scriptsize{$\r_{0}$}}}
\put(123,300){\includegraphics[height=5cm]{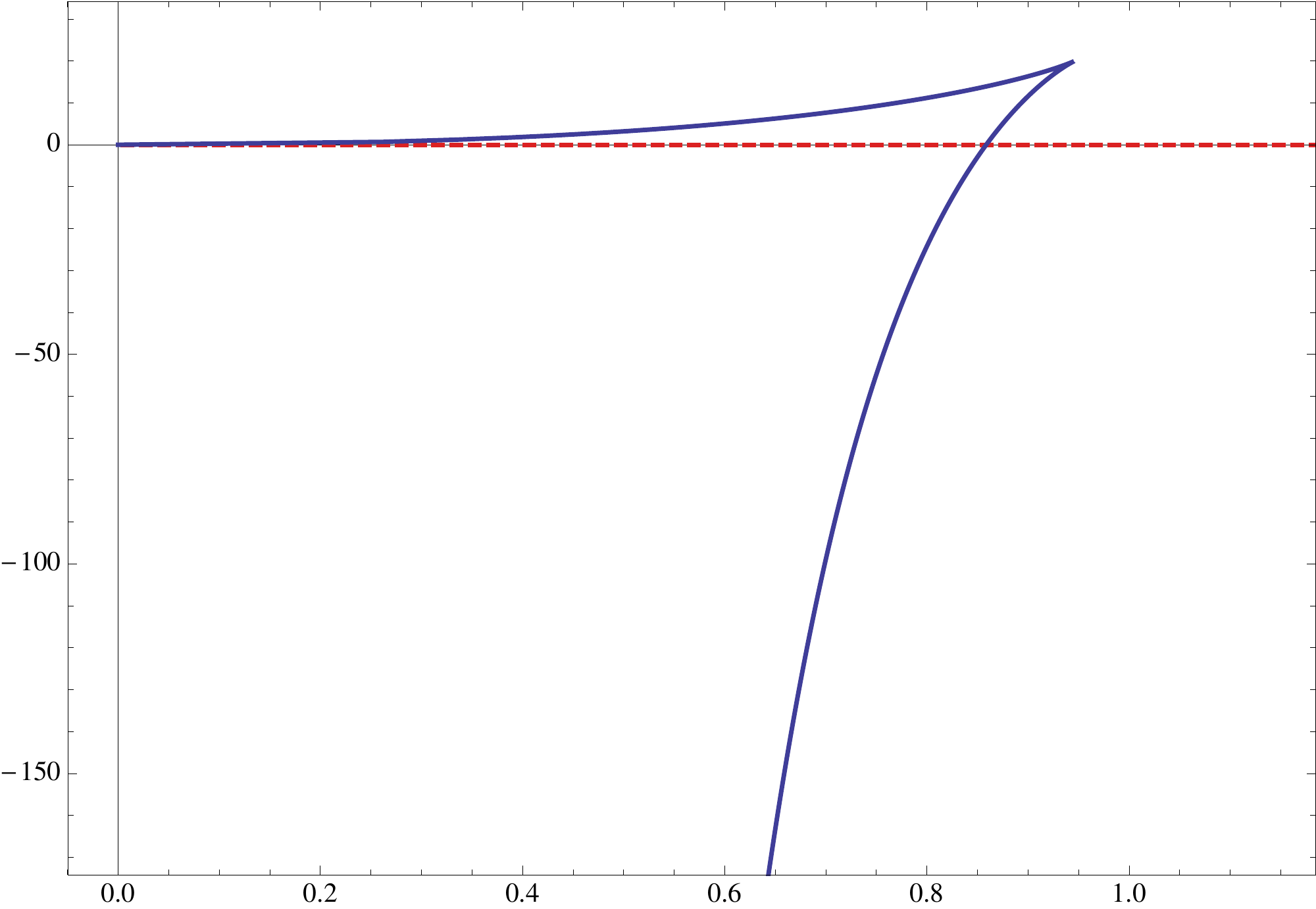}}
\put(126,438){{\scriptsize{$S$}}}
\put(332,303){{\scriptsize{$L$}}}
\put(-116,150){\includegraphics[height=5cm]{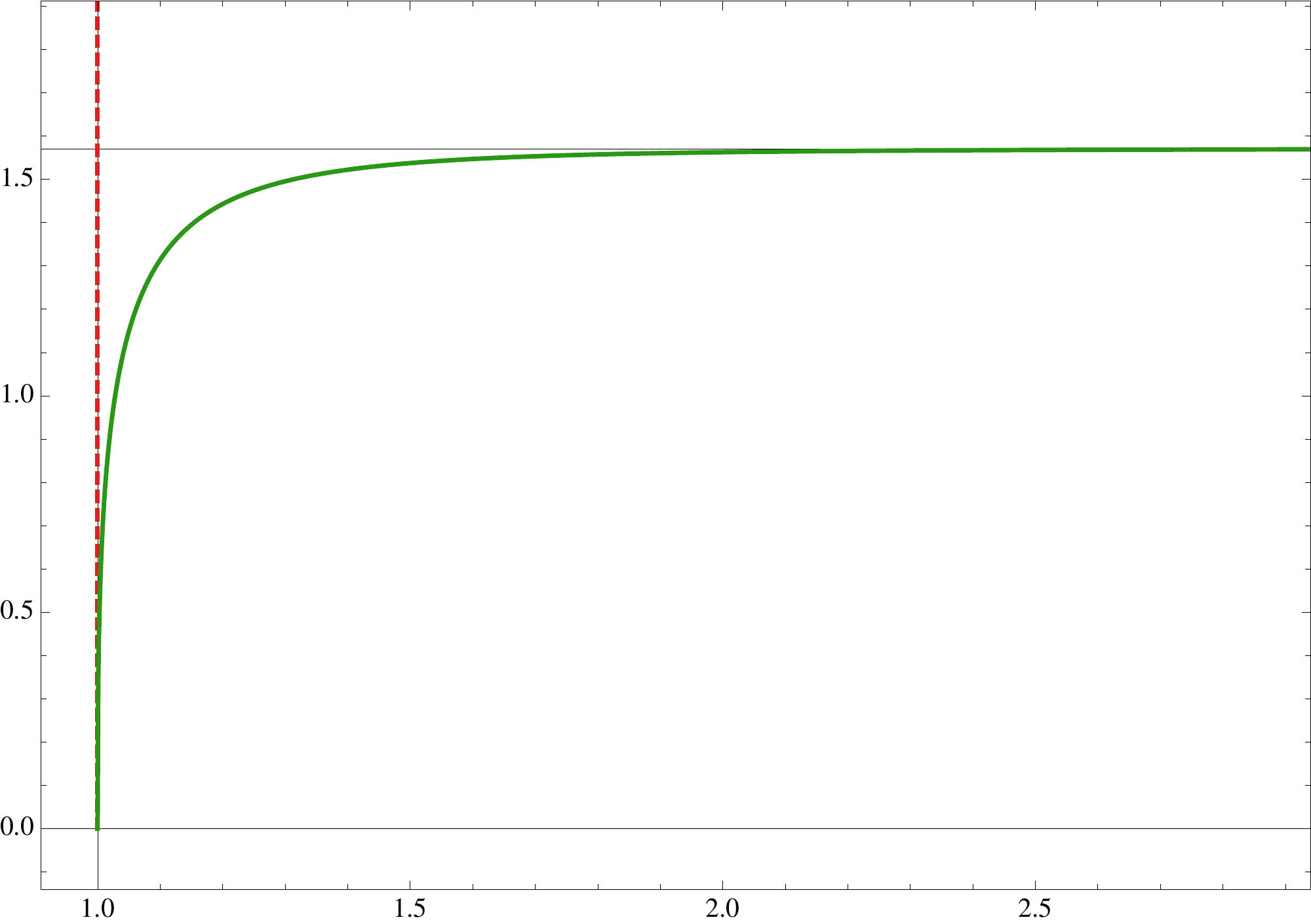}}
\put(-117,287){{\scriptsize{$L$}}}
\put(89,153){{\scriptsize{$\r_{0}$}}}
\put(130,150){\includegraphics[height=5cm]{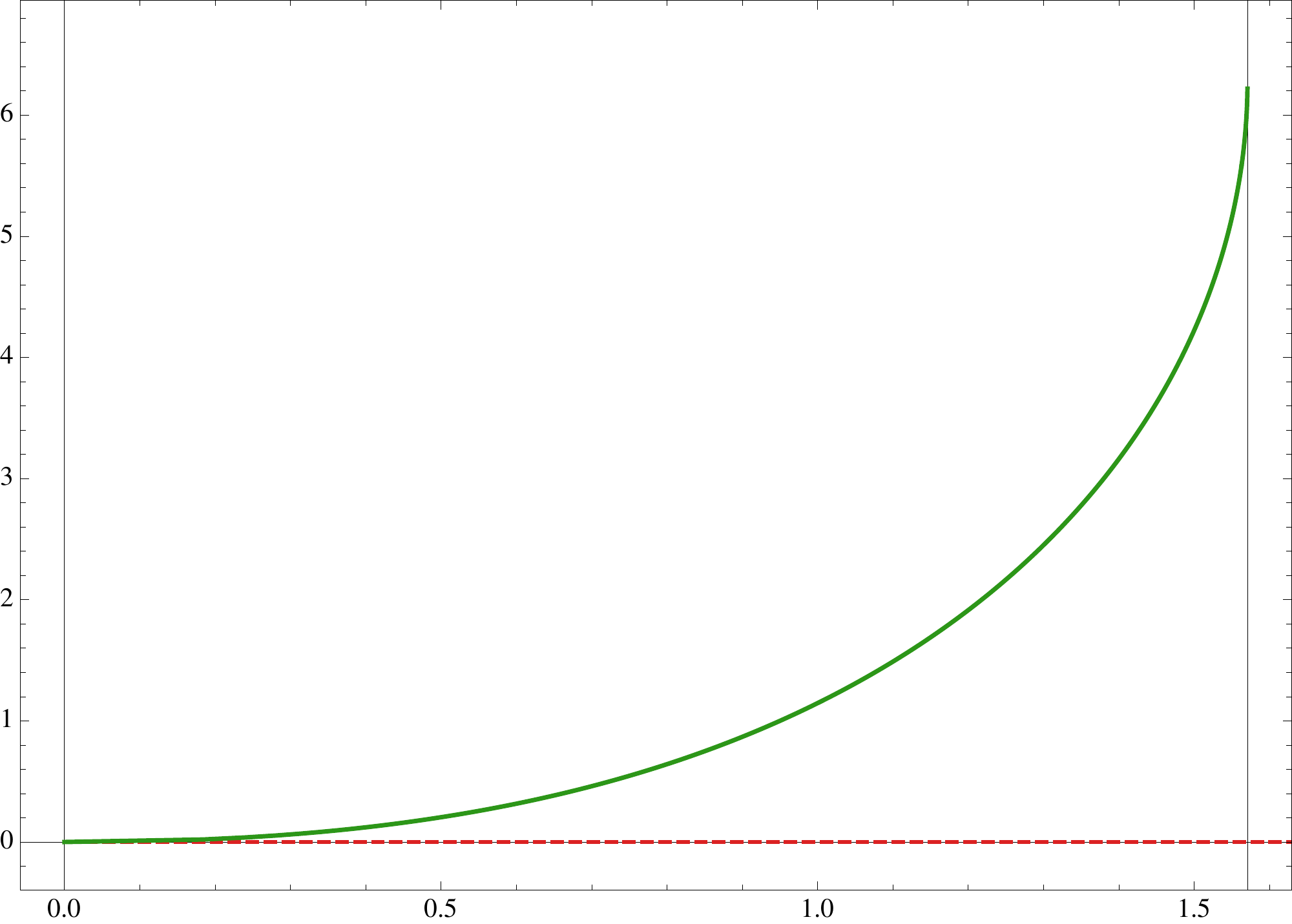}}
\put(126,287){{\scriptsize{$S$}}}
\put(332,153){{\scriptsize{$L$}}}
\put(-112,0){\includegraphics[height=5cm]{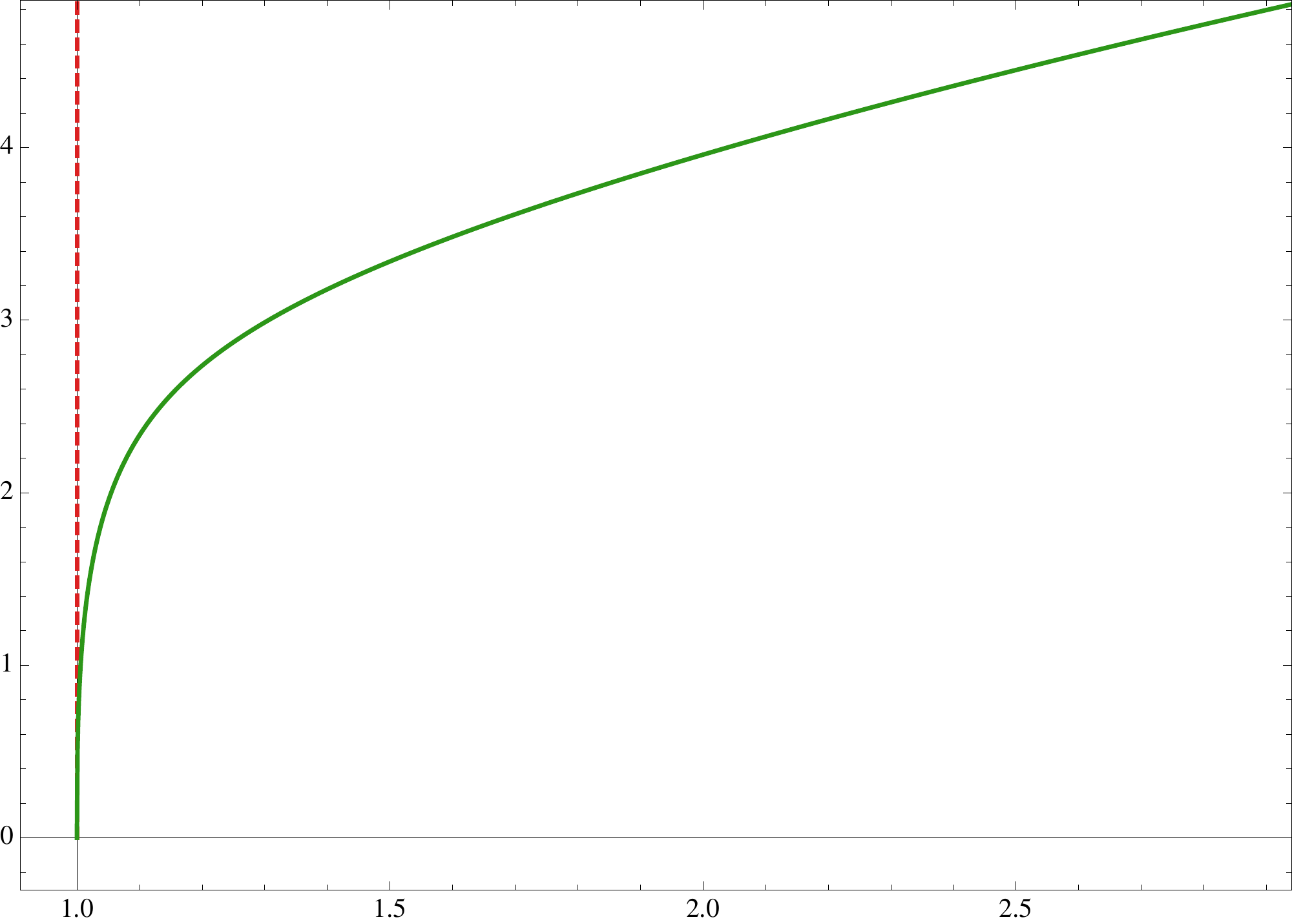}}
\put(-117,137){{\scriptsize{$L$}}}
\put(89,3){{\scriptsize{$\r_{0}$}}}
\put(126,0){\includegraphics[height=5cm]{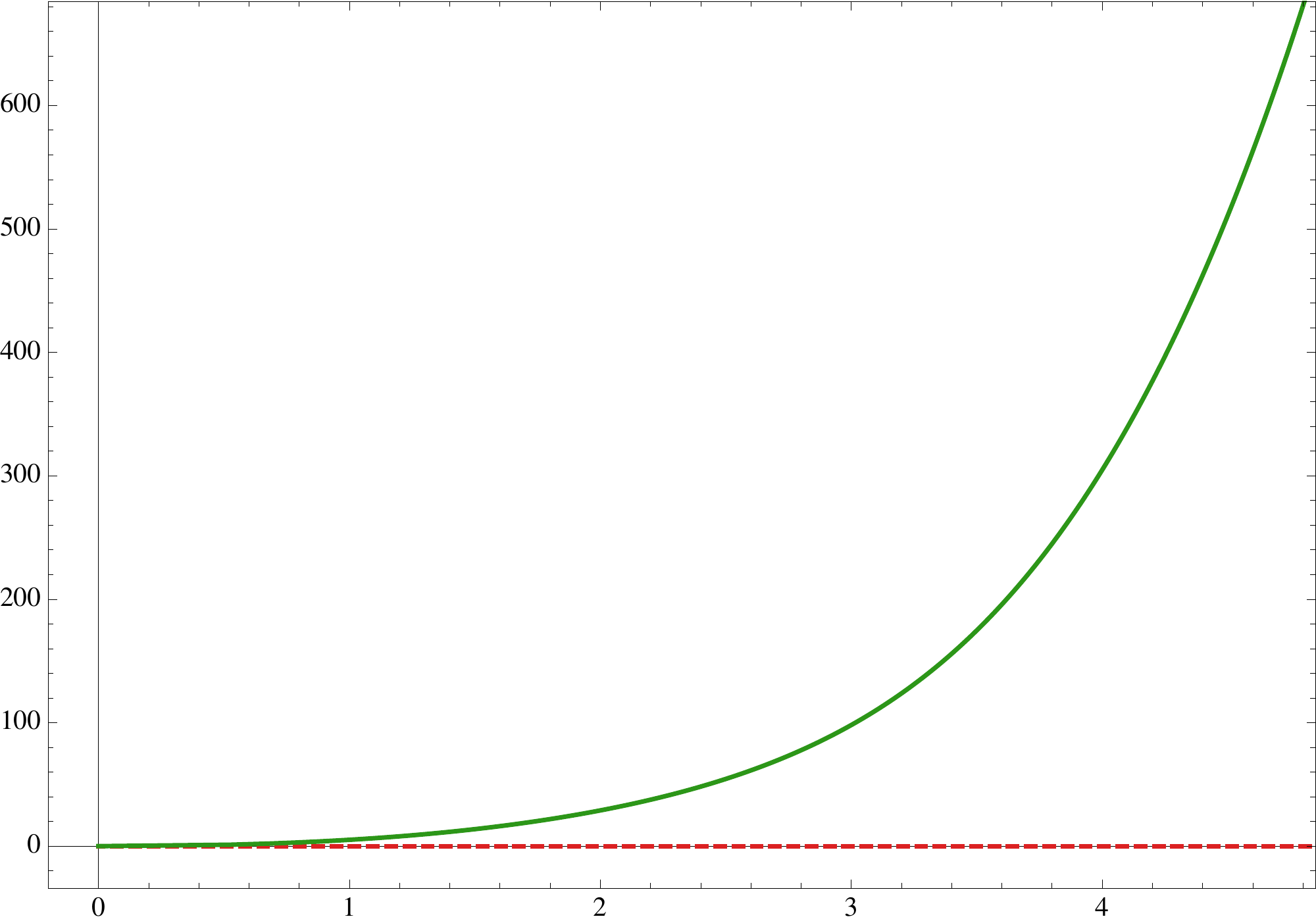}}
\put(126,137){{\scriptsize{$S$}}}
\put(332,3){{\scriptsize{$L$}}}
\end{picture}
\caption{The function $L(\rho_0)$ and $S(L)$ in the near extremal D$p$ brane backgrounds for $p=3,4,5,6$ moving down the page.
           The location of the horizon was set to $\rhol=1$ in the figures.
           The dashed red line is the disconnected solution.
           The D$3$ and D$4$ branes shows a phase transition behavior while in the D$5$ and D$6$ branes there is no phase transition.}
\label{FDbranes}
\end{center}
\end{figure}

We conclude that for $p\leq4$ there will be a phase transition since $L(\rhoz)$ is a non-monotonic function, but for $p>4$ $L(\rhoz)$ is monotonic and there will be no phase transition.
In Figure \ref{FDbranes} we draw the functions $L(\rhoz)$ and $S(L)$ for the cases $p=3,4,5,6$.
We observe
a crossing (and hence
a phase transition)
between the connected and disconnected solutions for the EE, in the case
$p<5$. We also see
that the concavity of $S$ is
the correct one for $p\leq4$,
but does not satisfy eq.(\ref{concavityxx}) for $p>4$.

\subsubsection{Hard and Soft Walls}

\begin{figure}[ht]
\begin{center}
\begin{picture}(220,300)
\put(-115,150){\includegraphics[height=5cm]{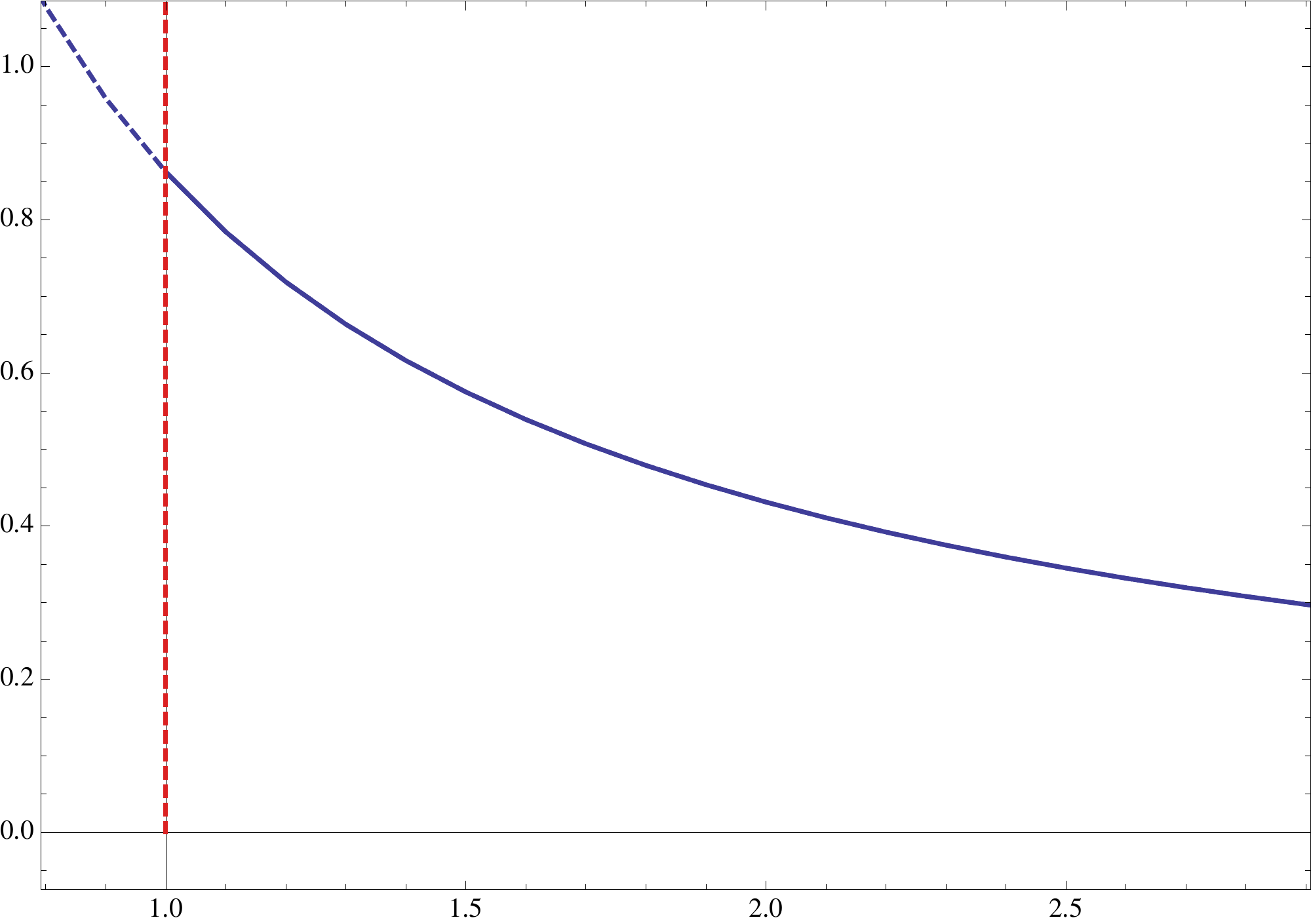}}
\put(-117,288){{\scriptsize{$L$}}}
\put(88,153){{\scriptsize{$\r_{0}$}}}
\put(123,150){\includegraphics[height=5cm]{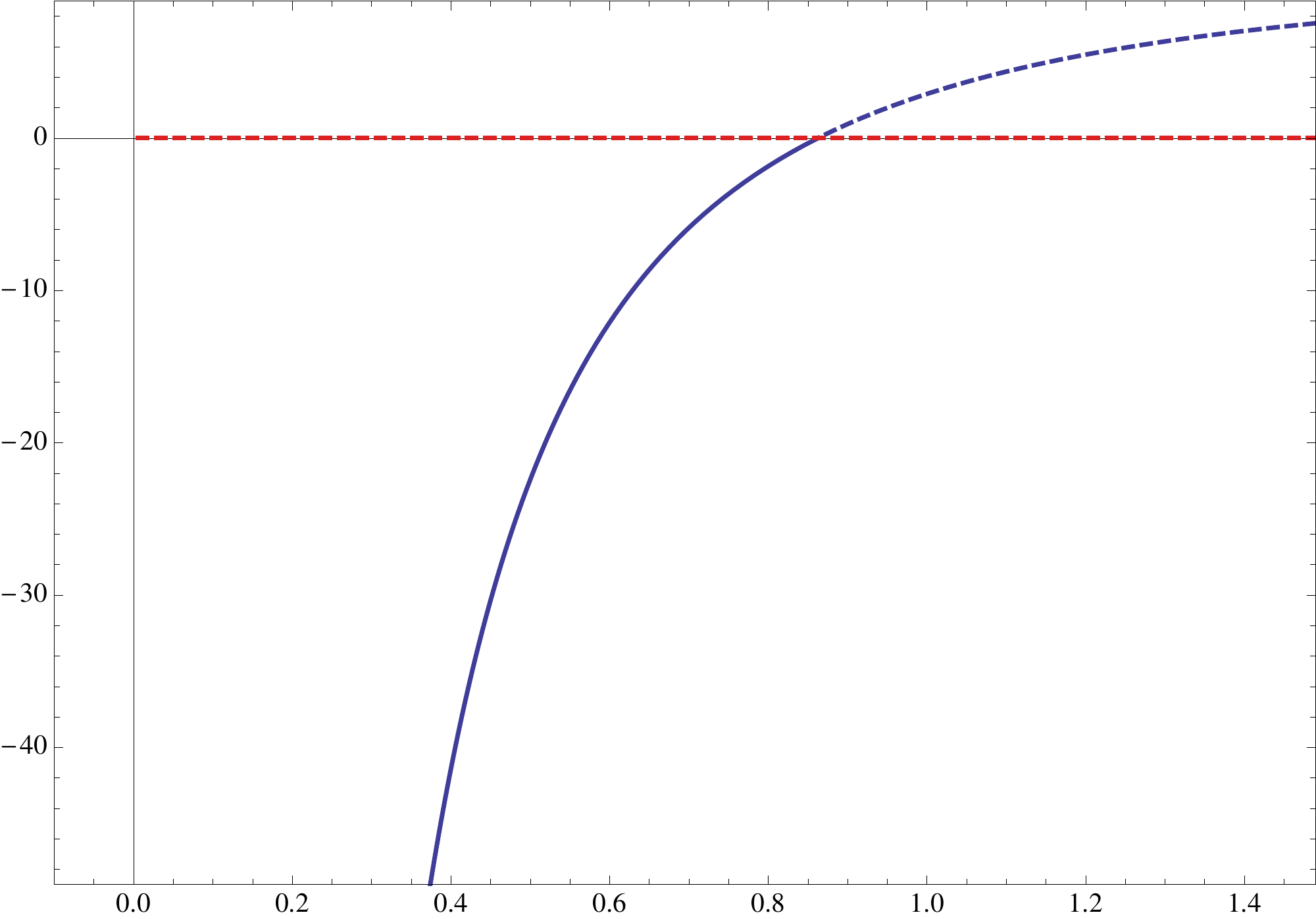}}
\put(124,288){{\scriptsize{$S$}}}
\put(329,153){{\scriptsize{$L$}}}
\put(-115,0){\includegraphics[height=5cm]{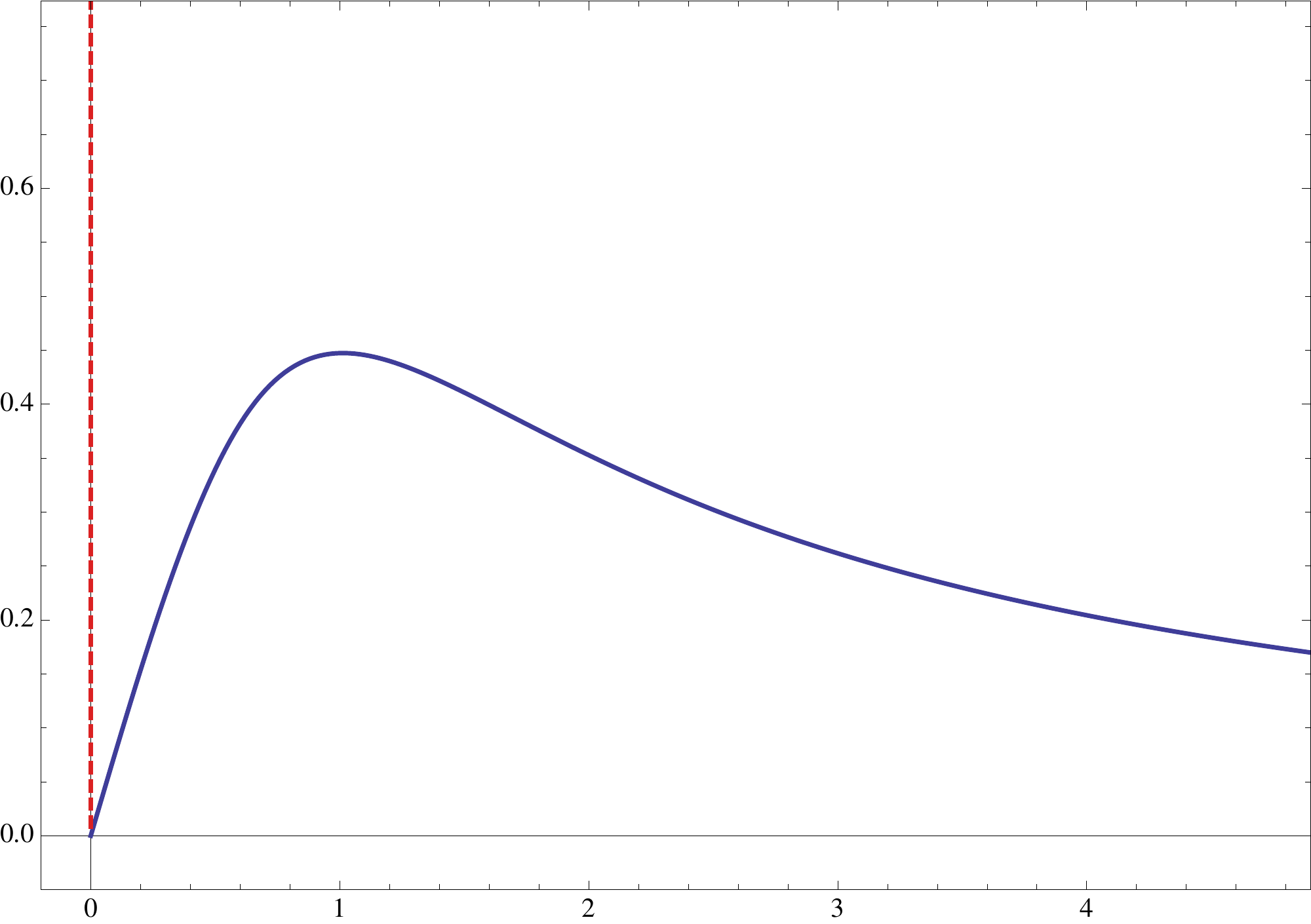}}
\put(-117,138){{\scriptsize{$L$}}}
\put(88,3){{\scriptsize{$\r_{0}$}}}
\put(119,0){\includegraphics[height=5cm]{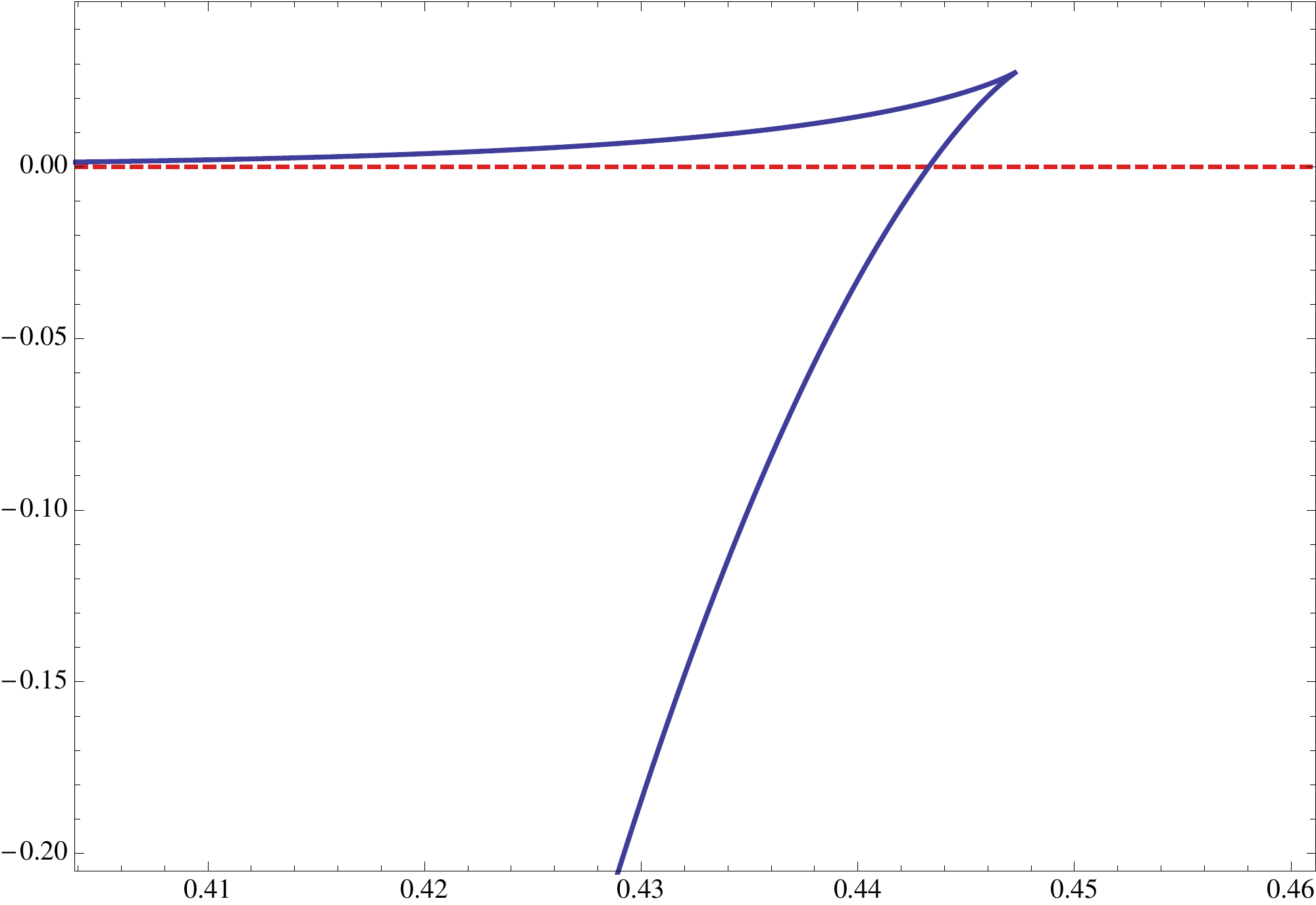}}
\put(124,138){{\scriptsize{$S$}}}
\put(329,3){{\scriptsize{$L$}}}
\end{picture}
\caption{The function $L(\rho_0)$ and $S(L)$ in the Hard (top row) and Soft Wall (bottom row) models.
          The location of the hard wall was set to $\rhol=1$ in the figures.
          The dashed red line is the disconnected solution and the dashed blue line represents the continuation of the $AdS$ solution beyond the hard wall.}
\label{FWalls}
\end{center}
\end{figure}

The Hard Wall model was proposed in \cite{Erlich:2005qh} as a holographic description of low-energy properties of QCD. It is described by the $AdS$ metric with the radial coordinate cut at some value $\rho=\rhol$.
The results for the entanglement entropy are shown in Figure \ref{FWalls}.

The authors of \cite{Karch:2006pv} have improved the Hard Wall model by cutting of the $AdS$ space smoothly, instead of a hard-wall cutoff in the IR.
The metric of the Soft Wall model is the same as the $AdS$ metric but there is a non-trivial dilaton
\begin{equation}
  e^{\phi} = e^{\frac{1}{\rho^2}}
\end{equation}
Then we have
\begin{eqnarray}
  \beta(\rho) &=& \frac{R^4}{\rho^4} \\
  H(\rho) &=& \left(\frac{8\pi^2}{3}\right)^2 R^4 \rho ^6    e^{-\frac{4}{\rho^2}}
\end{eqnarray}
The results are shown in Figure \ref{FWalls}. We see that the soft wall model admits a similar behavior to the D$3$ and D$4$ branes with a phase transition.

We cannot check the conditions for confinement in these examples. The Hard Wall background is just a cut $AdS$. The Soft Wall background does not admit the expansion \eqref{HIR} for the function $H(\rho)$ in the IR since it includes an exponential factor. Therefore, even though $\beta(\rho)$ diverges strongly, we still have a phase transition since the exponential decay of $H(\rho)$ takes-over the divergence of $\beta(\rho)$.

\subsubsection{Klebanov-Strassler}

The entanglement entropy of the background dual to a cascading supersymmetric gauge theory, the deformed conifold \cite{Klebanov:2000hb}, was analysed in details in \cite{Klebanov:2007ws}. The authors of \cite{Klebanov:2007ws} have shown that in this case there is a phase transition of the same form as in the D$3$ and D$4$ branes, as expected in a confining theory. We would like to demonstrate how this background follows the 
conditions for phase transition, which we derived above.

The supergravity solution of the deformed conifold is of the following
form \cite{Klebanov:2000hb,Loewy:2001pq}
\begin{equation}
ds^{2}=h^{-\frac{1}{2}}\left(\tau\right)dx_{1,3}^{2}+h^{\frac{1}{2}}\left(\tau\right)ds_{6}^{2}
\end{equation}
where $ds_{6}^{2}$ is the metric of the deformed conifold
\begin{eqnarray}
\nonumber ds_{6}^{2} &=& \frac{\epsilon^{4/3}}{2}K\left(\tau\right)
\left(\frac{1}{3K^{3}\left(\tau\right)}\left[d\tau^{2}+\left(g^{5}\right)^{2}\right]
+\cosh^{2}\left(\frac{\tau}{2}\right)\left[\left(g^{3}\right)^{2}+\left(g^{4}\right)^{2}\right] \right.  \\
&&  \qquad\qquad\qquad\qquad\qquad\qquad\qquad\:\:\:\: +\left. \sinh^{2}\left(\frac{\tau}{2}\right)\left[\left(g^{1}\right)^{2}+\left(g^{2}\right)^{2}\right]\right)
\end{eqnarray}
 $\epsilon$ is the energy scale and the functions $h\left(\tau\right)$
and $K\left(\tau\right)$ are given by\begin{equation}
h\left(\tau\right)=\left(g_{s}M\alpha'\right)^{2}2^{2/3}\epsilon^{-8/3}I\left(\tau\right),\end{equation}
 \begin{equation}
I\left(\tau\right)\equiv\int_{\tau}^{\infty}dx\frac{x\coth x-1}{\sinh^{2}x}\left(\sinh2x-2x\right)^{\frac{1}{3}},\end{equation}
 \begin{equation}
K\left(\tau\right)=\frac{\left(\sinh\left(2\tau\right)-2\tau\right)^{\frac{1}{3}}}{2^{\frac{1}{3}}\sinh\tau},\end{equation}
where $\tau$ is a dimensionless radial coordinate running from zero to infinity on the boundary.
Then we have,
\begin{eqnarray}
  \alpha &=& h^{-\frac{1}{2}}(\tau), \qquad \beta(\tau)=\frac{h(\tau)\epsilon^{4/3}}{6K^2(\tau)}, \\
  V_{int} &=& \frac{4\pi^3}{\sqrt{6}} h^{5/4}\epsilon^{10/3}K(\tau)\sinh^2(\tau), \\
  H(\tau) &=& e^{-4\phi}V_{int}^2 \alpha^3 = \frac{8\pi^6}{3}\epsilon^{20/3} h(\tau) K^2(\tau)\sinh^4(\tau).
\end{eqnarray}
\begin{figure}
\begin{center}
\begin{picture}(220,160)
\put(0,0){\includegraphics[height=5.35cm]{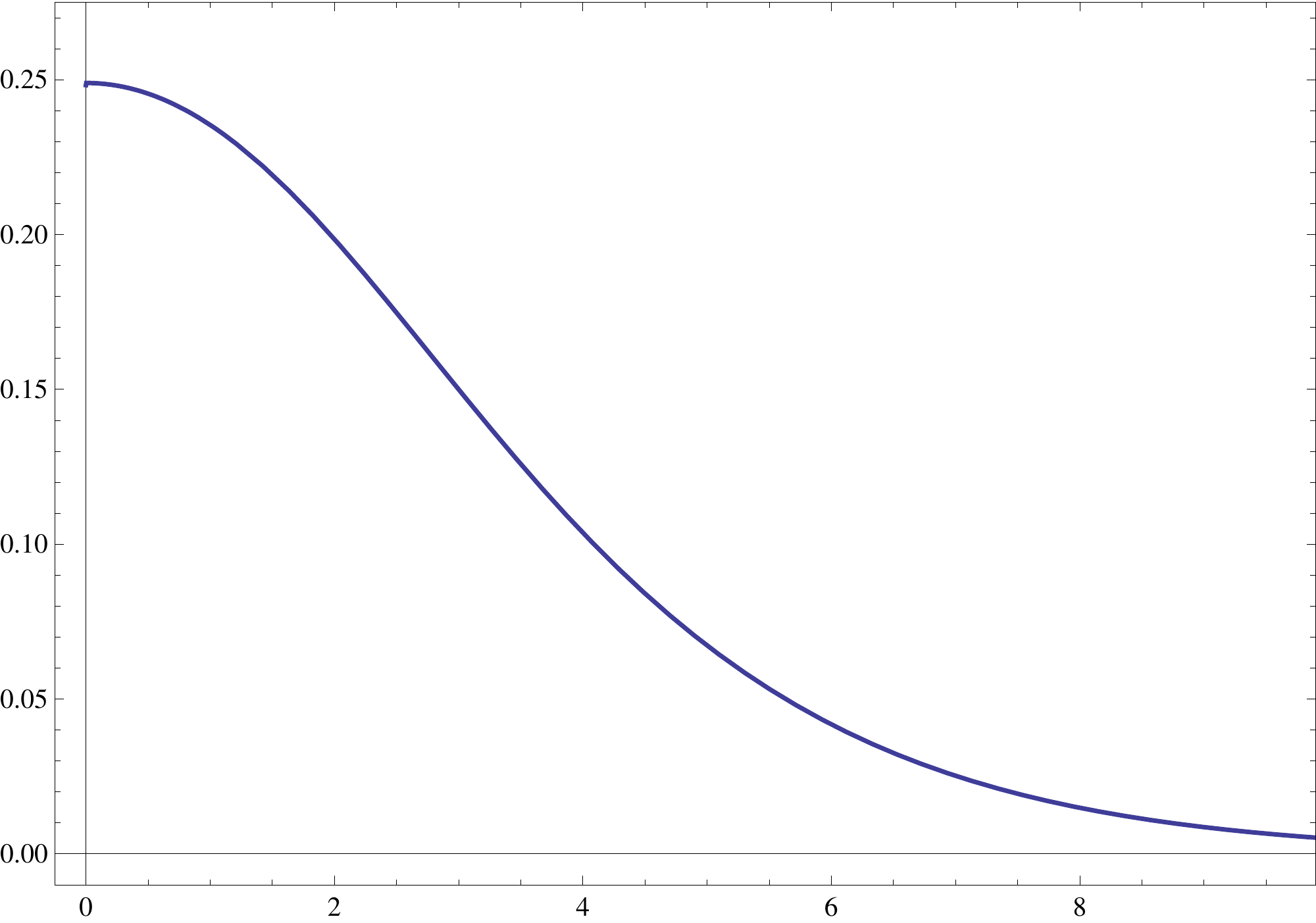}}
\put(2,148){{\scriptsize{$\beta$}}}
\put(220,3){{\scriptsize{$\tau$}}}
\end{picture}
\caption{The function $\beta(\tau)$ in the Klebanov-Strassler background. $\beta(\tau)$ saturates to a finite value at the origin $\tau=0$ and therefore meets the condition for a phase transition.}
\label{FKS}
\end{center}
\end{figure}

The function $\beta(\tau)$ approaches a finite value in the far IR $\tau=0$, and therefore we have in this case $t=0$, corresponding to a monotonically increasing $L(\rhoz)$ in the IR, in accordance with \eqref{t}.
Near the boundary, i.e. at large $\tau$, the $h\left(\tau\right)$ and $K\left(\tau\right)$
function takes the form
\begin{eqnarray}
  H(\tau\ll 1) &=& \epsilon^4(g_sM\alpha ')^2\pi^6 \tau e^{2\tau} \\
  \beta(\tau) &=& \epsilon^{-\frac{4}{3}}(g_sM\alpha ')^2 2^{-\frac{4}{3}}\tau e^{-\frac{2}{3}\tau}
\end{eqnarray}
In this region the functions does not admit the power expansion we have assumed in \eqref{boundaryexpansion} and therefore we cannot directly check the conditions we have found, but a direct computation \cite{Klebanov:2007ws} shows that $L(\rhoz)$ indeed goes to zero close to the boundary and therefore there is a phase transition.
The intuition is that $\beta(\tau)$ decays exponentially fast close to the boundary, and therefore meets the requirement for sufficiently
strong decay ($j>2$) of eq. \eqref{j}.

\subsection{Confinement and Phase Transitions}

We are now in a position to compare the conditions 
for confinement on the Wilson loop, see eq. \eqref{WLconditions} 
derived in \cite{Kinar:1998vq} and the conditions for 
a phase transition in the entanglement entropy as were 
suggested in \cite{Klebanov:2007ws} and further developed in the previous subsection. On physical grounds, since both observables are probes of confinement, we expect both conditions to coincide.
We will not be able to prove the last statement, but we will give a flavour of why it should be true in some examples. On the other hand, we will emphasise a puzzle which arises in other cases. The solution for this puzzle will be the aim of rest of the paper.

Let us start with the conditions in the IR.
The conditions on the Wilson loop, eq.\eqref{WLconditions} are really
a statement about the IR and therefore we will compare them
to the condition we derived for the entanglement entropy in the IR, see
eq.\eqref{t}. The condition of eq.\eqref{t}
means that $\beta(\rho)$ should diverge slower than
$\frac{1}{(\rho-\rhol)^2}$ in order to observe a phase transition in the EE.
Using eq.\eqref{g} we relate the divergence of
$\beta$ to the divergence of $g$, which is one of the conditions
for linear confinement in the Wilson loop, see eq.\eqref{WLconditions}
(remember that since $\alpha(\rhol)$ is the string tension it must be finite, and therefore does not play a role in the discussion about divergences). If we take the case of the D$p$ brane on $S^1$ ---
see eq.\eqref{DbraneMetricx} --- as an example, we see that close to $\rho=\rhol$
\begin{equation}
  g^2(\rho) = \left(\frac{\rhol}{7-p}\right)\frac{1}{\rho-\rhol}+\dots
\end{equation}
where ``$\dots$" stand for sub-leading finite corrections. This divergence is in
agreement with the condition on $\beta$ in eq.\eqref{t},
for any value of $p$.
A possible violation of \eqref{t} would correspond
to a stronger divergence of $g$. We are not aware of such examples.
Based on the intuition of backgrounds with compact circles,
such a case would correspond to a situation with two compact
circles with a topology of a cigar, where the tip
of both cigars located in the same radial position.
It would be interesting to try to rule out that case
(or alternatively find one such example), but we leave this for future work.
When $\beta$ approaches a finite value at $\rho=\rhol$,
the corresponding confinement-condition on the Wilson loop
is the first among the two in eq.\eqref{WLconditions},
which is that $\alpha$ has a minimum at $\rho=\rhol$.
In this case it seems that the maximum of
$\beta(\rho)$ at $\rho=\rhol$ (as in Figure \ref{FKS})
corresponds to the minimum of $\alpha(\rho)$,
via the relation $\beta\sim\frac{1}{\alpha^2}$, see eq.\eqref{g}.

While we have presented an intuition (but not a proof) of the equivalence
between the conditions on the EE and the Wilson loop in
the IR, there is a puzzle concerning the UV.
First, we note that there are no UV conditions on the Wilson loop to obey confinement. As far as the Wilson loop is concerned, the only condition we demand is linearity at long distances (IR).
On the other hand, in order to observe a phase transition in the EE,
the background also has to satisfy the condition in the UV, eq.\eqref{j}.
This last condition is not satisfied in certain confining
backgrounds --- D$p$ branes on $S^1$ with $p>4$ ---
as discussed in the previous subsection.
The question that arises is
\begin{itemize}
\item{{\it Question:}} Why does certain cases show linear confinement in the Wilson loop but do \emph{not} show a phase transition in the entanglement entropy?
\end{itemize}
Answering this question will take us to a nice detour that will include:
Area and Volume law behaviors for Entanglement Entropy, local and non-local QFTs, theories with
a UV cutoff and theories that are UV-completed approaching a near conformal
point.
This material will be carefully presented in Section \ref{newmaterial}. Then,
we will apply all that information to answer what happens in
a `trademark' model of Confinement in 4-d QFT, namely the twisted
compactification of D5 branes on a two-cycle of the resolved conifold (
with or without fundamental-matter fields).
We will introduce a cutoff that will recover the phase transition
in the EE and solve the concavity problem
mentioned above.
Understanding how to UV-complete
these non-local QFTs, nicely recovering the phase transitions
without appealing to cutoff-effects,
will be the goal of Section \ref{sectionabsence}.

\section{Volume and Area Laws, UV-cutoffs and Confinement}\label{newmaterial}
The goal of this section is to set-up the elements that will
allow us to answer the last question posed above.
We will briefly analyse the case of $AdS_5\times S^5$, then quickly
move into NS5 and D5 branes, where
an important role will be played by the non-locality
of the associated field theory. The analog to the confining
Witten's model, but with
D5 branes that wrap $S^1$ will close the analysis of
this section.

\begin{figure}[ht]
\begin{center}
\begin{picture}(220,170)
\put(-110,-20){\includegraphics[height=6.5cm]{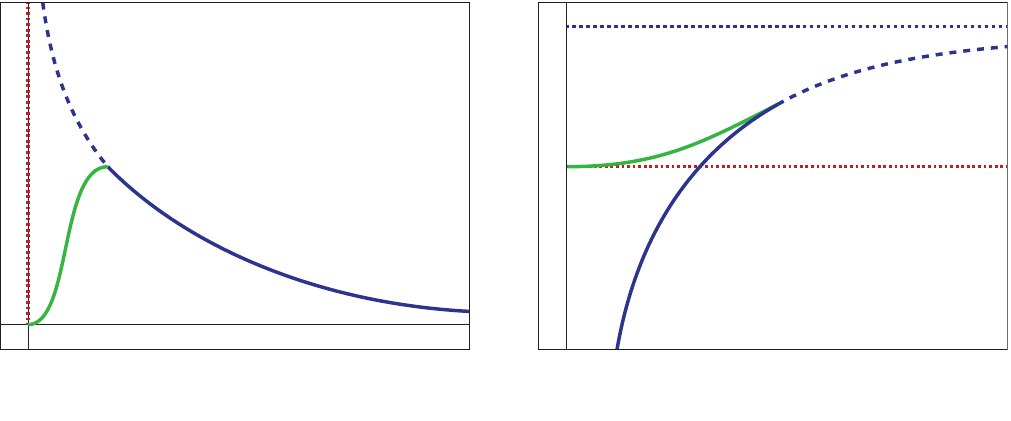}}
\put(-118,160){{\scriptsize{$L$}}}
\put(90,3){{\scriptsize{$\rho_{0}$}}}
\put(118,160){{\scriptsize{$S$}}}
\put(330,3){{\scriptsize{$L$}}}
\put(198,83){{\scriptsize{$L_{c}$}}}
\end{picture}
\caption{Here are cartoons of the change in $L$ and $S$ when we have a theory with confinement.  The navy blue lines (solid and dashed) in both plots represent the behaviour of the connected part of conformal solutions like that of $AdS_{5}\times S^{5}$ (i.e. $L(\rho_{0})\sim\rho_{0}^{-1}$), the green line is the {\it{unstable}} branch introduced by confinement (like in the soft-wall model).  The dotted red and navy lines represent the disconnected solution.  We can see that in the confining case there is a phase transition at the point $L_{c}$.}
\label{Fig:confinement}
\end{center}
\end{figure}
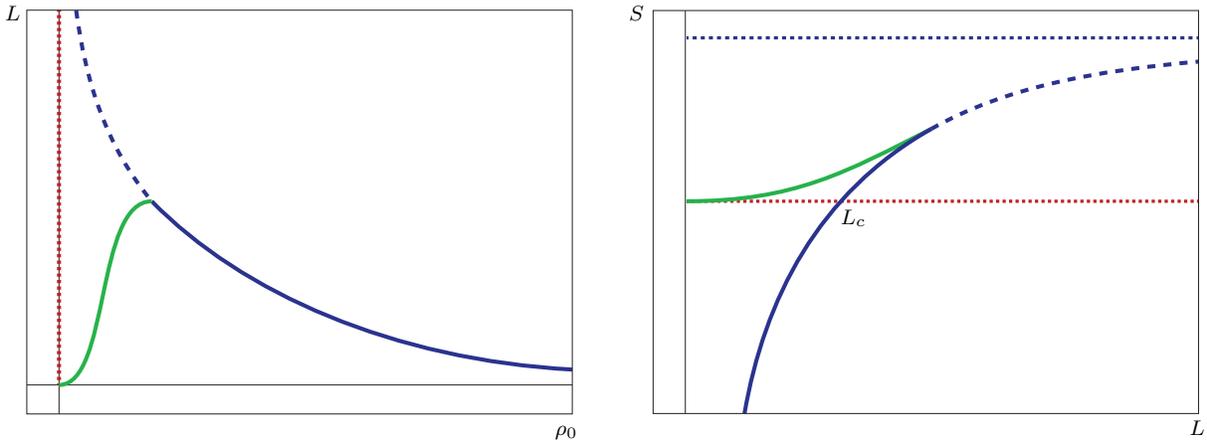

Let us discuss the well understood case of $AdS_{5}\times S^{5}$, as it will
be a basis for comparison for more complicated cases.
In $AdS_{5}\times S^{5}$ we found
that the connected solution is always the minimal solution
for the EE
--- and is always preferred to the disconnected solution for all values of $L$ which has a higher EE --- see Figure \ref{FAdS}.
The connected solutions asymptote the disconnected ones from below for large $L$.
For a local field theory, the EE follows what is called a ``Heisenberg-like"
relation, such that $L(\rho_{0})\sim \rho_{0}^{-1}$ for some region (typically for large $\r_0$)
of the minimal solution.  This type of behaviour can be seen in Figure \ref{Fig:confinement}
by considering only the navy blue lines (both solid and dashed).  We should further note that they have
the correct concavity for stability, see eq.(\ref{concavityxx}). The $AdS_{5}\times S^{5}$ has the
usual Area Law for the EE when we analyse the divergent parts.

Introducing confinement in this theory can be thought of as an effect on the IR region of the corresponding $AdS_{5}\times S^{5}$ plot.
This effect can be modelled by a soft-wall solution as we have already seen. We replace the IR region with the usual
$L\sim \rho_{0}^{-1}$ behaviour for
the connected solution, with an {\it{unstable}} branch (c.f. the green
lines in Figure \ref{Fig:confinement}) \footnote{Notice that this
solution does not satisfy the criteria for the  concavity
given in eq.(\ref{concavityxx}).}.
This has the effect of moving the disconnected
branch down such that it now meets the {\it{stable}} connected branch at a
finite value of $L=L_{c}$ (see the right panel of Figure \ref{Fig:confinement}).
Thus within backgrounds of this type we find that there exists a critical value $L_{c}$, such that for $L>L_{c}$,
the minimal solution is now the disconnected branch, and for $L<L_{c}$ we still have the original
$AdS$-like connected behaviour in the UV ({\it{stable}} branch).
The presence of the {\it{unstable}} branch will occur for all theories we will study that exhibit confinement (for zero temperature at least).
Again these solutions follow the usual Area Law for the divergent parts of the EE.
Something peculiar occurs if we now consider the case of D5/NS5 branes,
as discussed in \cite{Barbon:2008ut}.  In this case, we find that the separation of the connected branch is
constant, given by $L(\rho_0)=\pi R/2$ where $R^2=\alpha' g_s N_c$.
Thus there is an infinite number
of connected solutions which are parameterised by the depth to which
they probe but all have the same fixed value of $L$
(as in the left panel of Figure \ref{Fig:shape}).

The authors of \cite{Barbon:2008ut}
argued that the solutions for smaller
values of $L<L_{c}$ (using the approximation of a capped
cylinder similar to the one we will discuss shortly),
are exactly those that must live near
the UV-cutoff, with the contribution to the EE coming
from the cap, which present a Volume-Law, once the divergent part of the EE
is considered.

Another interesting
case occurs in the context of D5 branes wrapped on $S^{1}$, which as explained,
is a model of a confining $4+1$-d QFT.
Here we find that the connected branch of the EE is similar to that of
the IR {\it{unstable}} branch of the soft-wall
model, but that there is no {\it{stable}} branch as we move into the UV.
This would be akin to only keeping the green line in Figure
\ref{Fig:confinement}
(in this specific case the connected branch
asymptotically approaches that of the NS5/D5, i.e.
$L=\frac{\pi R}{2}$) --- see the third row in Figure \ref{FDbranes}.
The example of D5 wrapped on $S^1$ then presents the IR features of
a confining model (like a soft-wall model), but
the UV behavior of a non-local QFT (like flat NS or D5 branes). It
presents only an unstable branch and a 
disconnected one and the absence of a phase
transition in spite of
displaying a confining Wilson loop. We will appeal to cutoff-effects
to solve this issue. Below, we will clarify the details of this example.
Before that, we present a useful approximation to
some of the quantities involved in the calculations.

\subsection{A Useful Quantity}
As an aside, we would like to introduce a combination of background functions that (we have checked)
approximates
very well
the function $L(\rho_0)$ in all the cases studied in this paper.
This is useful, because in
the examples dealt with in the following sections, the functions
defining the background are only known numerically (or in semi-analytic expansions). Hence the integrals
defining $L$ and $S$ are very time-consuming.
Instead, the quantity
\beq
{\cal Y}(\r_0)= \left.2\pi \frac{H(\r)
\sqrt{\beta(\r)}}{H'(\r)}\right|_{\rho=\rho_0}
\label{yderho}
\eeq
can be seen to approximate very precisely the complicated integral
in \eqref{length} that
defines $L(\rho_0)$. The analog of this function also appeared
in studies of Wilson loops and other probes \cite{Faedo:2013ota},
\cite{Avramis:2006nv}.

\subsection{Study of the D$5$ branes on $S^1$ System}\label{StudyD5S1}
In this section, we will emphasize that the
presence of a phase transition for the entanglement entropy
in confining theories
is sensitive to the UV behaviour of the field theory. We will
make this point by considering the simplest confining field theory
in $4+1$ dimensions that one can construct by wrapping $N_c$ D5 branes
on a circle and imposing periodic (anti-periodic) boundary conditions
for the bosons (fermions) of this field theory.
This is in
 analogy with the example introduced by Witten
in \cite{Witten:1998zw} by double-Wick rotating a black-brane solution.

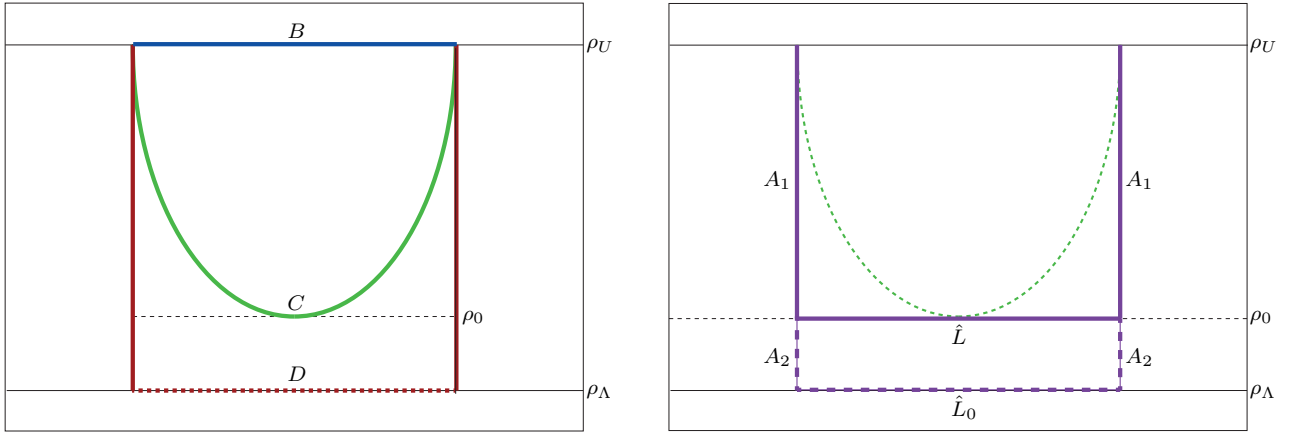
\begin{figure}[ht]
\begin{center}
\begin{picture}(220,170)
\put(-125,0){\includegraphics[height=5.75cm]{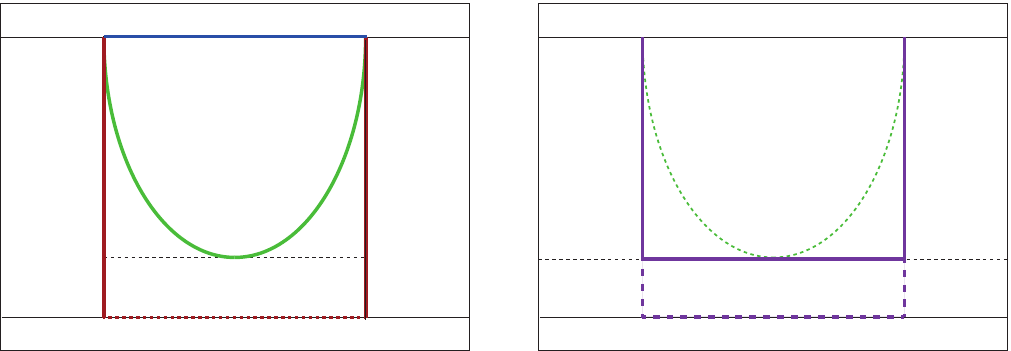}}
\put(95,145){{\scriptsize{$\rho_{U}$}}}
\put(95,14){{\scriptsize{$\rhol$}}}
\put(346,145){{\scriptsize{$\rho_{U}$}}}
\put(346,14){{\scriptsize{$\rhol$}}}
\put(48,42){{\scriptsize{$\rho_{0}$}}}
\put(346,42){{\scriptsize{$\rho_{0}$}}}
\put(-18,19){{\scriptsize{$D$}}}
\put(-18,149){{\scriptsize{$B$}}}
\put(-18,46){{\scriptsize{$C$}}}
\put(162,93){{\scriptsize{$A_{1}$}}}
\put(299,93){{\scriptsize{$A_{1}$}}}
\put(162,26){{\scriptsize{$A_{2}$}}}
\put(299,26){{\scriptsize{$A_{2}$}}}
\put(233,33){{\scriptsize{$\hat{L}$}}}
\put(233,6){{\scriptsize{$\hat{L}_{0}$}}}
\end{picture}
\caption{Here are cartoons of the types of solutions we shall be considering in the left
panel and details of the
approximation in the right panel.
In both, $\rho_{U}$ represents the UV boundary and $\rhol$ is the IR end of the space.
In the left panel, the red lines (including the dashed line at $\rhol$) represent the disconnected solution (D),
the green line represents a generic connected solution (C) which probes down to a depth $\rho_{0}$
and finally in blue are the solutions which live close to the boundary (B) and behave under the Volume Law for the EE.
In the right panel, we outline the various sections of the approximation.
The purple solid lines map out the approximation to the connected dashed green solution,
which we split into three parts: two vertical contributions labelled as $A_{1}$ and a horizontal
contribution labelled $\hat{L}$. The surface mapped out by the dashed purple lines, which is useful when we regularise our
approximation, consists again of three parts: the two vertical contributions labelled $A_{2}$ and the horizontal contribution labelled $\hat{L}_{0}$.}
\label{Fig:shape}
\end{center}
\end{figure}

Indeed, specialising the results of eq.(\ref{DbraneMetricx}),
we will have that the string frame metric  reads,
\bea
\frac{ds^2}{\alpha'}=\left(\frac{u}{R}\right)[dx_{1,4}^2 + h(u) d\varphi_c^2] +
\frac{R}{u h(u)} du^2 +
R u d\Omega_3^2,\;\;\;\; R^2=g_s\alpha' N_c.\nonumber
\eea
where the functions $h(u)$ and the dilaton
are,
\bea
h(u)=1-\left(\frac{\Lambda}{u}\right)^2,\;\;\;\; e^{\phi}=g_s \alpha' \left(\frac{u}{R}\right).\nonumber
\eea
It is more convenient to change from the `energy' variable $u$ to the
radial variable $r=\alpha' u$. The background and
relevant functions for our
calculations are,
\bea
& & ds^2=\left(\frac{r}{R}\right)[dx_{1,4}^2 +\alpha' h(r) d\varphi^2] +
\frac{R}{r h(r)}dr^2 +
R r d\Omega_3^2,\;\;\;\; R^2=g_s\alpha' N_c.\nonumber\\
& & h(r)=1-\left(\frac{R_\Lambda}{r}\right)^2,\;\; e^{\phi}=g_s  \frac{r}{R},\;\;\; R_\Lambda
=\alpha' \Lambda.\label{rararara}\\
& & \alpha(r)= \frac{r}{R},\;\;\beta(r)=\frac{R^2}{r^2 h(r)},
\;\;  V_{int}^2= (4\pi)^4 l_{\varphi}^2 R^2h(r) r^4, \nonumber\\
& &
 H(r)=\frac{(4\pi)^4 R^2 l^2_{\varphi}}{g_s^4 }h(r) r^4.\;\;\;\; l_\varphi=
\sqrt{\alpha'}\oint d\varphi.\nonumber
\label{flatD5}
\eea
Using the approximation discussed in eq.(\ref{yderho}),
one can compute that the function
$L(r_0)$ asymptotes (from below) to a constant value
\beq
L(r_0\to \infty)\sim \mathcal{Y}(r\to \infty)= \lim_{r_0\to \infty}
\frac{\pi R \sqrt{r^2(r^2-R_\Lambda^2)}}{(2r^2- R_\Lambda^2)}=\frac{\pi}{2}
\sqrt{g_s \alpha' N_c}
\eeq
hence preventing any form of double-valuedness and phase transition.
See the third row in Figure \ref{FDbranes} for a plot of the calculation done with the
background in eq.(\ref{rararara}).
Notice again, that the connected solution has the wrong
concavity, hence it is unstable.
This would lead us to believe that the disconnected solution would always
be the minimal EE solution for all values of $L$.  With this case
in mind one can instead ask the question
\begin{itemize}
\item{ {\it Question:} Are there other solutions which have smaller EE that we should consider ?}
\end{itemize}
As we anticipated above, the answer goes like this: in \cite{Barbon:2008ut},
the authors discuss how non-locality affects EE calculations.
They argue that one should add a UV-cutoff and also consider solutions
which live close to it
(represented by $B$ in Figure \ref{Fig:shape}). These solutions can minimise the
EE
in cases which exhibit non-locality.
There is a difference between these solutions and the ones
we have discussed so far, in that these
solutions no longer follow the standard Area Law but instead
follow a ``Volume Law" for the divergent part of the EE.  This observation
was also made in other contexts by
\cite{Fischler:2013gsa},\cite{Karczmarek:2013xxa},\cite{Shiba:2013jja}.
From this insight we can try to understand in which cases
these new Volume Law solutions may be relevant to our question.

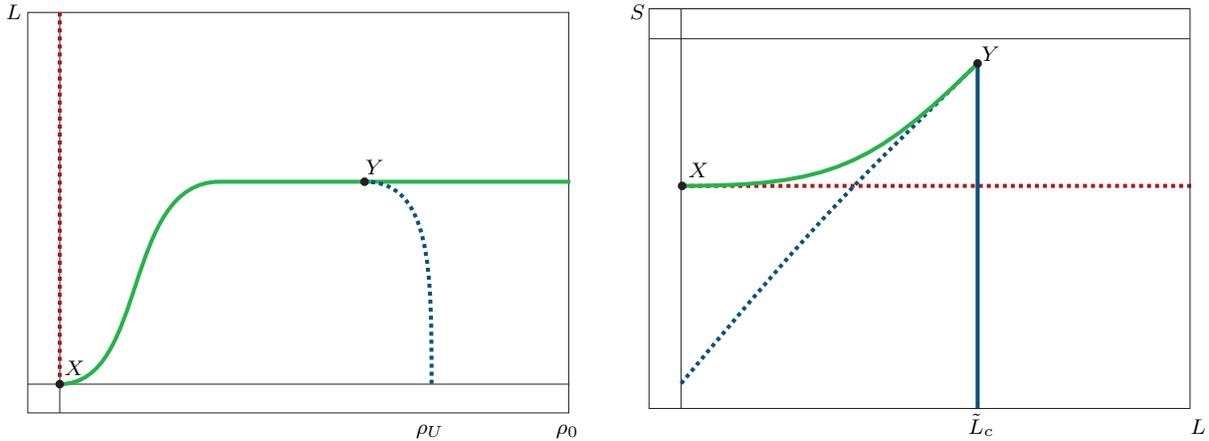
\begin{figure}[ht]
\begin{center}
\begin{picture}(220,170)
\put(-110,-20){\includegraphics[height=6.5cm]{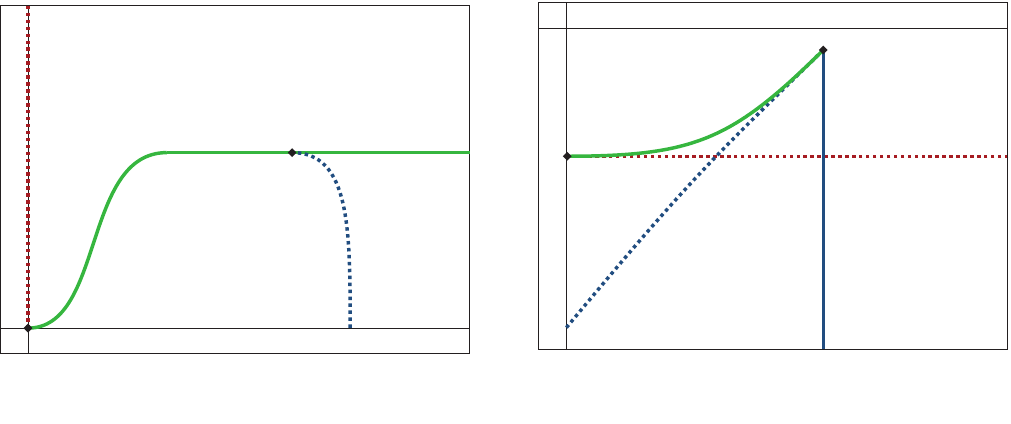}}
\put(-118,160){{\scriptsize{$L$}}}
\put(90,3){{\scriptsize{$\rho_{0}$}}}
\put(37,3){{\scriptsize{$\rho_{U}$}}}
\put(-96,25){{\scriptsize{$X$}}}
\put(18,101){{\scriptsize{$Y$}}}
\put(118,160){{\scriptsize{$S$}}}
\put(330,3){{\scriptsize{$L$}}}
\put(140,100){{\scriptsize{$X$}}}
\put(251,144){{\scriptsize{$Y$}}}
\put(246,3){{\scriptsize{$\tilde{L}_{c}$}}}
\end{picture}
\caption{Here are cartoons of the behaviour in a background like that of the D5 branes wrapped on an $S^{1}$.  The dashed red line represents the disconnected solution, and the green that of the confinement branch, which join at the point $X$.  With finite UV cutoff $\rho_{U}$, we would find something similar to the dashed navy blue branch in both plots for solutions near the cutoff scale.  If we increase the UV cutoff (meaning the point $Y$ moves to larger $\rho_{0}$), we find that the gradient of the UV branch becomes steeper, such that in the limit that we remove the cutoff completely, it becomes the vertical solid navy blue line and we reproduce exactly the extensive solutions.}
\label{Fig:UVcutoff}
\end{center}
\end{figure}

Furthermore, it is possible to have a phase transition
between these two types of behaviour
(Volume Law $\leftrightarrow$ Area Law).
We will find that the Volume Law behaviour is always linked
with the non-local UV behaviour of our theories.
So, we may wonder what this implies for
the case of the D5 branes wrapped on $S^{1}$ ?
In this case,
we are introducing a confinement scale and thus as discussed above (with the example of the soft-wall) we will introduce an {\it{unstable}} branch, joining our disconnected solutions, which have a degenerate point at one end of the confinement branch (point $X$ in the right panel of
Figure \ref{Fig:UVcutoff}),
to our near UV solutions,
which have a degenerate point at the other end of the
confinement branch (point $Y$ in the
right panel of Figure \ref{Fig:UVcutoff}).
The practicalities of realising this UV branch of solutions
will be discussed shortly.

Note that in the cartoon in Figure \ref{Fig:UVcutoff}, the introduction of solutions like $B$ (c.f. `short solutions' in Figure
\ref{Fig:shape}) mean that for the EE, there is now a phase transition between the disconnected
solutions and the extensive solutions at the point $\tilde{L}_{c}$.  In
the cases we shall consider,
we can argue that a transition to a region in the UV
where an extensive behaviour of the EE is the minimal solution, is a sign of
non-locality \cite{Barbon:2008ut},\cite{Fischler:2013gsa},\cite{Karczmarek:2013xxa},
but further to that, a sign that one may want to look to UV complete these theories in a non-trivial way,
if they are to be correct duals to ``nice" field theories. Let us now be more precise about the `short solutions'.

\subsection{Finding the `Short Configurations'}

Now we would like to motivate the existence of these new
`short configurations' explicitly and then use them as a completion
for some specific EE diagrams.  To this end we will use
a particular {\it approximation}.
We let our surface be rectangular in shape, such that the sides
follow the same path as the disconnected surface
from $\rho=\rho_{U}$ to $\rho=\rho_{0}$
 (we can take $\rho_{U}\to\infty$ later).
Then we connect the two vertical surfaces $A_1$ with the
horizontal surface $\hat{L}$ at constant $\rho_{0}$
(depicted as the solid purple lines in the right panel of
Figure \ref{Fig:shape}).  Initially, we shall rewrite the
EE of the disconnected solution $S_{D}$, by splitting it
into two parts joined at $\rho_{0}$,
\begin{equation}
S_{D}=2(A_1+A_2)+\hat{L}_0
\end{equation}
where $A_1$ is the contribution from
$\rho_{U}$ down to $\rho_{0}$, while
$A_2$ is the contribution from $\rho_{0}$ down
to the end of space $\rhol$.
Note that $\hat{L}_{0}$ would be a contribution from the horizontal
piece at the end of space which is
vanishing in the cases we consider.  From this, we can now write
the surface area of our
approximating surface as
\begin{equation}
S_{app}= 2 A_1+\hat{L}.
\end{equation}
It should be noted here  that the
approximate surfaces $S_{app}$ are {\it not} extremal surfaces
as they are not proper solutions of the equations of motion.  Nevertheless, we note that for a
fixed value of $\rho_{0}$, $A_1$ and $A_2$ are strictly constant, while $\hat{L} \propto L$, thus
when $L\to 0$ then $\hat{L}\to 0$. {\emph{This means that, no matter how small $A_2$ is,
there exists small enough values of $L$ such that we have $\hat{L}<2 A_2$, and thus $S_{app}<S_{D}$}}.
The existence of these configurations would indicate, that for small $L$ there will exist
solutions that have lower EE than the disconnected case, and
that there exists actual extremal solutions also with lower
EE than the disconnected one.


Now let us derive the precise formula for the area of $S_{app}$.
Starting with $\hat{L}$, we notice that
since the surface is volume filling in all but
the $\{x_1,\rho\}$ directions, we have
\begin{equation}
 \hat{L}=\int \prod_{i=1}^{8-d} \prod_{j=1}^{d} d\theta_i dx_j \sqrt{g_{\text{ind}}}\int du\sqrt{g_{\mu\nu}\dot{x}^{\mu}\dot{x}^{\nu}}.
 \end{equation}
Using the parametrisation $x^{\mu}=\{\hat{\r},u\}$, with
$u\in\{-\frac{L}{2},\frac{L}{2}\}$, we can easily deduce that
\begin{equation}
 \hat{L}\propto V_{int}\a^{\frac{d}{2}}e^{-2\phi}|_{\r=\r_0}L
=\sqrt{H(\rho_0)} L
\end{equation}
The two sides $A_{1}$ are given by
\begin{equation}
  2 A_1 = 2 \int _{\rho_{0}}^{\rho_{U}} d\rho \sqrt{\beta(\rho)H(\rho)}
\end{equation}
Note that $A_1$ is divergent for $\r_U \to\infty$.
Thus we renormalise using the same approach
as for the extremal solutions; we subtract the
disconnected surface area $S_{D}=
2\int _{\rhol}^{\rho_{U}} d\rho \sqrt{\beta(\rho)H(\rho)}$.
Thus overall we have\footnote{Note that we have added the extra multiplicative factor to take care of the sharp edges of
our surfaces that otherwise make the surface a worse
approximation to the actual minimal solutions.}
\begin{equation}
S_{app}(L)=\frac{2}{\pi} V_{int}\a^{\frac{d}{2}}e^{-2\phi}|_{\rho_0}
L-2\int _{\rhol}^{{\r_0}} d\rho  \sqrt{\beta(\rho)H(\rho)}.
\end{equation}
Notice that this last formula is an approximation of the expression in \eqref{S}.
We now study what happens in the examples of the last section.

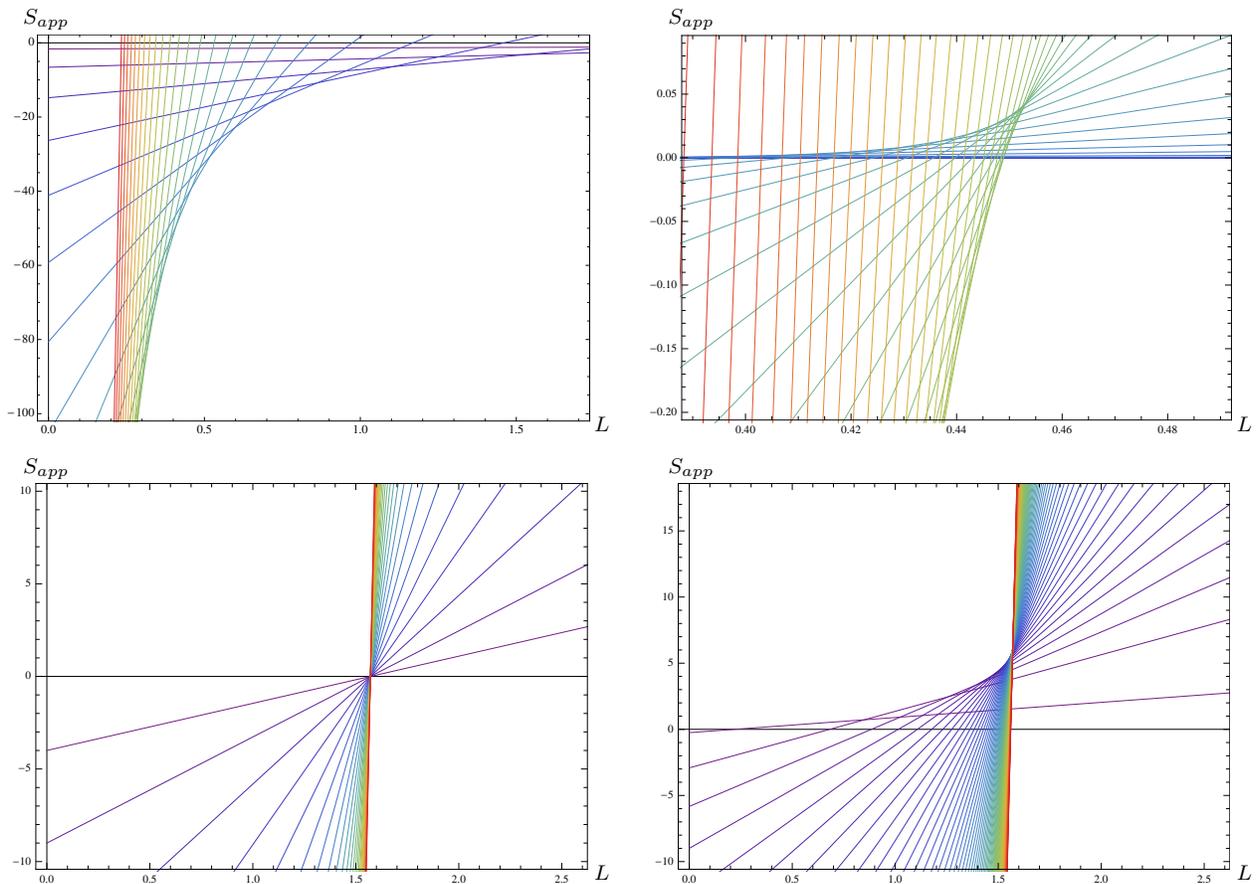
\begin{figure}[ht]
\begin{center}
\begin{picture}(220,310)
\put(-122,0){\includegraphics[height=5.35cm]{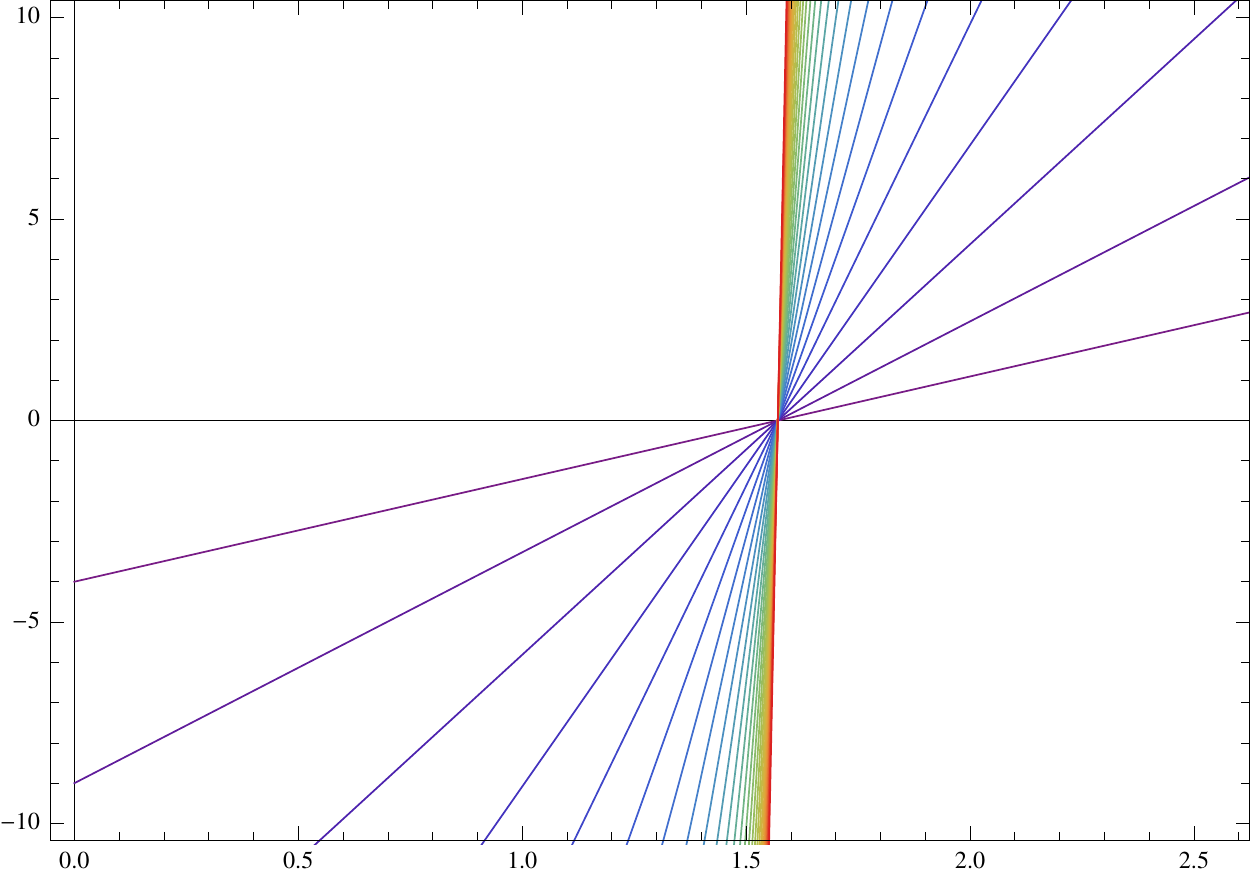}}
\put(121,0){\includegraphics[height=5.35cm]{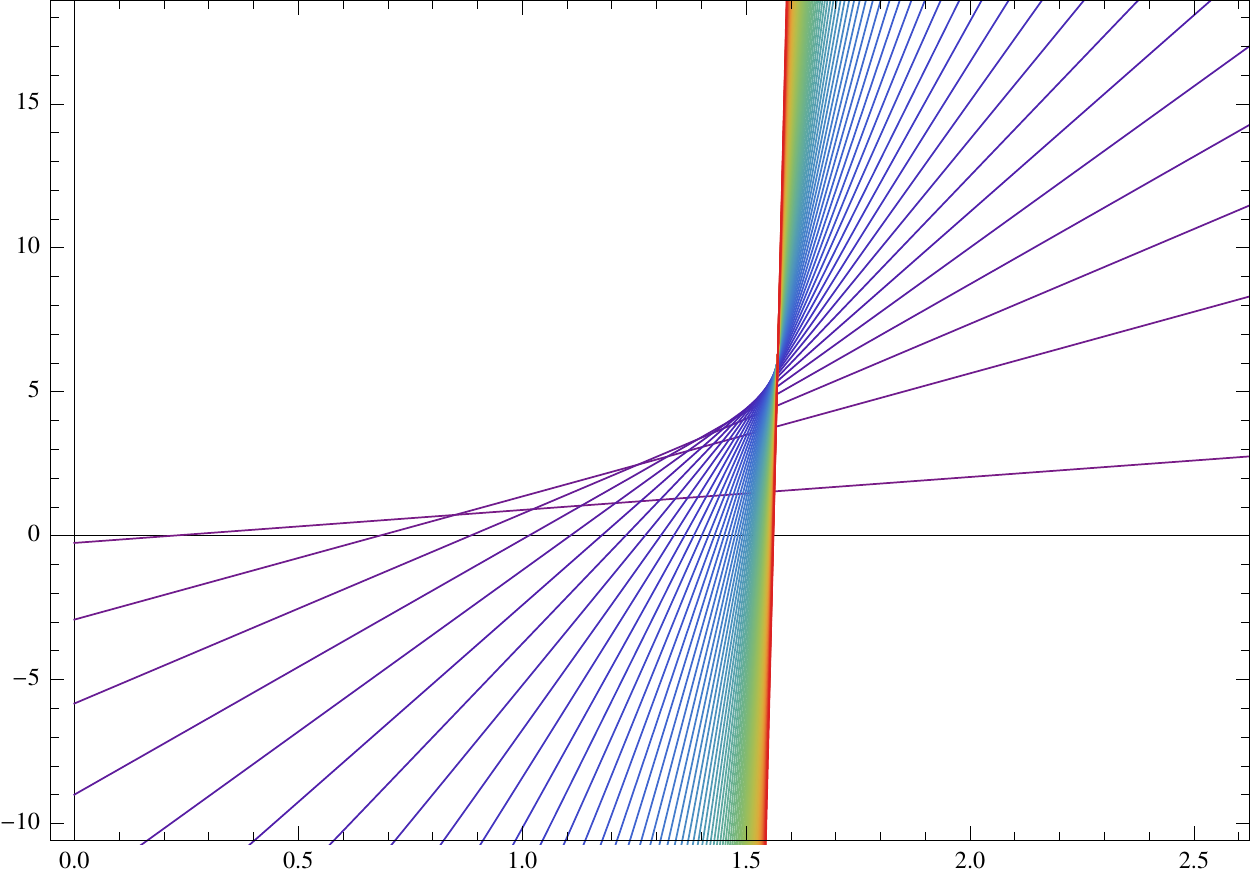}}
\put(-124,170){\includegraphics[height=5.35cm]{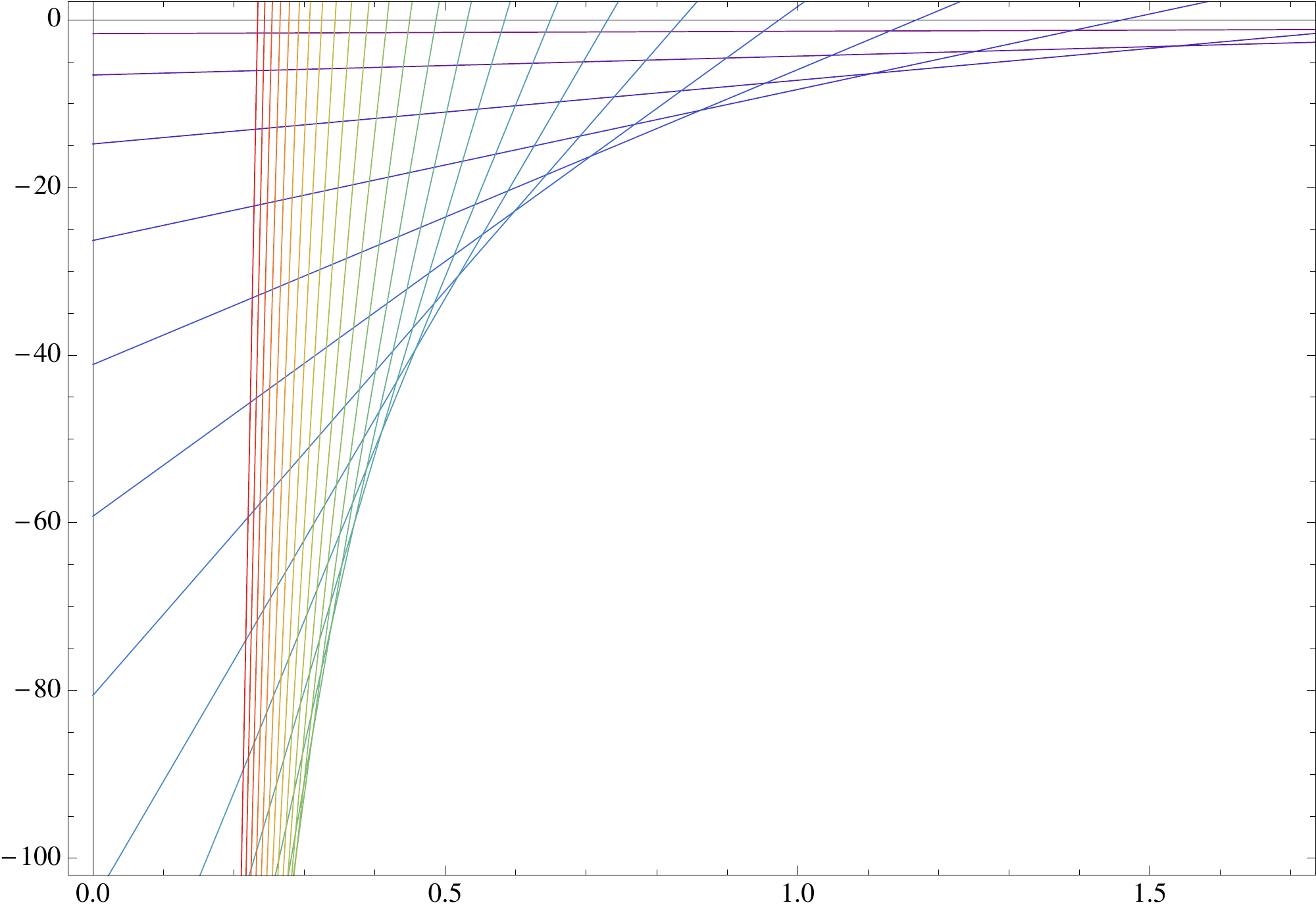}}
\put(119,170){\includegraphics[height=5.35cm]{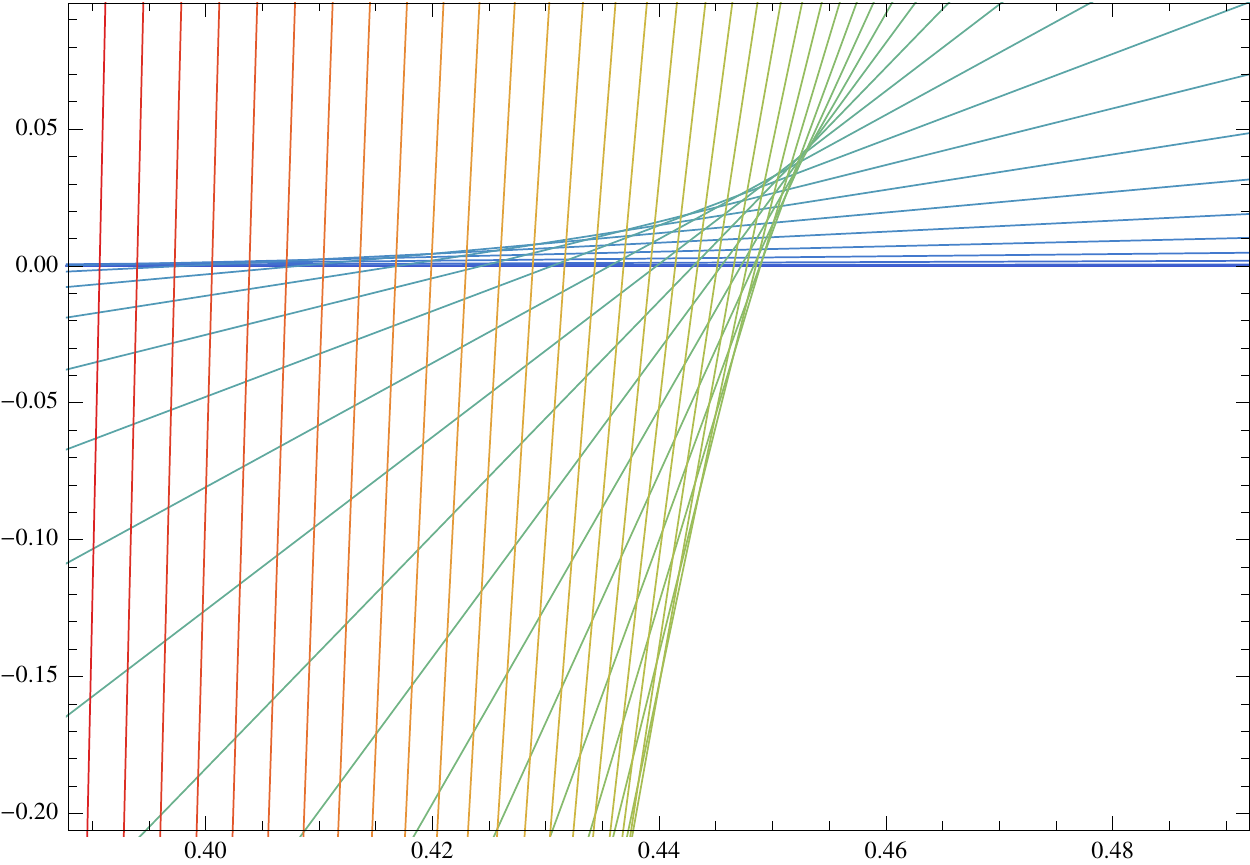}}
\put(-118,156){{\scriptsize{$S_{app}$}}}
\put(98,2){{\scriptsize{$L$}}}
\put(-118,326){{\scriptsize{$S_{app}$}}}
\put(98,172){{\scriptsize{$L$}}}
\put(126,326){{\scriptsize{$S_{app}$}}}
\put(341,2){{\scriptsize{$L$}}}
\put(126,156){{\scriptsize{$S_{app}$}}}
\put(341,172){{\scriptsize{$L$}}}
\end{picture}
\caption{In the above we have plotted a number of the surfaces $S_{app}(L)$. The colour scheme is such that purple lines are surfaces with $\rho_{0}$ approaching $0$, and the red solutions which have $\rho_{0}$ approaching $\rho_{U}$. The top-left panel is that of $AdS_{5}\times S^{5}$, then in the top-right the Soft-Wall, the bottom-left is flat D5 branes and the bottom-right is D5 branes wrapped on $S^{1}$.}
\label{Fig:surfaces}
\end{center}
\end{figure}

If we take $\r_{0}\to \rhol$ we will have that
$V_{int}\alpha^{d/2} e^{-2\Phi}\to 0$ --- as we observed,
it happens in all our models that $H_{EE}(\rhol)\to 0$---see eq.(\ref{MEE}).
We then recover the disconnected solutions
and due to our renormalisation scheme,
it is easy to see that the solution will always sit on top of the $L$-axis
of our $S_{app}(L)$ plots.
This is a feature of all backgrounds studied here.
In the other limit $\rho_{0}\to\rho_{U}$
our surface becomes the $S_{app}$-axis of the
$S_{app}(L)$ plot, and we have a smooth interpolation in
between --- see the top left of Figure
\ref{Fig:surfaces}.
Note that these lines map out the actual connected solution and thus
approximate this case very well.
A quantity that proves to be very useful is the point at which the surfaces cross the $L$-axis. We know that $S_{app}=0$ whenever
\begin{equation}
L=\frac{\pi \int _{\rhol}^{\rho_{0}} d\rho
\sqrt{\beta(\rho)H(\rho)}}{V_{int}\a^{\frac{d}{2}}e^{-2\phi}|_{\rho_0}}
\equiv T(\hat{\r})
\end{equation}
Thus in cases
like that of $AdS_{5}\times S^5$,
$T(\hat{\r})$ will be monotonically decreasing
function with $T$ varying between $(\infty,0]$
as $\rho_{0}$ varies between $[0,\infty)$.

If we now study
the soft-wall case (see the top-right panel of Figure \ref{Fig:surfaces}),
we can see that the solutions again near the UV look similar
to that of the $AdS_{5}\times S^{5}$ case above,
as expected.  The difference lies in the solutions near the IR,
where the surfaces initially begin to move into the
positive plane of the $S_{app}$ plot, meaning they have higher
EE than the disconnected solution.
A phase transition
appears as one would expect.  This change can also be seen
in the behaviour of $T(\rho_{0})$.
Indeed, now $T\to\{0,0\}$ as $\rho_{0}\to\{0,\infty\}$.
Additionally $T(\rho_{0})$ is an increasing function for small
$\rho_0$ (below the confinement scale)
and a decreasing function above it.


Now let us discuss backgrounds which exhibit non-locality.
First we look at the case of D5 branes as studied
in \cite{Barbon:2008ut} and depicted in the bottom-left panel of Figure
\ref{Fig:surfaces}.  We see that the surfaces all
intercept the axis at $L=\pi/2$ (as we have set $R=1$)
with increasing gradient as we move towards the UV.
Thus in the limit we would expect to find a vertical line at $L=\pi/2$.
 This agrees with the expectations of our discussion above.

Finally, we move to the D5 wrapped on $S^{1}$ that mostly
occupied us in this section.
The associated plot can be seen in the
bottom-right panel of Figure \ref{Fig:surfaces}.
Here we find that the surfaces initially move up into
the positive plane of the $S_{app}$ plot
and they then asymptote the same value as the flat
D5 solution leading again to a vertical line at $L=\pi/2$
(choosing $R=1$) in the UV.
The two D5 cases discussed differ in a subtle point.
While the phase transition in the two cases is always
between Area $\leftrightarrow$ Volume Law behaviour,
in the flat D5 case, all the connected solutions sit at the transition point.

These `short configurations' living at the cutoff
appear and play an important role, every time
we have a non-local QFT. They are {\it needed} in order to
avoid having only a connected-unstable and the disconnected solutions.
The short configurations imply the existence of a phase
transition between connected-stable and disconnected branches of the EE
\footnote{Notice that we are taking
limits in a given order; first $\rho_0\to \rho_U$
followed by $\rho_U\to \infty$. This is very
reminiscent of the treatment in
\cite{Faedo:2014naa}}.

In the next section, we will study another example of this `cutoff effect'
when studying the behavior of the EE in one of the `trademark'
models of confining field theory (but with a non-local UV).
We will also learn that a similar job as the one done by the cutoff
can be done by a UV-completion of the QFT. 
This gives a well-behaved EE-phase transition,
with an area law for the divergent part of the EE, etc.

\section{The Absence of Phase Transitions in (some) Confining Models}\label{sectionabsence}

The reader might object that the example of D5 wrapped on $S^1$ discussed
above is not such a good model for a confining field theory.
In principle theories in $4+1$ dimensions have strongly coupled UV
behavior, their IR tends to be weakly coupled and what we observed
is just an effect of these features, in contradiction with us
imposing the model to be confining.
One may imagine
that for duals to field theories in $3+1$ dimensions
the phase transition should reappear.

Below, we will analyse
this claim, by first studying the case of a dual QFT obtained
by wrapping $N_c$ D5 branes on a two-cycle of the resolved conifold
\cite{Maldacena:2000yy}. We will discover that the behaviour
in the UV is not much different from the one of the flat-D5 brane
just analysed above. The reader may argue that this is due to
the fact that at
energies high enough, the dual QFT becomes higher dimensional, an infinite
set
of KK-modes coming from the compactification of the D5
branes on the two-cycle indicate the
higher dimensional character of the QFT.
This is of course correct, but the point is subtle. Indeed, as
Andrews and Dorey
\cite{Andrews:2006aw} have
shown (in the perturbative regime), the field theory is completely
equivalent to four-dimensional $N=1^*$ Yang-Mills,
expanded at a particular point
of its Higgs branch, which is a well-defined 4-d QFT.
Also, the same sort of KK-modes appear if we compactify a 
stack of D4 branes on $S^1$ and  
in that confining model the phase transition is present,
see \cite{Klebanov:2007ws}
 and section \ref{relationsWEE}.
We will then carefully calculate the
entanglement entropy for this QFT based on wrapped D5 branes.

Let us start by writing explicitly the metric describing
D5 branes wrapping a two-cycle inside the resolved conifold.
In string frame, we have a (dimensionless) vielbein,
\begin{align}
&e^{x^i}= \frac{e^{\frac{\Phi}{2}}}{\alpha' g_s}dx^i\,    ,\qquad e^{\r}
=  e^{\frac{\Phi}{2}+k}d\r\,  ,
\qquad e^{\theta}=  e^{\frac{\Phi}{2}+h}d\theta\,  ,
\qquad e^{\varphi}= e^{\frac{\Phi}{2}+h} \sin\theta\, d\varphi \,, \nonumber\\
& e^{1}=  \frac{1}{2}e^{\frac{\Phi}{2}+g}(\tilde{\omega}_1 +a\, d\theta)\,
,\;\; e^{2}=\frac{1}{2}e^{\frac{\Phi}{2}+g}(\tilde{\omega}_2
-a\,\sin\theta\, d\varphi)\,,\;\; e^{3}= \frac{1}{2}e^{\frac{\Phi}{2}+k}
(\tilde{\omega}_3 +\cos\theta\, d\varphi)\,. \nonumber
\end{align}
The quantities $\tilde{\omega}_i,\; i=1,2,3$ are the left-invariant
forms of $SU(2)$. In the string frame we have a metric,
\begin{equation}
\begin{aligned}
ds_{str}^2	&= \alpha' g_s   \sum_{i=1}^{10} (e^{i})^2\,.
\end{aligned}
\end{equation}
The background is completed with a  dilaton $\Phi(\r)$ and a RR three-form
that we will not need to write here --- see, for example
Section II in the paper
\cite{Elander:2011mh}
for a complete
description of the system.

It is sometimes
more efficient, to `change the basis'  to describe
the background and RR fields from the functions $[a, h, g, k, \Phi]$
to another set of
functions $[P, Q, \tau, Y, \Phi]$. This is useful
because in terms of the second set, the BPS equations decouple.
It is then possible to solve the
non-linear ordinary BPS equations, so that everything is left in terms
of a function $P(\r)$, that satisfies a non-linear
ordinary second order differential
equation.
The `change of basis' is explicitly given in
eq.(3) of the paper \cite{Elander:2011mh}.
We summarise it here for future reference. After having solved
for $[Y, Q, \tau, \Phi]$ and choosing integration constants to avoid
singularities, we have
\bea
& & 4 e^{2h}=\frac{P^2-Q^2}{P\coth(2\r)-Q},\;\; e^{2g}=P\coth(2\r)-Q,\;\;
2e^{2k}=P',\;\; ae^{2g}\sinh(2\r)=P,\nonumber\\
& & e^{4\Phi-4\Phi_0}= e^{-(2h+2g+2k)}\sinh(2\r)^2, \;\;\;\;
Q(\r)=N_c(2\r\coth(2\r)-1).
\label{changePQ}
\eea
The function $P(\r)$ satisfies a second order non-linear differential equation,
sometimes called `master equation' in
the bibliography \cite{HoyosBadajoz:2008fw}.
Different
solutions to the master
equation have been discussed and classified in
\cite{HoyosBadajoz:2008fw}, \cite{Gaillard:2010qg},
\cite{Conde:2011aa}.

The functions needed to calculate the entanglement
entropy in this string-frame background are,
\bea
\a=e^{\Phi},
\quad \b=\alpha' g_s e^{2k},
\quad V_{int}^{2}=(2\pi)^6(\alpha' g_s )^5e^{4h+4g+5\Phi+2k},
\;\;H=(2\pi)^6 (\alpha' g_s)^5 e^{4\Phi+4g+4h+2k}.
\label{MNEEquants}
\eea
One simple solution of this `master' equation
for the function $P(\r)$, leading to a smooth background,
is known analytically and given by $P(\r)= 2 N_c \r$. This is the solution
in \cite{Maldacena:2000yy}.
One can check by replacing in
eq.(\ref{changePQ}) that the dilaton behaves
as $e^{4\Phi}\sim \frac{e^{4\r}}{\r}$.
We obtain a behavior similar
to the one around eq.(\ref{rararara}), considering the change in the
radial variable between one description and the other
(that for large radius is $\r\sim\log r$), both are examples
of `linear dilaton' backgrounds.

Indeed, calculating the entanglement entropy $S(\r_0)$,
the separation between regions $L(\r_0)$ and then plotting parametrically
$S(L)$, we find
that there is no minimal solution present,
only the disconnected and the unstable connected ones.
The latter exists over a finite range for
$0<L<\frac{\pi}{2}\sqrt{g_s\alpha' N_c}$
with $S_c(0)=0$ and $S_c(\frac{\pi}{2}\sqrt{g_s\alpha' N_c})=\infty$.
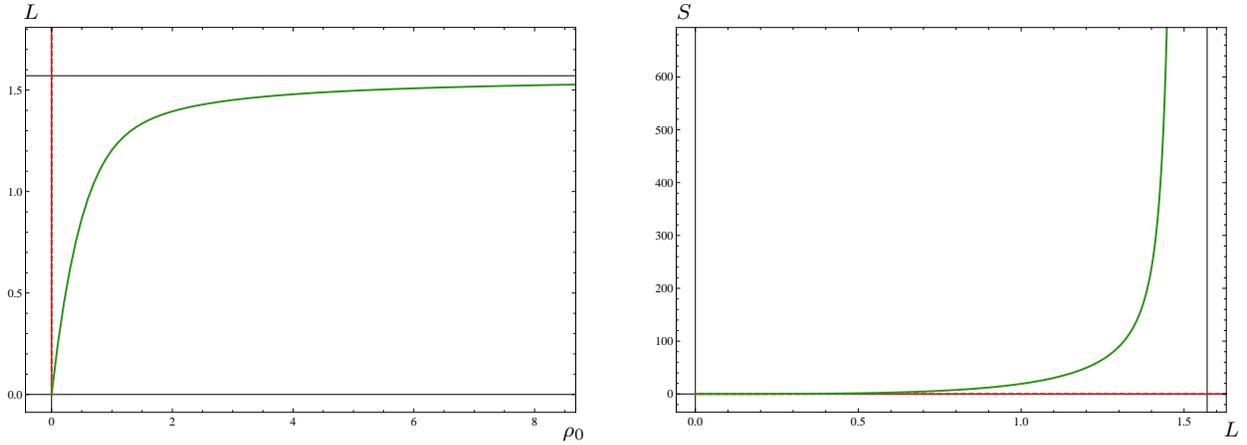
\begin{figure}[h]
\begin{center}
\begin{picture}(220,160)
\put(-125,0){\includegraphics[height=5.35cm]{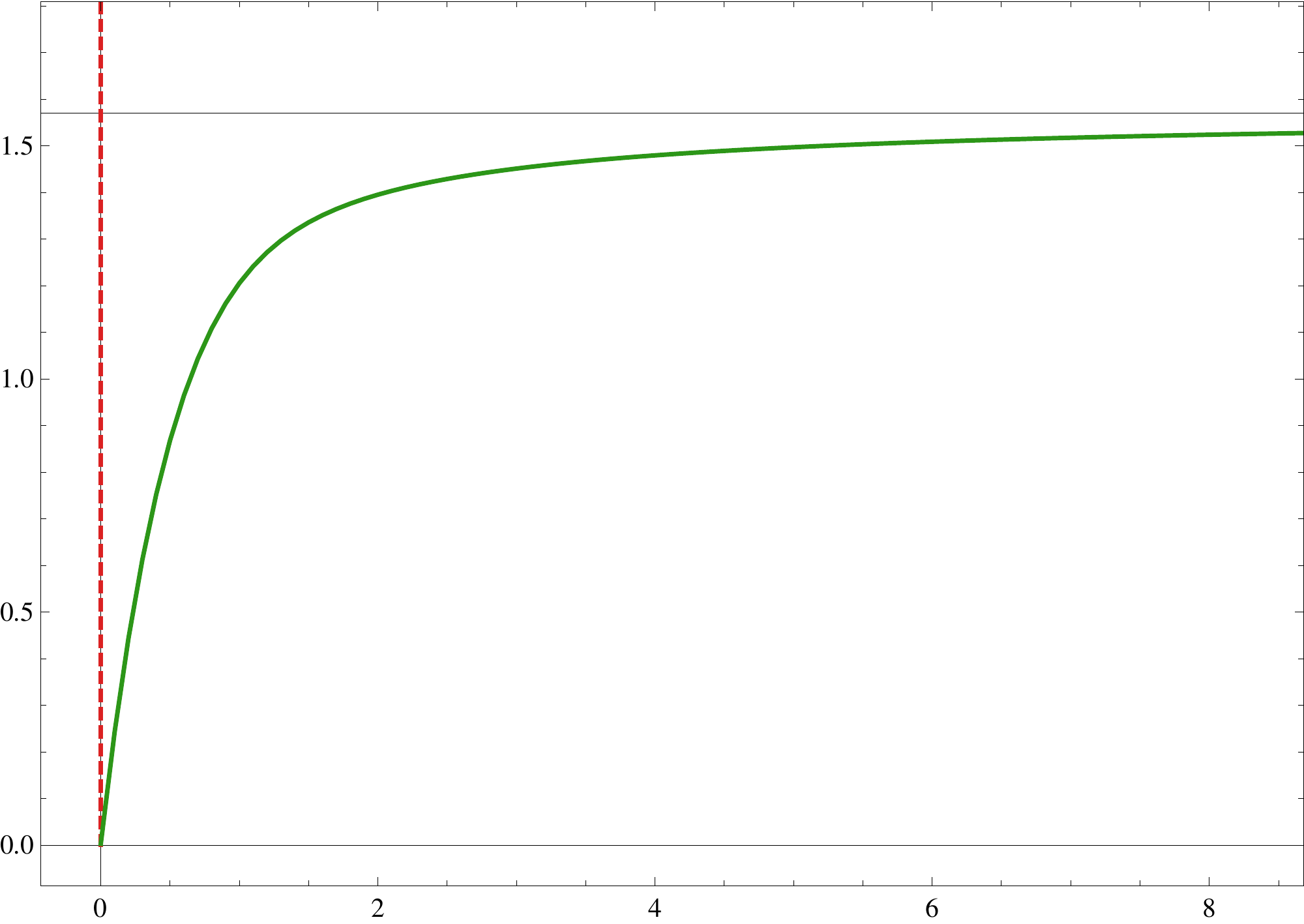}}
\put(-119,155){{\scriptsize{$L$}}}
\put(85,-3){{\scriptsize{$\r_{0}$}}}
\put(120,0){\includegraphics[height=5.35cm]{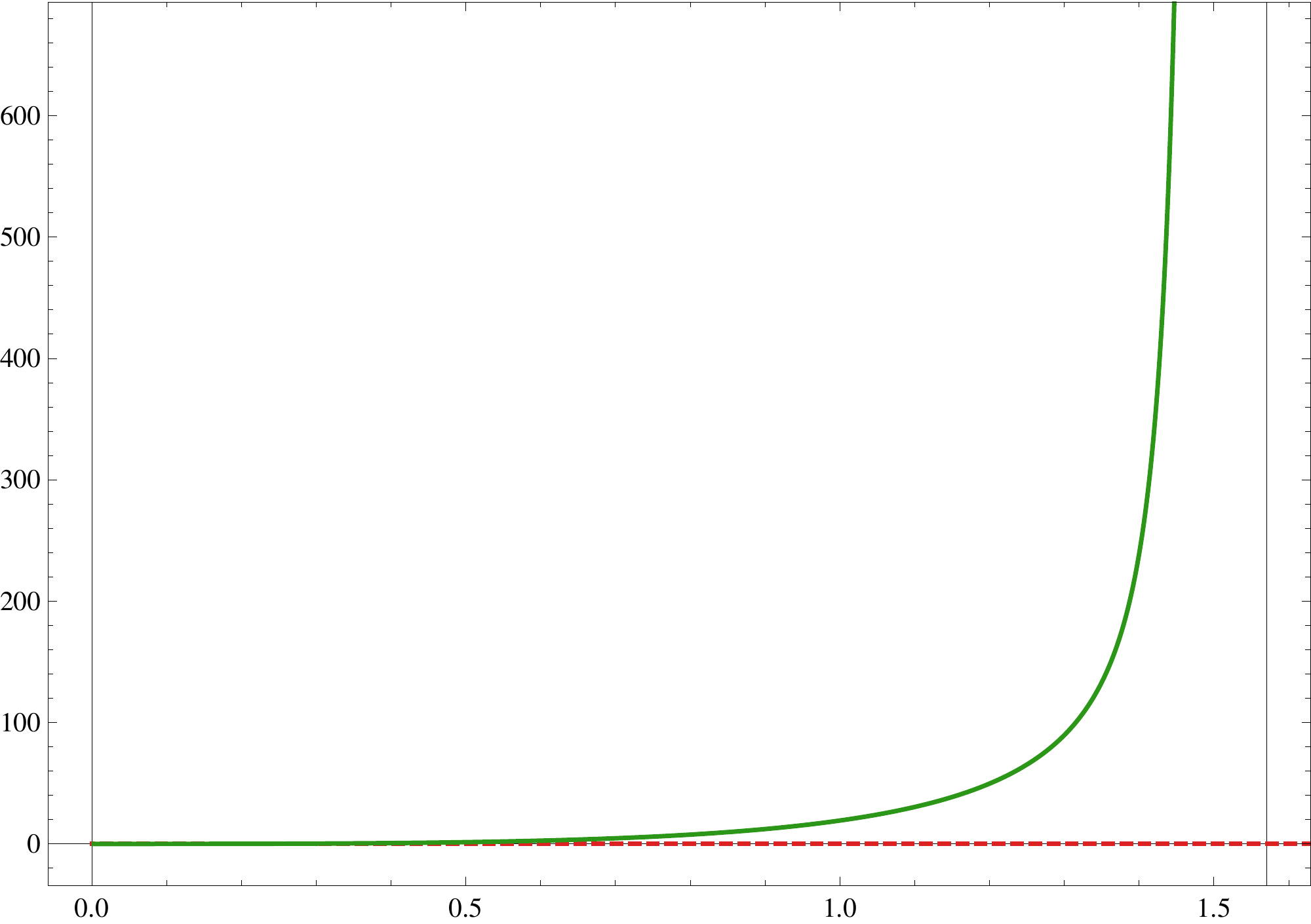}}
\put(128,155){{\scriptsize{$S$}}}
\put(335,-3){{\scriptsize{$L$}}}
\end{picture}
\caption{The system of D5s on a two-cycle --- Here we plot $L(\rho_0)$ and $S(L)$.}
\label{Fig:MN}
\end{center}
\end{figure}
Using the approximation of eq.(\ref{yderho}),
we find,
\bea
& & L(\r) \sim {\cal Y}(\r)= \sqrt{\alpha' g_s}\pi
\frac{e^{k}}{2\Phi' +2 h'+2g'+k'}= \label{LinfMN}
\\
& & {\cal Y}(\r)= \frac{\pi\sqrt{\alpha' g_s P' }}{\sqrt{2}}
\frac{(P^2-Q^2)}{\big(2P^2 \coth(2\r) + PP' - QQ' -2 Q^2\coth(2\r)\big)}.
\nonumber
\eea
Which for the exact  solution
\bea
P=2 N_c\r,\;\;\; Q= N_c(2\r \coth(2\r)-1),\nonumber
\eea
gives the approximated asymptotics for the function $L(\r_0)$,
\bea
L(\r_0\to\infty)\sim \frac{\pi\sqrt{\alpha' g_s N_c}}{2}
(1-\frac{1}{4\r_0}),\;\;
L(\r_0\to 0)\sim \frac{\pi\sqrt{\alpha' g_s N_c}}{2}\rho_0.\nonumber
\eea
notice that the Heisenberg-like relation
$L\sim \rho_0^{-1}$ is violated here.
Also note that the entropy scales as
\bea
\frac{G_{10}S_c}{V_2}\sim (\alpha' g_s)^3N_c^{\frac{3}{2}}.\nonumber
\eea
Like in the example of the compactified
D5 branes on $S^1$ above, we see that the connected configuration
is unstable --- not satisfying the concavity
condition of eq.(\ref{concavityxx}).
We do not see the possibility of a phase transition. But here is where the `short configurations'
and the effects of the UV cutoff enter to cure the problem
\footnote{As an aside, it should be noted that the `finite size' effect
reflected in the non-zero value of $L(\r_0\to\infty)$, was observed
also for the Wilson loop when calculated in this sort of backgrounds in
\cite{Casero:2006pt}, \cite{Warschawski:2012sx},
\cite{Conde:2011rg}.}.

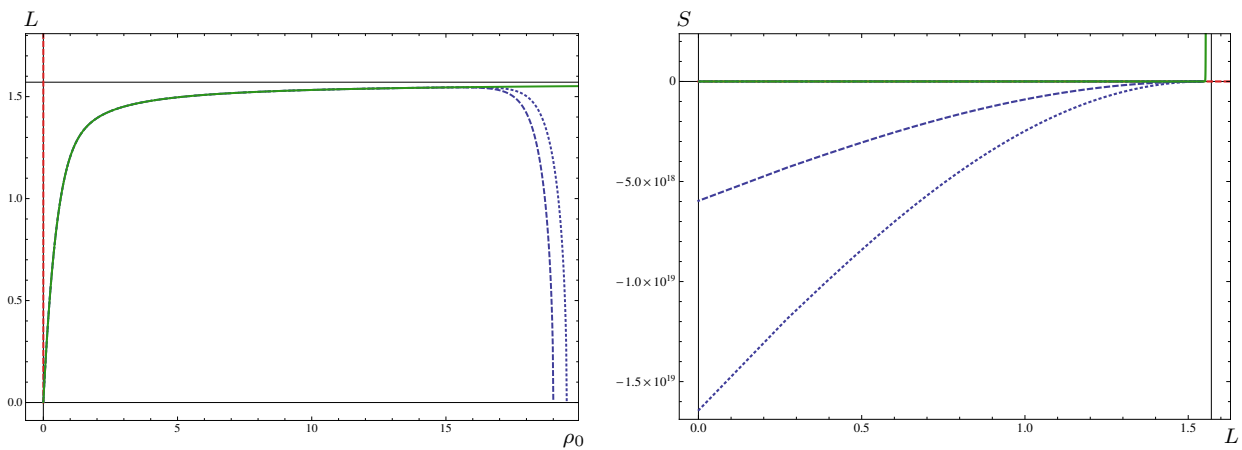
\begin{figure}[h]
\begin{center}
\begin{picture}(220,160)
\put(-125,0){\includegraphics[height=5.35cm]{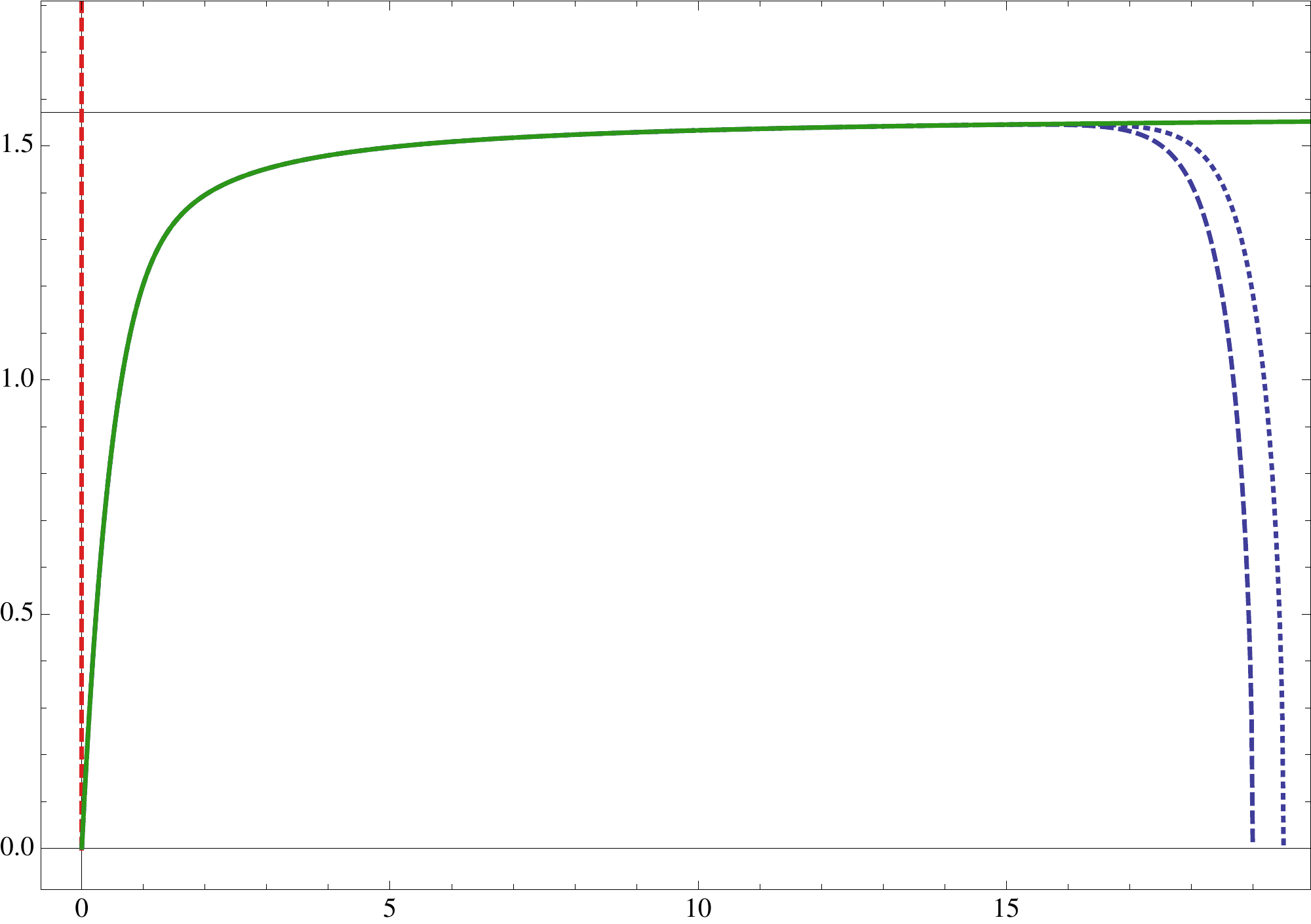}}
\put(-119,155){{\scriptsize{$L$}}}
\put(85,-3){{\scriptsize{$\r_{0}$}}}
\put(105,0){\includegraphics[height=5.35cm]{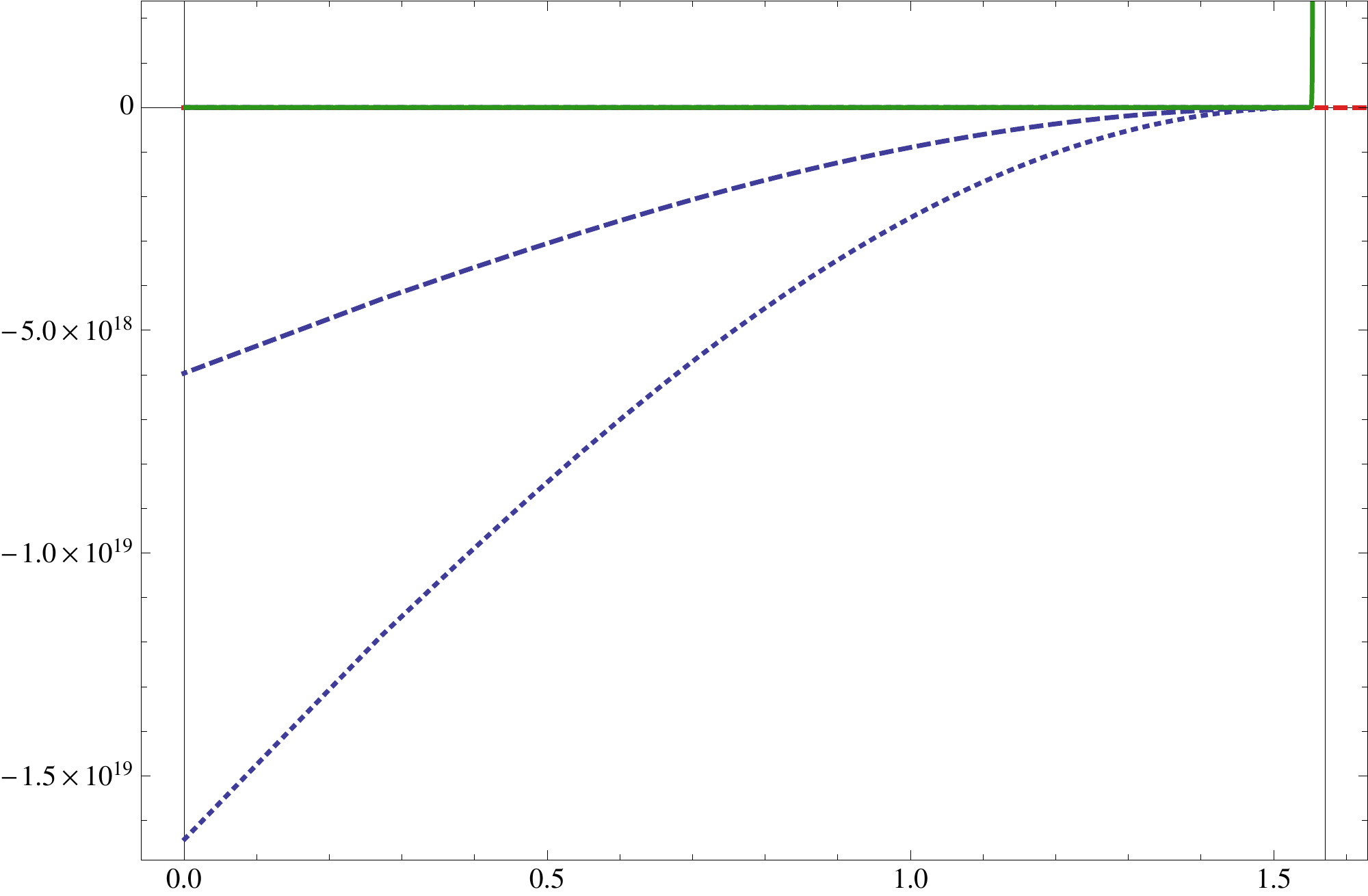}}
\put(128,155){{\scriptsize{$S$}}}
\put(335,-3){{\scriptsize{$L$}}}
\end{picture}
\caption{The system of D5s on a two-cycle --- Here we plot $L(\rho_0)$ and $S(L)$ again, but this time introducing a UV cutoff.  The green line is the solution without UV cutoff, the dashed blue line is with the cutoff at $\rho_{U}=19$, the dotted blue is with the cutoff at $\rho_{U}=19.5$, and finally the dashed red line represents the disconnected solution. Notice that increasing the value of $\rho_{U}$ leads to an increase in the gradient in the visible branch in the $S(L)$ plot in the right panel.}
\label{Fig:MNapp}
\end{center}
\end{figure}
In Figure \ref{Fig:MNapp} we show the result of considering these short configurations
at the UV-cutoff. We see that these are the {\it correct} configurations
to consider as they avoid the instability issue.
The phase transition in $S(L)$ is recovered, as it corresponds
to a confining model.

\subsection{More on Non-Locality}
To make the point of the
`non-locality' clearer, we will consider another solution
describing D5 branes compactified on the two-cycle
of the resolved conifold (the field theory
has different operators driving the dynamics).
This second solution is well described in
the papers
\cite{HoyosBadajoz:2008fw},
\cite{Gaillard:2010qg} --- see Section 4 of the paper
\cite{Conde:2011aa} for a good summary.

In contrast with the simple analytic
solution --- $P=2N_c\r$; this second solution
is not know analytically. Indeed, only a large radius/small radius expansion
is known. A numerical interpolation between both asymptotic
behaviours is quite easy to obtain.
The large and small radius expansions for the function $P(\r)$ read,
\bea
& & P(\r\to 0)\sim h_1\r + \frac{4(h_1^2 - 4N_c^2)}{15h_1}\r^3+
\frac{16}{1575h_1^3}(3h_1^4 -4h_1^2 N_c^2 -32N_c^4)\r^5+\mathcal{O}(\r^7)\nonumber\\
& & P(\r\to \infty)\sim c e^{4\r/3} + \frac{N_c^2}{c}e^{-4\r/3}
(4\r^2-4\r +\frac{13}{4})+\mathcal{O}\left(e^{-8\r/3}\right)
\label{zzxxaa}\eea
where $h_1, c$ are two integration constants of the `master' equation mentioned above.

We see that the small radius expansion is quite similar to the one of the
exact solution $P=2N_c\r$.
Indeed, for the constant $h_1=2 N_c$ we recover the exact solution.
On the other hand, the large radius expansion of the solution for
the function $P(\r)$ is quite
different from the linear behaviour of the exact solution. These differences and similarities
suggest that the dynamics of the QFT dual to the
second solution is actually afflicted
by an irrelevant operator. In the paper
\cite{Elander:2011mh}, this point was made precise (see also
the discussion in \cite{Conde:2011aa}). The operator can be seen to be of dimension eight.
The situation is not so different from the case of
`keeping the constant factor' in the warp factor of the D3 branes
solution $\hat{h}=1+ \frac{L^4}{r^4}$. Indeed, it can be shown that
the factor `$1$' makes the background
 of $N_c$ D3 branes
dual to $N=4$-SYM with a dimension eight operator inserted.
In order to UV-complete
this QFT one needs to insert back the whole tower of string modes.

\begin{figure}[h]
\begin{center}
\begin{picture}(220,160)
\put(-125,0){\includegraphics[height=5.35cm]{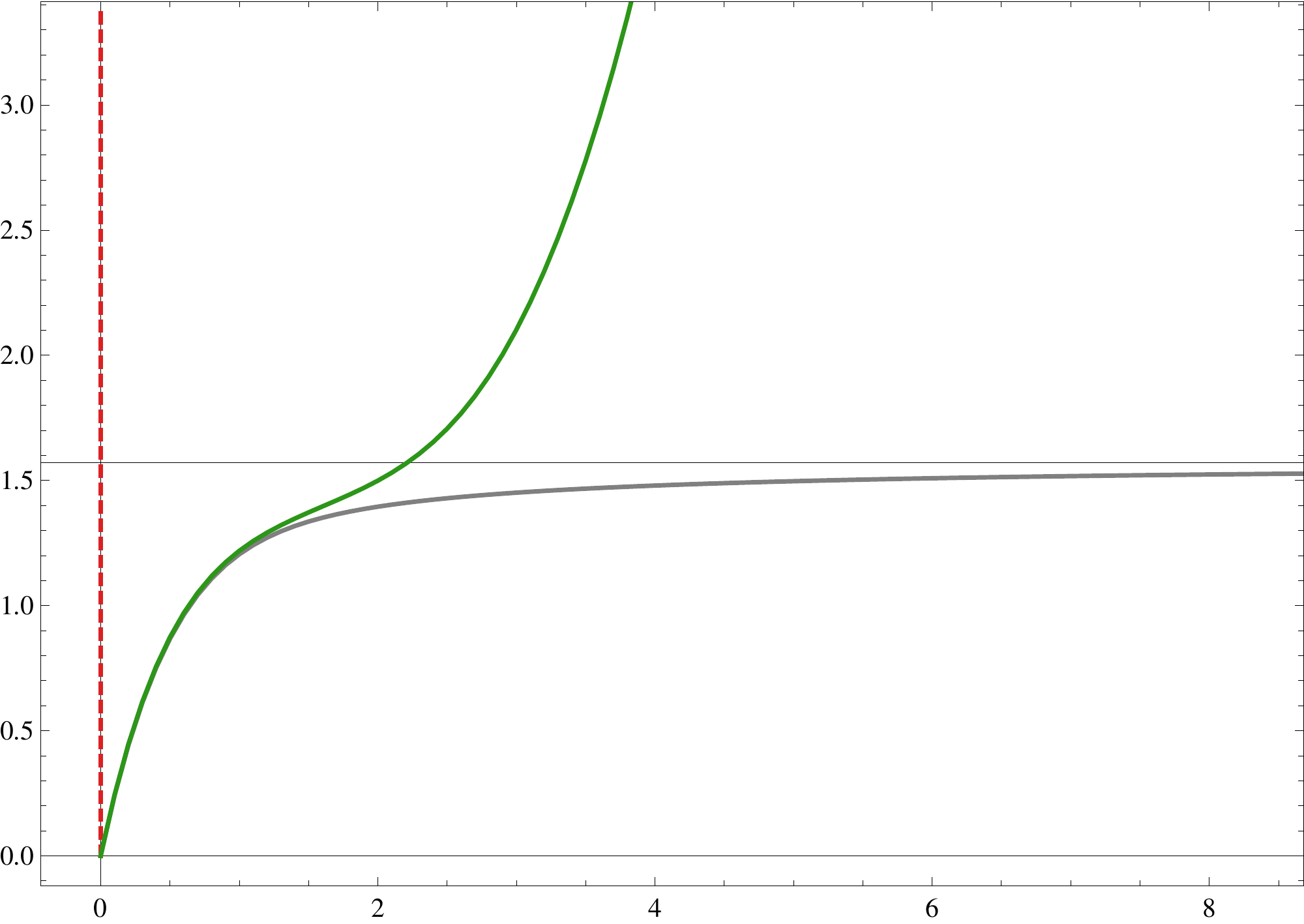}}
\put(-119,155){{\scriptsize{$L$}}}
\put(85,-3){{\scriptsize{$\r_{0}$}}}
\put(116,0){\includegraphics[height=5.35cm]{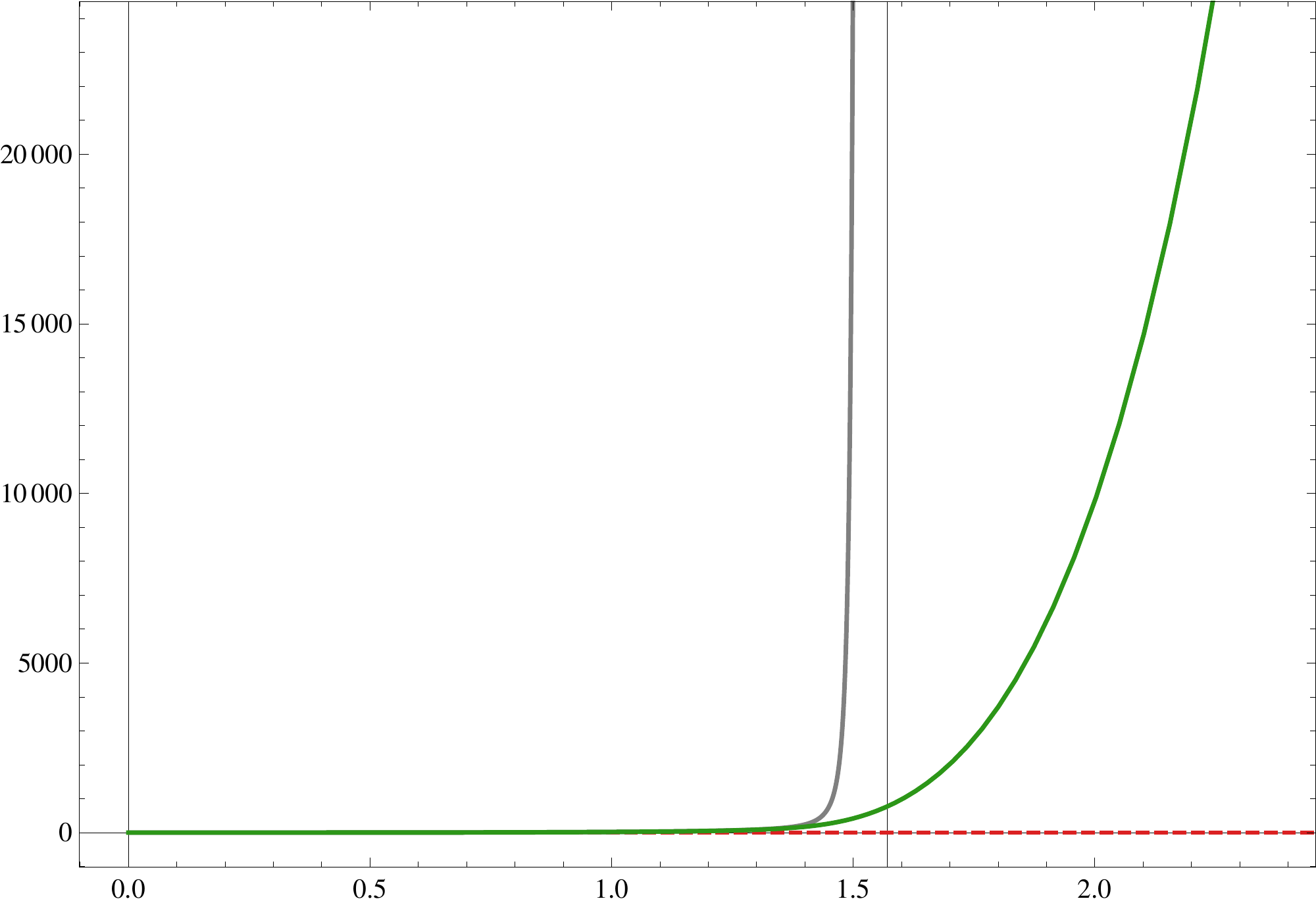}}
\put(128,155){{\scriptsize{$S$}}}
\put(335,-3){{\scriptsize{$L$}}}
\end{picture}
\caption{The system of D5s on a two-cycle but with exponential behaviour in $P$ ($h_{1}=\frac{203}{100} N_{c}$) --- Here we plot $L(\rho_0)$ and $S(L)$.  The grey line is the linear $P$ solution ($h_{1}=2 N_{c}$) for comparison.}
\label{Fig:MNunrot}
\end{center}
\end{figure}
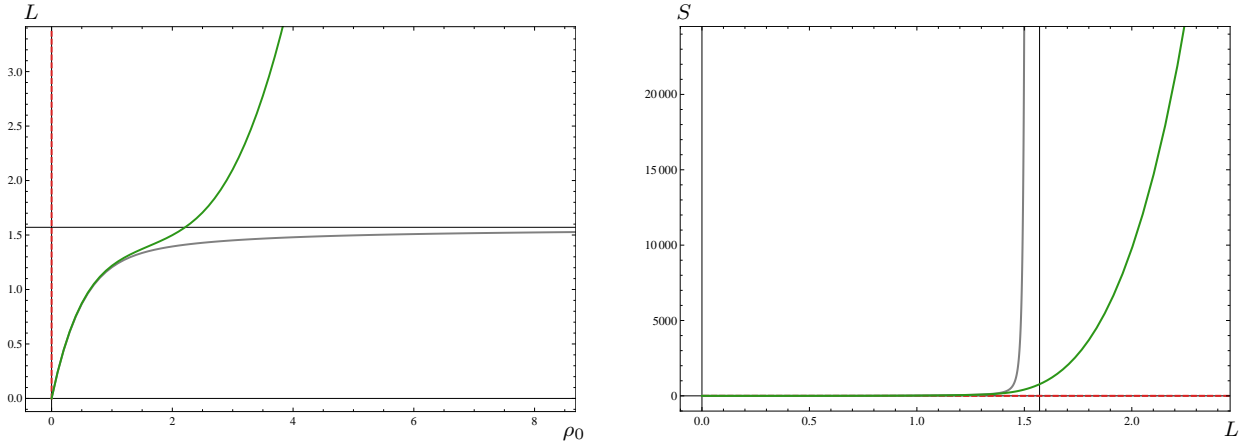

The field theory dual to the background obtained with
the solution $P=2N_c \r$ is afflicted by
{\it less severe non-localities} than the field theory dual to the second
semi-analytical solution of eq.(\ref{zzxxaa}). We can see how the
entanglement entropy reflects this. We recalculate
the numerics for the entanglement entropy $S(\r_0)$
and separation $L(\r_0)$ (we do this for a sample numerical solution
of the form given by eq.(\ref{zzxxaa}), where the irrelevant operator is
inserted with a  small coefficient $h_1=2N_c+\epsilon$).
The result is described in Figure \ref{Fig:MNunrot},
showing that $L(\r_0)$
deviates even more from the needed
double valuedness.
The saddle point solution does exist
for all values of $L$ and there is not
any other connected solution.
Note that whilst $L(\r_0)$ vanishes linearly for small values of
$\r_0$, it grows exponentially in the UV,
$L(\r_0\to\infty)\sim e^{2\r_0/3}$. Of course, this departs even
more from the Heisenberg-like scaling $L\sim \rho_0^{-1}$ characteristic
of local field theories.
All these results
can be obtained from eq.(\ref{LinfMN}) with the solution
in eq.(\ref{zzxxaa}) and  are reflected by Figure \ref{Fig:MNunrot}.

As described above, even when
the string dual shows a confining Wilson loop or area law behaviour
\footnote{A subtle point is that the Wilson loop cannot
be strictly calculated with the solution of eq.(\ref{zzxxaa}).
This is due to the boundary condition for the strings at
infinity, that cannot be satisfied. See \cite{Nunez:2009da}.}, the system
is not showing a phase transition in the entanglement entropy.
All this relies on the UV properties of the field theory. The same phenomena
can be displayed in other systems with qualitative similar field
theoretic high energy behaviour, see Appendix \ref{app:ZZ}.

One may wonder if all hope for observing a phase transition
in the entanglement entropy
(without appealing to the above discussed regulating 
cutoff and short solutions)
is lost.
In the next section, we will discuss a possible UV completion of the
field theory on the compactified D5 branes. This completion, in terms of
an inverse Higgs mechanism is discussed in the papers
\cite{Maldacena:2009mw}, \cite{Elander:2011mh}
and \cite{Gaillard:2010qg}. To this we turn now.
\section{Recovering the Phase Transition: The Baryonic Branch}
\label{recoverbb}
As we discussed above, the non-local UV-properties of the field theory dual
to a stack of $N_c$
D5 branes compactified on the two-cycle of the resolved conifold,
needs of the introduction of a UV-cutoff (and associated effects)
for the existence of a phase transition for the entanglement entropy
(even when the Wilson loop displays an area law). A cleaner way of recovering
the phase transition that the Klebanov-Strassler  system displays
\cite{Klebanov:2007ws},
will be to appeal to the connection between both systems. This connection
is well understood and discussed in the papers
\cite{Maldacena:2009mw}, \cite{Elander:2011mh}
and \cite{Gaillard:2010qg}. A simple U-duality (or equivalently, a rotation
of the $SU(3)$-structure, characterising the D5 brane solution)
connects both backgrounds and this constitutes
the string/geometric version of an
inverse Higgs mechanism; see the discussion in the paper
\cite{Elander:2011mh}.

There is a simple and subtle point we wish to emphasise.
In performing this U-duality or rotation of the $SU(3)$-structure, there
is a constant to be chosen.
Choosing it to a precise value amounts
to UV-completing with the precise matter content such that the UV of the
resulting QFT is as healthy as the Klebanov-Strassler QFT.
See the discussion in \cite{Elander:2011mh} for a careful
explanation of this point. In other words, choosing this constant
in the appropriate way implies that the warp factor of the background
asymptotes to zero and the switching-off of the dimension-eight
irrelevant operator
discussed in the previous section.

The new background, generated by the U-duality or the rotation
on the $SU(3)$-structure forms is described by a (dimensionless)
vielbein, in string frame,
\bea
& & E^{x^i}= e^{\frac{\Phi}{2}}\hat{h}^{-1/4}\frac{dx^i}{\alpha' g_s}, \;\;
E^{\r}=  e^{\frac{\Phi}{2}+k} \hat{h}^{1/4}d\r, \;\;
E^{\theta}=  e^{\frac{\Phi}{2}+h} \hat{h}^{1/4}d\theta,
\;\; E^{\varphi}= e^{\frac{\Phi}{2}+h}\hat{h}^{1/4}
\sin\theta\, d\varphi,  \\
& &E^{1}=  \frac{1}{2}e^{\frac{\Phi}{2}+g} \hat{h}^{1/4}
(\tilde{\omega}_1 +a\, d\theta)\,  ,
E^{2}=\frac{1}{2}e^{\frac{\Phi}{2}+g}\hat{h}^{1/4}
(\tilde{\omega}_2 -a\,\sin\theta\, d\varphi)\,,
E^{3}= \frac{1}{2}e^{\frac{\Phi}{2}+k}
\hat{h}^{1/4}
(\tilde{\omega}_3 +\cos\theta\, d\varphi)\,.\nonumber
\label{bbvielbein}
\eea
The function $\hat{h}$ is,
\beq
\hat{h}=1- e^{2\Phi-2\Phi(\infty)}.
\label{warphh}\eeq
The crucial choice of constant mentioned above
reflects the fact that the warp factor $\hat{h}$
vanishes at large radius. Notice also
that this expression for $\hat{h}$ implies that
this U-duality or rotation of the $SU(3)$-structure can only be performed
in the case in which the dilaton is stabilised at large radius,
namely $\Phi(\infty)$ is a finite value.
In other words, we cannot use this for the
analytical solution $P=2N_c\r$, that has linearly growing dilaton.
Only the solutions described around eq.(\ref{zzxxaa})
can be used.

We only quote the background vielbeins
after the U-duality, from which the full background can be obtained.
In this case the fields $H_3, F_5, F_3, \Phi$ are generated and the solution
is precisely the dual to the Baryonic Branch of the Klebanov-Strassler QFT
\cite{Butti:2004pk}.

The string frame metric is,
\begin{equation}
\begin{aligned}
ds_{str}^2	&= \alpha' g_s \sum_{i=1}^{10} (E^{i})^2\,,
\end{aligned}
\end{equation}
and the background functions are again
determined by the functions $P$, $Q$ as in eq.(\ref{changePQ}),
where $P$ satisfies the same differential `master' equation as before, solved by the function $P(\r)$ in
eq.(\ref{zzxxaa}). In other words, both solutions
are `the same'; this U-duality or rotation of $SU(3)$-forms
is a solution generating technique.

The  quantities  needed for the calculation
of the entanglement
entropy are,
\bea
& \a=e^{\Phi}\hat{h}^{-1/2} ,
\quad \b=g_s\alpha' e^{2k}\hat{h},
\quad V_{int}^{2}=
(2\pi)^6 (\alpha' g_s )^5 e^{4h+4g+5\Phi+2k}\hat{h}^{5/2},\nonumber\\
&  H=(2\pi)^6 (\alpha' g_s )^5 e^{4\Phi+4g+4h+2k} \hat{h}.
\label{BBEEquants}
\eea
We see that the difference to eq.(\ref{MNEEquants}) is the presence
of the factor $\hat{h}$ defined in eq.(\ref{warphh}).
This warp factor has a large radius asymptotic
given by --- see Section 4 of \cite{Conde:2011aa},
\beq
\hat{h}\sim \frac{3N_c^2}{8c^2}e^{-8\r/3} (8\r-1)+....
\label{warpbbzz}
\eeq
and it is precisely this decaying behaviour that will bring back
our phase transition. Indeed, we can calculate using eq.(\ref{rararara}),
that the small and large
asymptotics of
the function $L(\r_0)$ vanish,
\bea
& & L(\r)\sim {\cal Y}(\r)= 2\pi\sqrt{\alpha' g_s}
\frac{e^{k}\sqrt{\hat{h}}}{(4\Phi' +4g'+4h'+2k')+\frac{\hat{h}'}{\hat{h}}},\\
& & L(\r_0\to 0)\sim \r_0,\;\;\; L(\r_0\to \infty)\sim e^{-2\r_0/3}.\nonumber
\label{Lrho0BB}
\eea
As anticipated, the IR behavior for $L(\r_0)$ is the same as the one
for the analytic solution $P=2N_c\r$ or the one of eq.(\ref{zzxxaa});
this is because the warp factor $\hat{h}$ is constant for small
radial coordinate. The
UV behaviour instead, is quite different and driven by the factors
of $\hat{h}$ in  eq.(\ref{BBEEquants}).
Notice also, that in a convenient radial variable
$r=e^{2\r/3}$, we have
\beq
L(r_0\to\infty)\sim \frac{1}{r_0},
\label{locality}
\eeq
This is a signal of `locality' according to \cite{Barbon:2008ut}.
The UV-completion provided by the baryonic branch field theory has recovered
locality. What about the phase transition in the EE?

\begin{figure}[h]
\begin{center}
\begin{picture}(220,160)
\put(-125,0){\includegraphics[height=5.35cm]{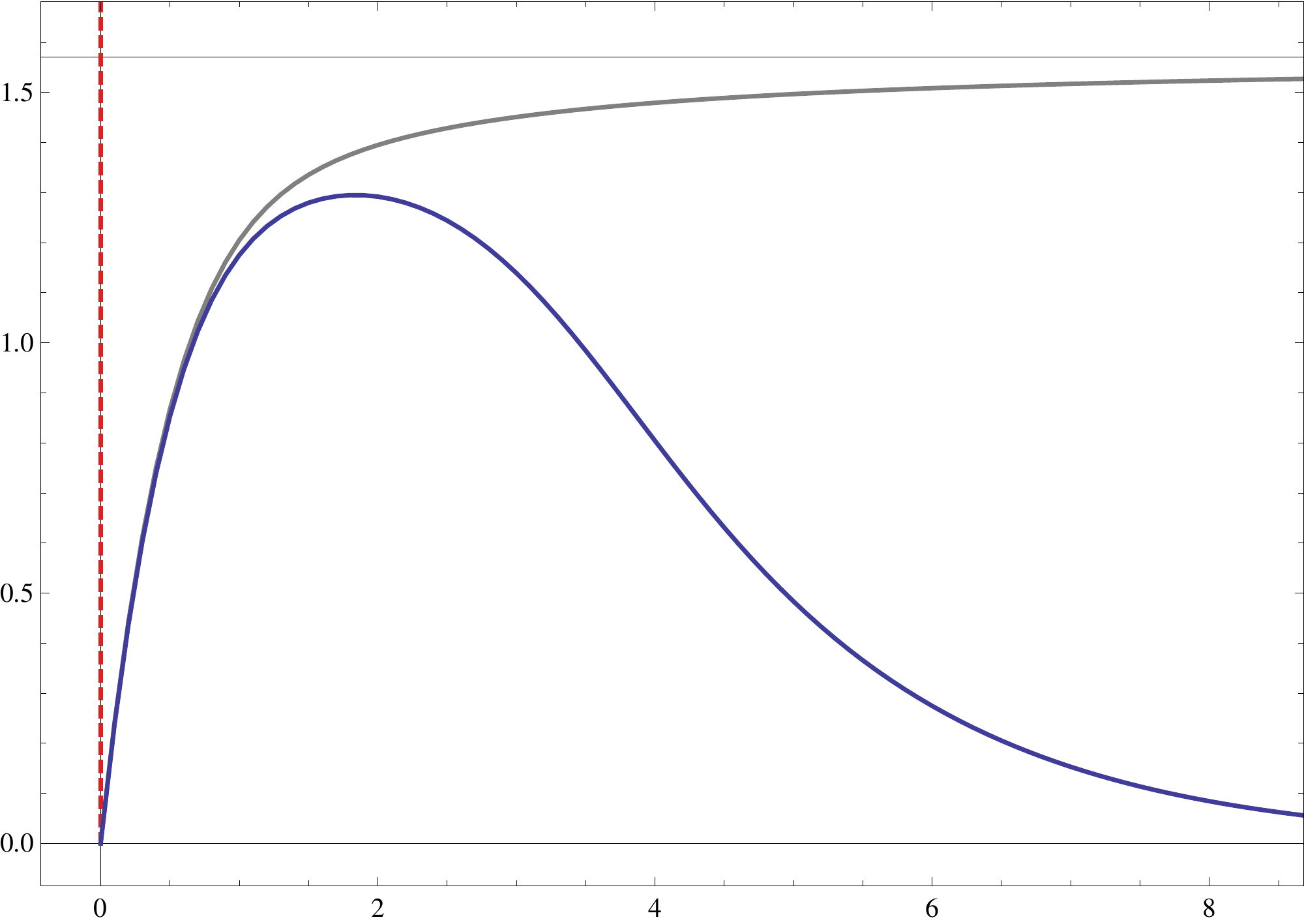}}
\put(-119,155){{\scriptsize{$L$}}}
\put(85,-3){{\scriptsize{$\r_{0}$}}}
\put(116,0){\includegraphics[height=5.35cm]{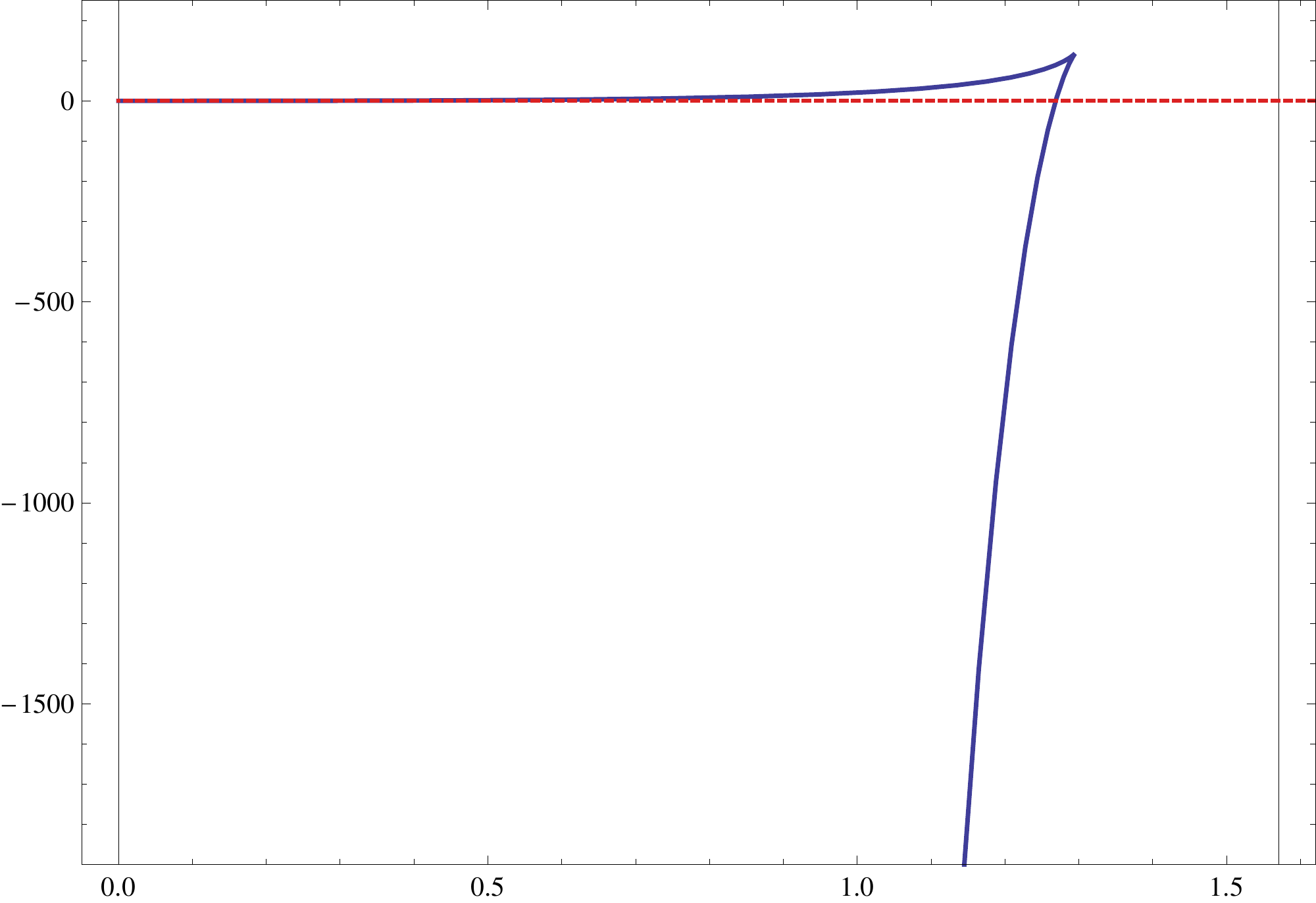}}
\put(128,155){{\scriptsize{$S$}}}
\put(335,-3){{\scriptsize{$L$}}}
\end{picture}
\caption{A typical solution on the Baryonic Branch of Klebanov-Strassler ($h_{1}=\frac{203}{100} N_{c}$) --- Here we plot $L(\rho_0)$ and $S(L)$.  The grey line is the linear $P$ solution ($h_{1}=2 N_{c}$) for comparison.}
\label{Fig:MNrot2}
\end{center}
\end{figure}
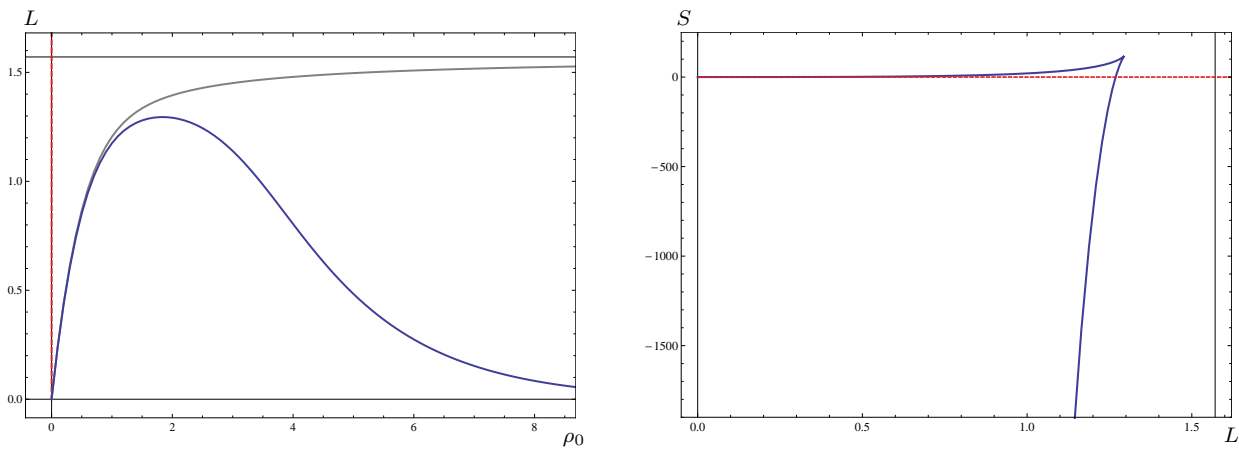

Calculating with the expressions
of eq.(\ref{BBEEquants}), the entanglement entropy
associated with the Baryonic Branch of the KS-field theory gives the result
displayed in Figure \ref{Fig:MNrot2}.

\begin{figure}
\begin{center}
\begin{picture}(220,160)
\put(0,0){\includegraphics[height=5.35cm]{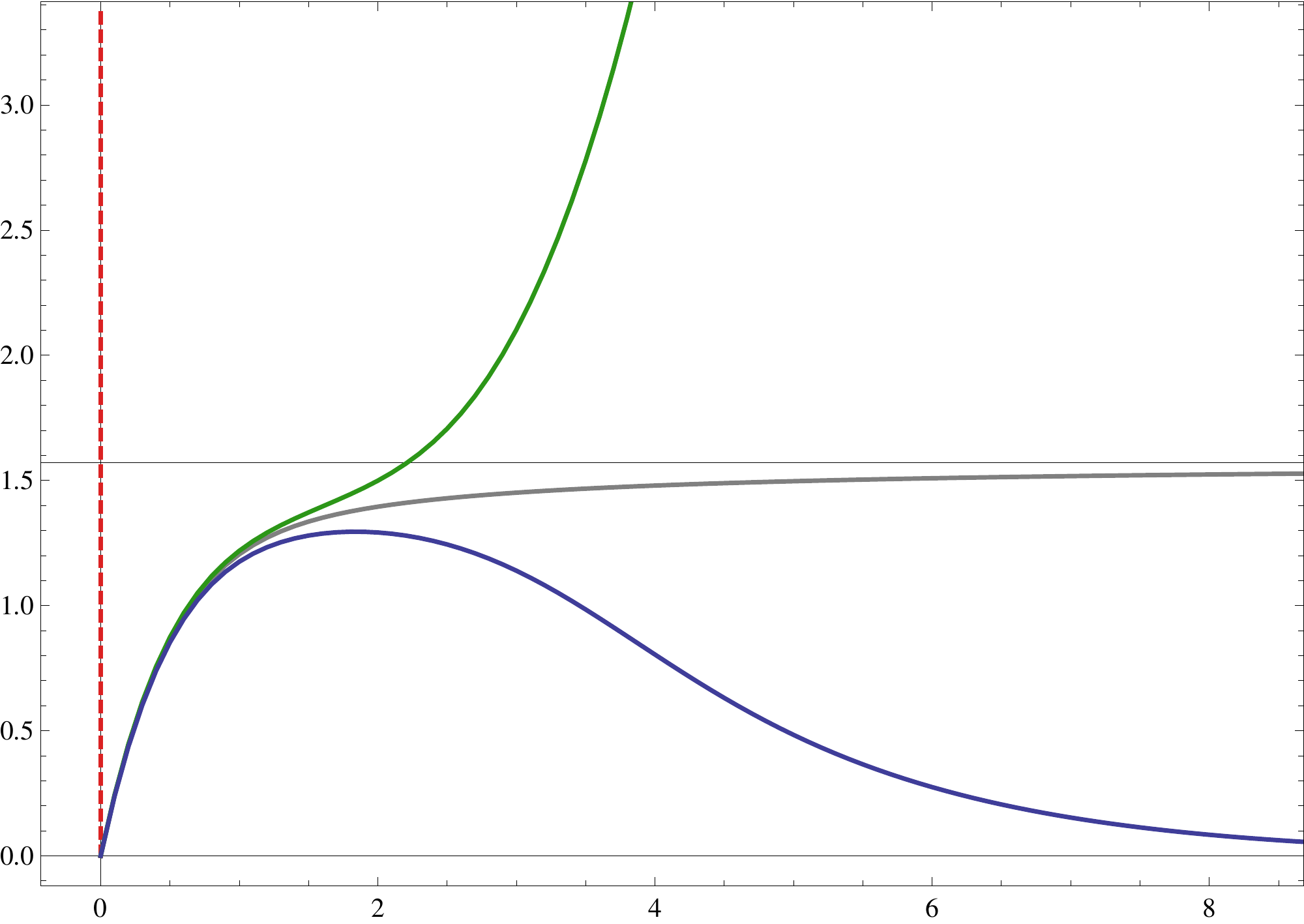}}
\put(-1,148){{\scriptsize{$L$}}}
\put(218,3){{\scriptsize{$\rho_{0}$}}}
\end{picture}
\caption{These are various plots of $L(\r_0)$ for comparison.
The grey graph is the solution $P=2N_c\r$ with linear dilaton. The green and blue lines represent numerical solutions before and after the U-duality
respectively (solved for the same value of $h_1=\frac{203}{100}N_{c}$ with asymptotically constant dilaton.}
\label{Fig:Comp}
\end{center}
\end{figure}

As one can see, here we get the nice
phase transition behaviour, also  expected from the calculation
in the paper \cite{Klebanov:2007ws}. The intuitive description
of what is going in terms of a Van der Waals gas analogy is consistent
with this result as well.
Figure \ref{Fig:Comp} serves as a good summary of the discussion in the
previous and the present sections.

We have settled the problem of recovering the phase transition.
We would like now to do a  couple of calculations that will lead to an improved understanding of the Klebanov-Strassler system
(in nice agreement with the
ideas presented in \cite{Dymarsky:2005xt}). We will move our field theory
to a mesonic branch. We will do this first in a way, that as was argued in
\cite{Gaillard:2010qg}, implies that towards high energies the evolution of the
QFT is described in terms of Seiberg dualities, but more importantly successive
Higgsings that change the matter content very fast. This quick growth
of  the matter content is equivalent to the addition of another
irrelevant operator (of dimension-six in this case,
different from the dimension-eight one
we have discussed above). This
was discussed in detail in \cite{Gaillard:2010qg}
and \cite{Conde:2011aa}.

As  a consequence of the insertion of this new irrelevant operator, with the
added non-localities to the QFT, we will loose
the phase transition achieved (in nice agreement with the
discussion of the previous section). We will then explain
how to get the phase transition back, with a precise way of
switching off that irrelevant operator.
We turn to that now.
\subsection{Losing our Phase Transition: Adding Sources} \label{losing}
As described  above, we will move our Klebanov-Strassler QFT from the
baryonic branch to a mesonic branch. In order to do this, we need to de-tune
the ranks of the two gauge groups --- that in the baryonic branch
is $SU(k N_c)\times SU(k N_c +N_c)$. The unbalance is achieved by adding matter
that in the dual string theory is represented by D5 branes with
induced D3 brane charge. This will change the
gauge group to $SU(k N_c +n_f)\times SU(k N_c + N_c +n_f + \frac{N_f}{2})$.
The $N_f $-D5 and $n_f$-D3 branes
are added as sources; namely the background is a solution to
the equations of motion of Type IIB Supergravity plus
the Born-Infeld Wess-Zumino action for the sources. The associated
solutions were
discussed in detail in
\cite{Gaillard:2010qg}, \cite{Conde:2011aa}. One characteristic is
that the sources need to be added with a profile that vanishes close to $\r\to 0$,
in order to avoid curvature singularities (see the discussion in
\cite{Nunez:2010sf}, \cite{Conde:2011rg} and \cite{Conde:2011aa}).
A profile that preserves the same SUSY as the background
and avoids all singularities, can be derived to be \cite{Conde:2011rg},
\bea
\mathcal{S}(\r)=N_f \tanh(2\r)^4.\nonumber
\eea
Also, this profile can be `translated' according to
\bea
\mathcal{S}(\r)=N_f\, \Theta(\r-\r_*)
\tanh(2\r-2\r_*)^4
\label{sigmoid}
\eea
and still preserve SUSY and avoid singularities everywhere.

The formalism used in deriving these backgrounds
runs parallel to the one described above, in the sense
that a `change of basis' and a second order differential
`master' equation can be written for a single function $P(\r)$ that
in this case will contain the effect of the ($n_f, N_f$) D3-D5 sources.
A rotation of the $SU(3)$-structure forms is used
to generate the solutions dual to the mesonic branch of the KS field theory.
See Section 3 of the paper \cite{Conde:2011aa} for a clear summary
of the set-up and its subtleties.

In the papers \cite{Gaillard:2010qg},
\cite{Conde:2011aa} a solution encoding the effect of the sources
was found. The large radius asymptotics
for the warp factor $\hat{h}$ in eq.(\ref{warphh}) is given in
eq.(2.25) of the paper \cite{Conde:2011aa}, using
the radial coordinate $r=e^{2\r/3}$ we have,
\beq
\lim_{r\to\infty}\hat{h}\sim
\frac{N_f r^2+ 3N_c^2\log r}{r^4},
\label{warprr}
\eeq
This deviates (by what seems to be the addition of an
irrelevant operator of
dimension-six)
from the cascading behaviour of the Klebanov-Strassler QFT.

The field theoretic logic is again, that
the number of D3 sources, $n_f$, grows very fast --- as was discussed in
\cite{Conde:2011aa}
\beq
n_f\sim \mathcal{S}(\r)(\sinh(4\r) -4\r)^{1/3}\sim e^{4\r/3},
\label{rapido}
\eeq
and this rapid growth of the gauge group ranks going
to higher energies (due to Higgsing every time  a source
D3 is crossed) implies that the
QFT looses the 4-d character of the KS system.

Aside from this, the
calculation of the Wilson loop in the solution mentioned
above will produce an area-law behaviour, indicating confinement.
Following the logic of the previous sections, we should expect that
in spite of the confining behaviour, a phase transition in the
entanglement entropy is wiped-off by the irrelevant operator (the presence of
this irrelevant can also be associated with finite-size
effects in computing the Wilson loop).

\begin{figure}[h]
\begin{center}
\begin{picture}(220,160)
\put(-120,0){\includegraphics[height=5.35cm]{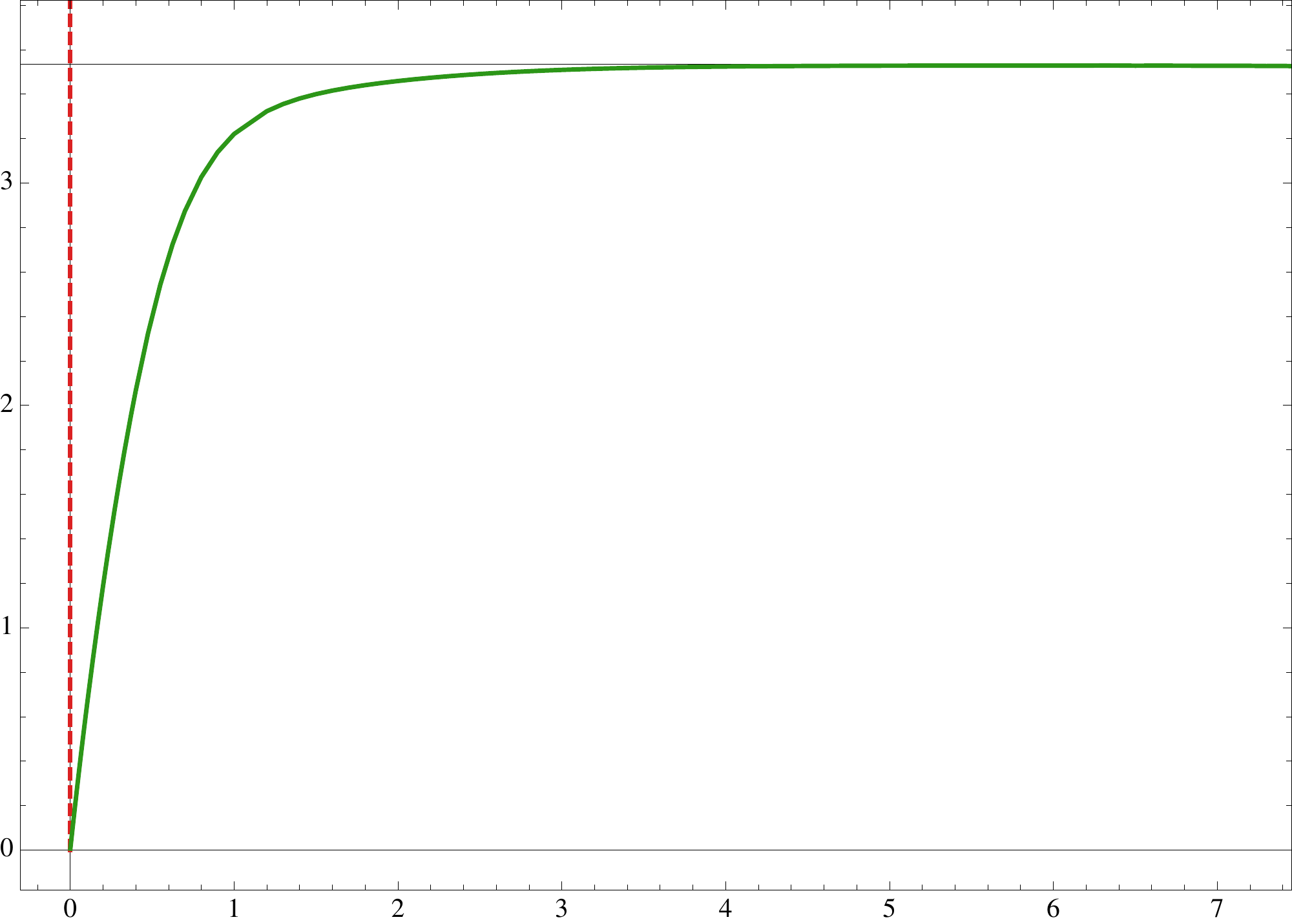}}
\put(-119,155){{\scriptsize{$L$}}}
\put(85,-3){{\scriptsize{$\r_{0}$}}}
\put(115,0){\includegraphics[height=5.35cm]{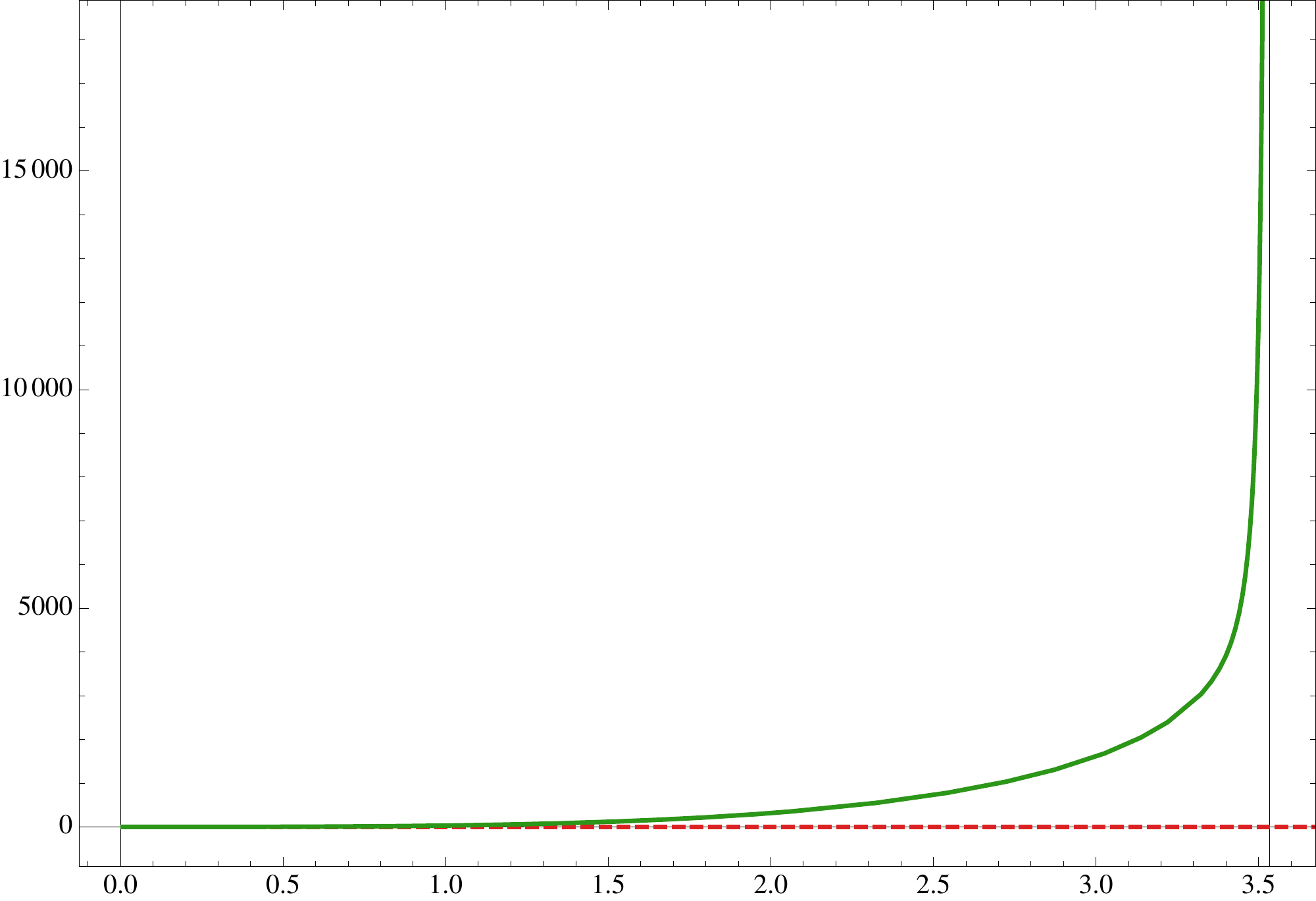}}
\put(128,155){{\scriptsize{$S$}}}
\put(335,-3){{\scriptsize{$L$}}}
\end{picture}
\caption{Plots of $L(\r_0)$ and $S(L)$ for a solution with $\mathcal{S}\rightarrow N_f$ in the UV and
$N_c=4$, $N_f=9$.  No phase transition is present. Note that we have chosen
a value of $h_1$ such that we have hardly any linear behaviour in $P$.}
\label{Fig:nophase}
\end{center}
\end{figure}
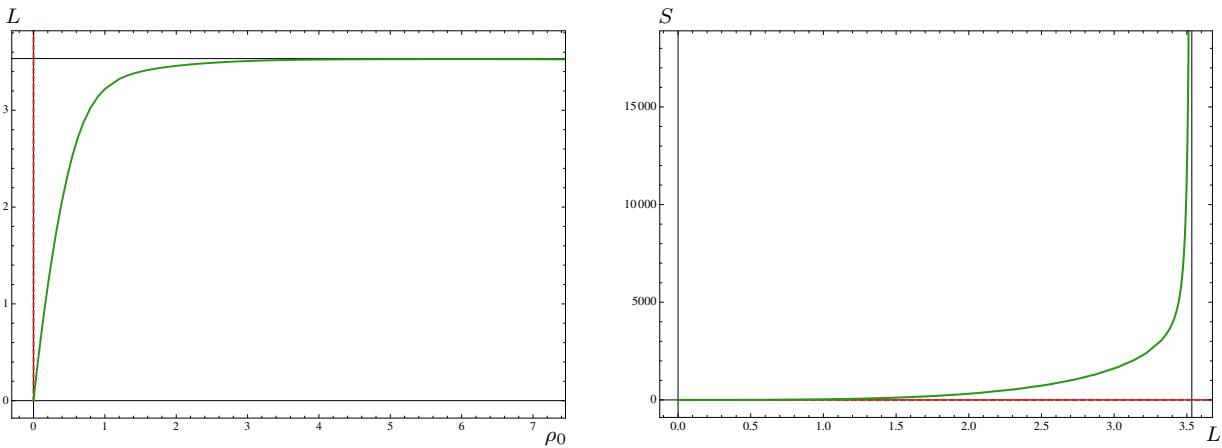

This is indeed the case as one can explicitly calculate.
While the small radius asymptotics for $L(\r_0\to 0)\sim \r_0$
is unchanged,
we obtain that
\bea
L(\r_0\to \infty)={\cal Y}(\r\to\infty)
\sim \frac{3\pi \sqrt{\alpha' g_s N_f}}{8}.\nonumber
\eea
Figure \ref{Fig:nophase} illustrates this point, with
 $\r_*=0$, so no phase transition, unless of course a UV-cutoff
is introduced and the study of `short configurations' sorts
out the problem as in the examples above.

This brings us to the conclusion that the field theory may be in a
mesonic branch with good IR properties,
but that we should look to somewhat `localise' the sources
to avoid this `too-fast growth of degrees of freedom' expressed in eq.(\ref{rapido}),
if we wish to recover
the high energy 4-d behaviour the baryonic branch was
displaying (which is UV completing the system).
We do this in the next section.

\section{Getting Back the Phase Transition(s): Sources with a Decaying Profile} \label{getback}
As discussed above, the lost phase transition in the entanglement
entropy (without introducing the UV-cutoff), together with the asymptotic form of the
warp factor in eq.(\ref{warprr}),
strongly indicates that the QFT is not behaving as a 4-d QFT, in the sense of
`locality' being lost.
It was understood in
\cite{Conde:2011aa}
that this is due to a very rapid growth of degrees of freedom.

In backgrounds where the profile for the sources is $\mathcal{S}\sim \tanh(2\r)^4$,
this is reflected by the fact that the flow to the UV of the QFT is
described by a superposition of Seiberg dualities --- the logarithmic
term in eq.(\ref{warprr}) --- and a Higgsing,
represented by the term quadratic in the radial variable.
Another interplay between Higgsing and Seiberg cascade was previously
observed in \cite{Aharony:2000pp}.
In the particular solution with source profile
$\mathcal{S}\sim \tanh(2\r)^4$, the rate of Higgsing becomes too fast at high energies,
the
UV of the
field theory does not behave like a 4-d QFT. This is reflected in the entanglement
entropy, which does not display
the nice phase transition achieved in Section
\ref{recoverbb}.

We would like this mesonic branch solution
of the KS field theory to behave like a 4-d QFT.
In order to do this, we will slow-down the growth of degrees of
freedom, by proposing a
{\it phenomenological} profile for the sources. This is phenomenological
in the sense that is not derived from first principles (as a kappa symmetric
embedding of sources with this profile).
Nevertheless,
the profile we will propose
has the following properties \cite{Conde:2011aa}:
\begin{itemize}
\item{The background still satisfies BPS equations, suggesting
SUSY preservation.}
\item{The energy density of the sources $T_{00}$ is
positive definite for profiles
that decay at most as fast as the one we will propose.}
\item{The central charge of the dual QFT when calculated with this profile
is a monotonic and growing function.}
\end{itemize}
The profile we will adopt, following
\cite{Conde:2011aa}
is,
\beq
\mathcal{S}(\r)=N_f \tanh(2\r)^4 e^{-4\r/3}.
\label{bump}
\eeq
Notice that now, one can find a new background solution, where
the sources are somewhat `localised'. The dual QFT is in a  mesonic branch
as explained in
\cite{Conde:2011aa}, where the
background solution was explicitly written.
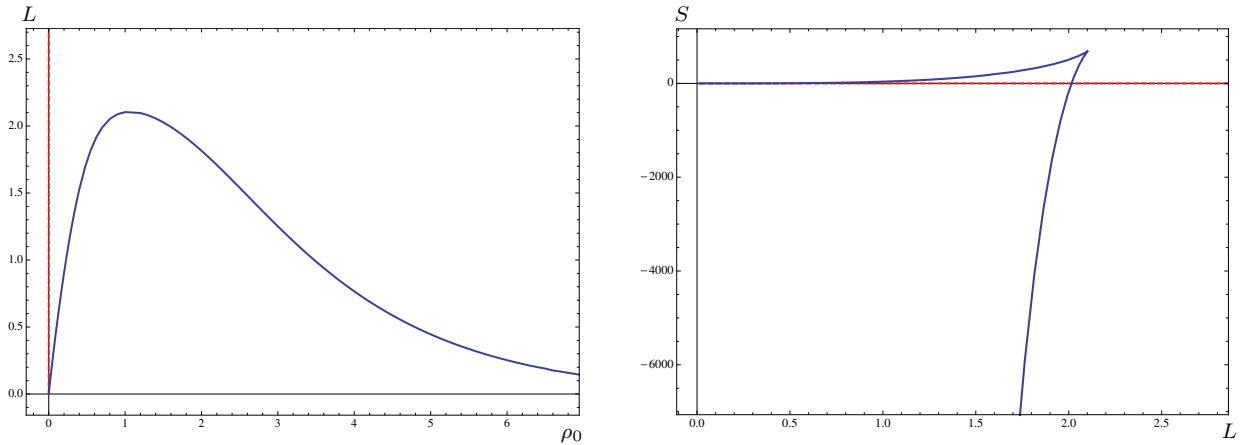
\begin{figure}[h]
\begin{center}
\begin{picture}(220,160)
\put(-124,0){\includegraphics[height=5.35cm]{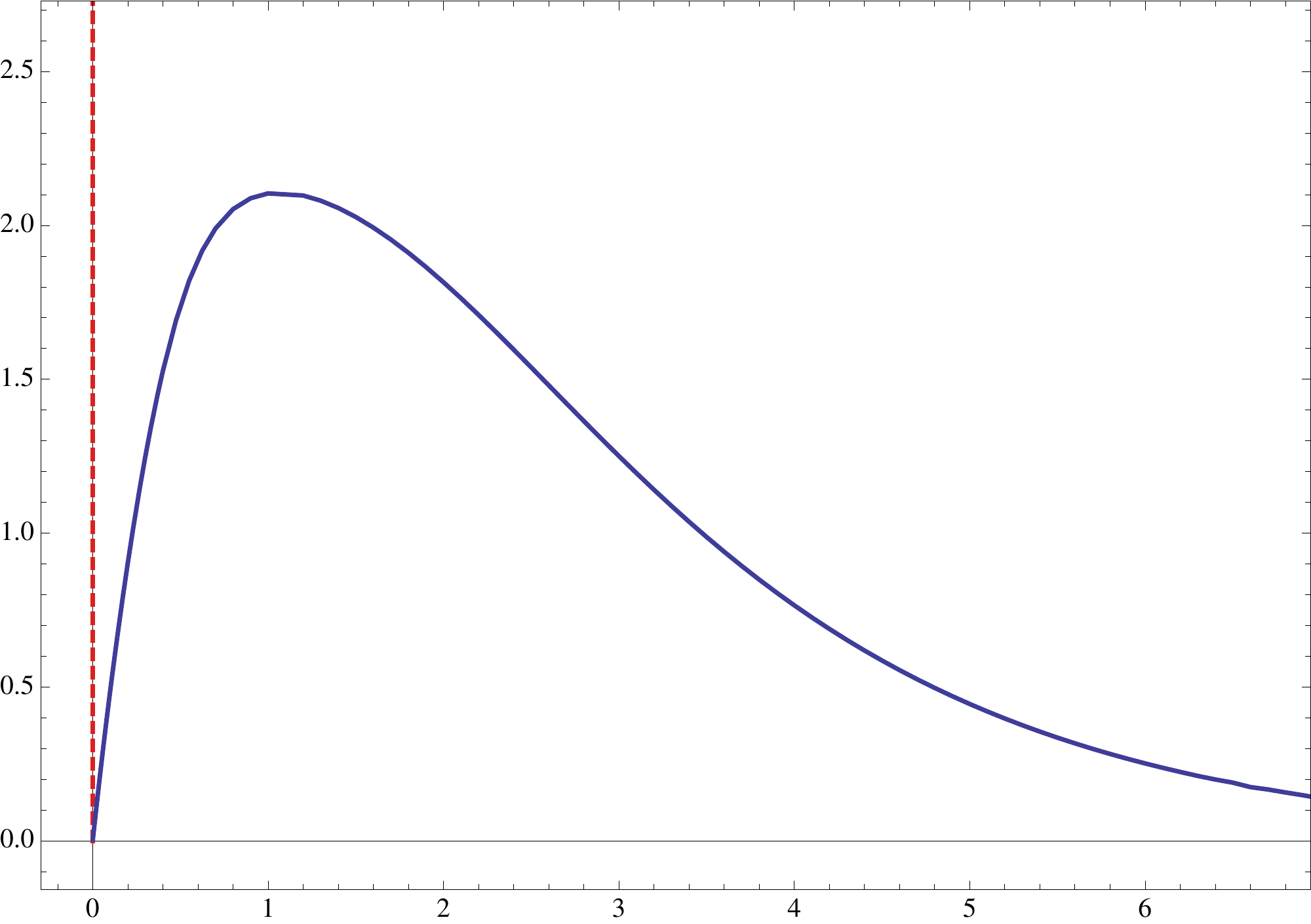}}
\put(-119,155){{\scriptsize{$L$}}}
\put(85,-3){{\scriptsize{$\r_{0}$}}}
\put(115,0){\includegraphics[height=5.35cm]{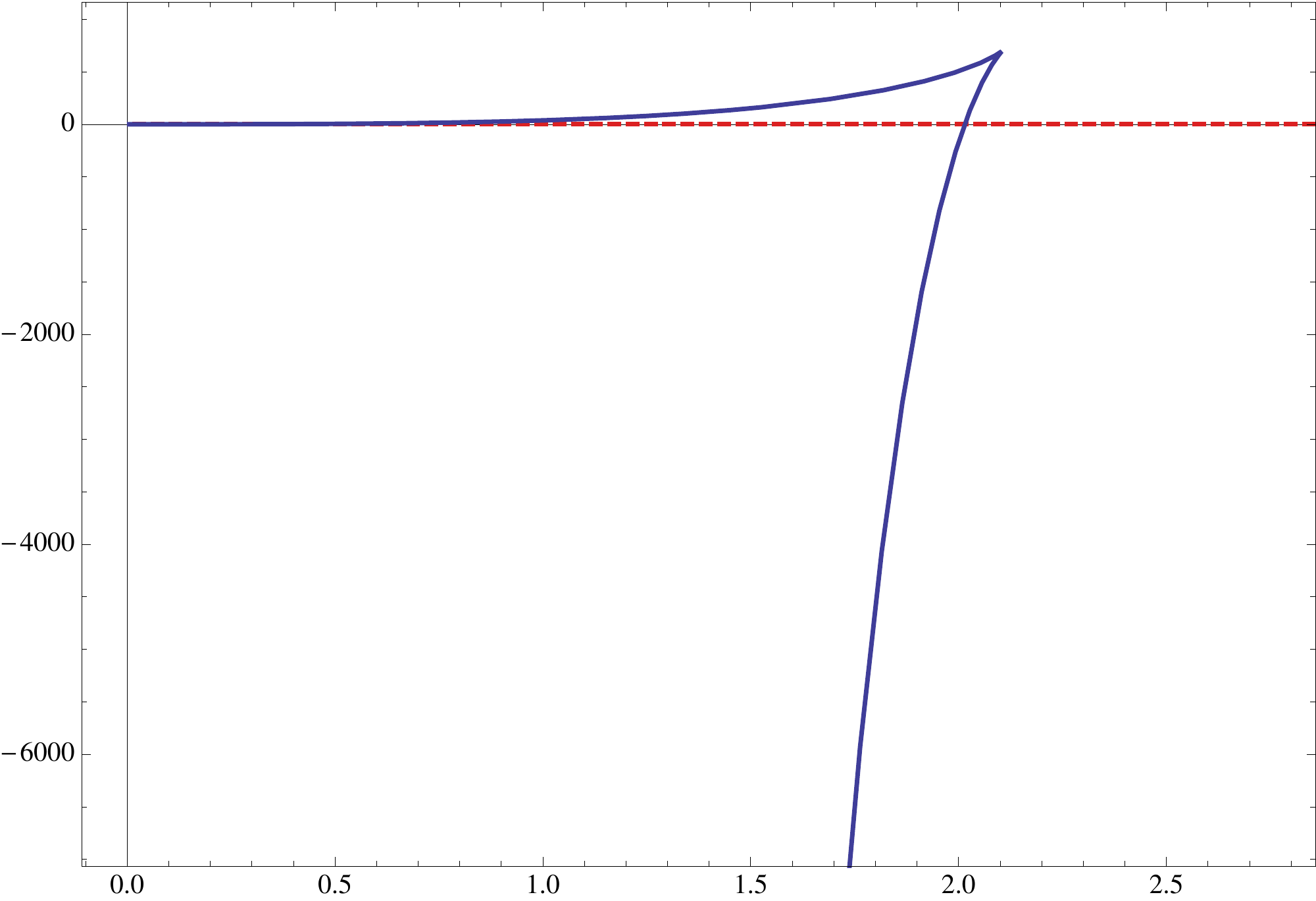}}
\put(128,155){{\scriptsize{$S$}}}
\put(335,-3){{\scriptsize{$L$}}}
\end{picture}
\caption{Plots of $L(\r_0)$ and $S(L)$ for a solutions with $\mathcal{S}\rightarrow 0$ in the UV and
$N_c=4$, $N_f=9$.  A single first order phase transition is present. Note that we have chosen
a value of $h_1$ such that we have hardly any linear behaviour in $P$.}
\label{Fig:phase}
\end{center}
\end{figure}

The relevant functions for the calculation
of the entanglement entropy are those in
eq.(\ref{BBEEquants}). The explicit expressions for the asymptotics
(and numerical solutions)
of all
the participating functions are described in Section 4 of
\cite{Conde:2011aa}. In particular
\bea
& & \hat{h}(\r\to \infty)\sim \frac{3}{8c^2}e^{-8\r/3}
\Big[N_c^2(8\r-1)+ 2 c N_f - 4 N_cN_f \mathcal{S}_{\infty}   \Big]+....\nonumber\\
& & \mathcal{S}_{\infty}=\int_0^\infty d\r\, \mathcal{S} \tanh(2\r)^2.\nonumber
\eea
Comparing this with eq.(\ref{warpbbzz}) we observe that the functional
decay of the warp
factor is the same. The cascade of Seiberg dualities is still present
(in this radial coordinate is represented by the term $N_c^2\r$).
The constant terms
represent the effects of the sources, that even when they are very suppressed
at large values of $\r$ still contribute.
Indeed, using the radial coordinate $r\sim e^{2\r/3}$, they
contribute to the warp factor as $\hat{h}\sim \frac{N_f}{r^4}$,
which was expected for a localised stack $N_f\sim n_f$ D3 branes.

It is not hard to believe that if we follow the full numerical calculation
with the expressions of eq.(\ref{BBEEquants}), we will find a phase transition
in the entanglement entropy. This is indeed the case,
the plots of Figure \ref{Fig:phase}, make this point concrete.

The function $L(\r_0)$ is well approximated by
the function ${\cal Y}(\r_0)$
and satisfies a Heisenberg-like relation,
as shown in eq.(\ref{Lrho0BB}), with the numerical differences
of the case induced by the factors of $N_f$ and $\mathcal{S}_{\infty}$ in the warp
factor $\hat{h}$, and in the other background functions. Still,
we have $L(r_0\to\infty)\sim 1/r_0$, in agreement with the
`locality criteria' proposed in \cite{Barbon:2008ut}.

An interesting observation is that since the sources can be translated
at a given point $\r_*$ as discussed above
\beq
\mathcal{S}(\r)=N_f\, \Theta(\r-\r_*) \tanh(2\r-2\r_*)^4 e^{-4\r/3}
\eeq
we can introduce another scale to the system,
represented by $\r_*$, the point where
the source become `activated'. Numerical solutions
(that are a bit more time-expensive to find) can now show a
double phase transition, as can be seen in Figure \ref{Fig:twophase} (for further discussion --- see Appendix C).

This is a phenomena that probably was observed in other contexts,
but we are unaware of them in the bibliography. There might be
gas, generalisation of
Van der Waals' one
which includes two different interactions of similar strength between
the particles composing the gas. This could lead to a double or
even multiple first order transitions.
\begin{figure}[h]
\begin{center}
\begin{picture}(220,160)
\put(-124,0){\includegraphics[height=5.35cm]{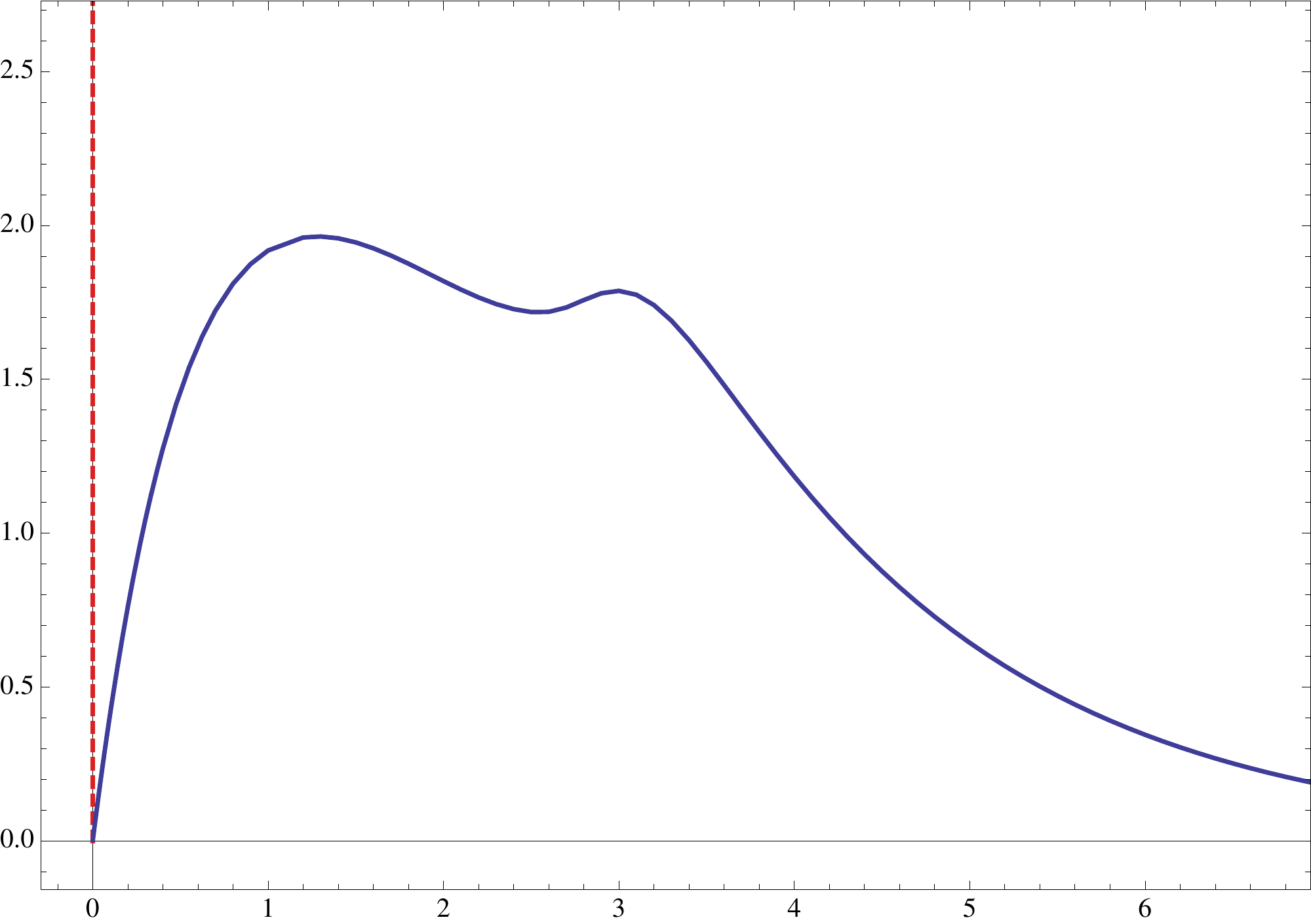}}
\put(-119,155){{\scriptsize{$L$}}}
\put(85,-3){{\scriptsize{$\r_{0}$}}}
\put(112,0){\includegraphics[height=5.35cm]{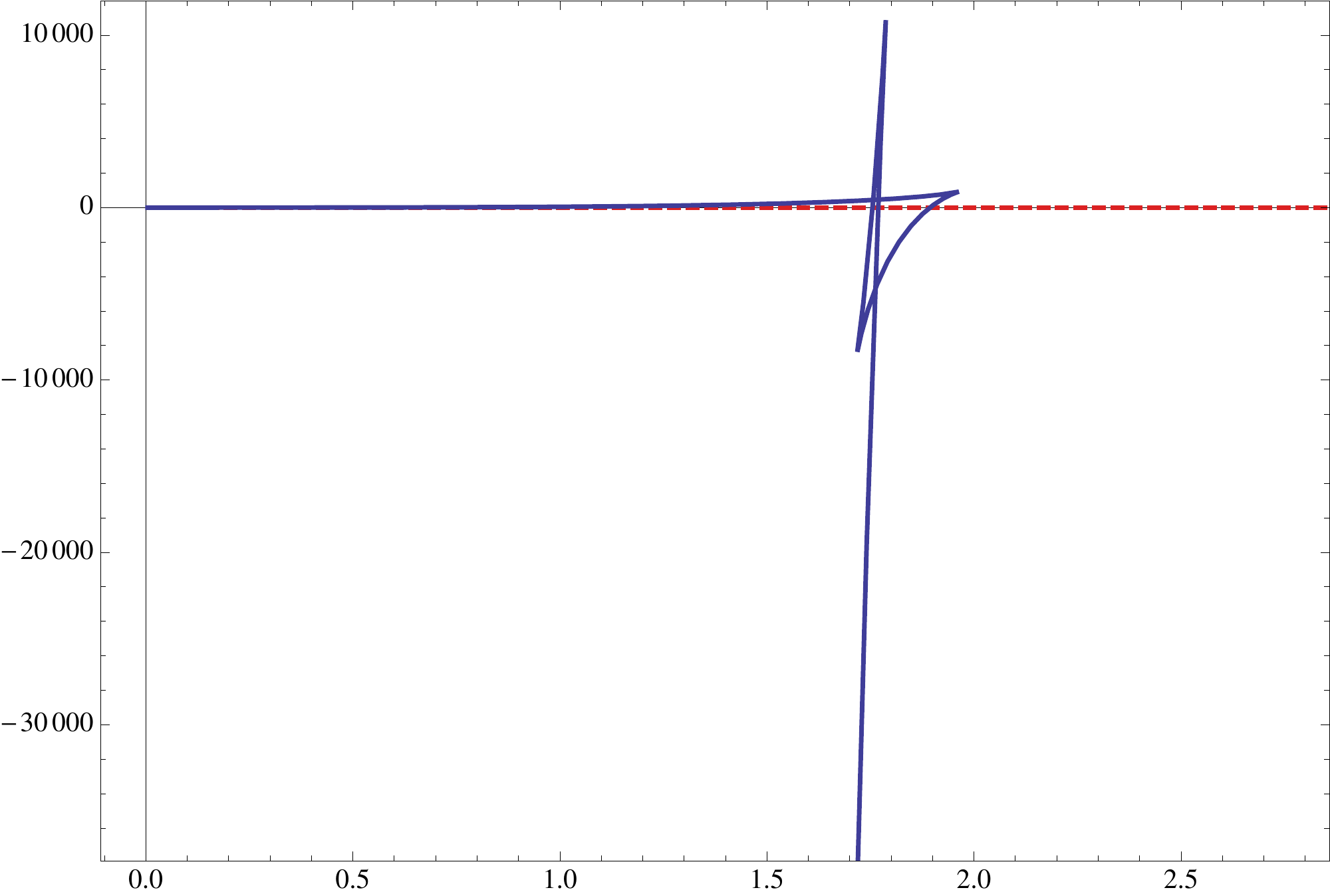}}
\put(128,155){{\scriptsize{$S$}}}
\put(335,-3){{\scriptsize{$L$}}}
\end{picture}
\caption{Plots of $L(\r_0)$ and $S(L)$ for a solutions with $\mathcal{S}\rightarrow 0$ in the UV and
$N_c=4$, $N_f=9$ but $\r_*=\frac{25}{10}$.  Two first order phase transitions are present. Note that we have chosen
a value of $h_1$ such that we have hardly any linear behaviour in $P$.}
\label{Fig:twophase}
\end{center}
\end{figure}
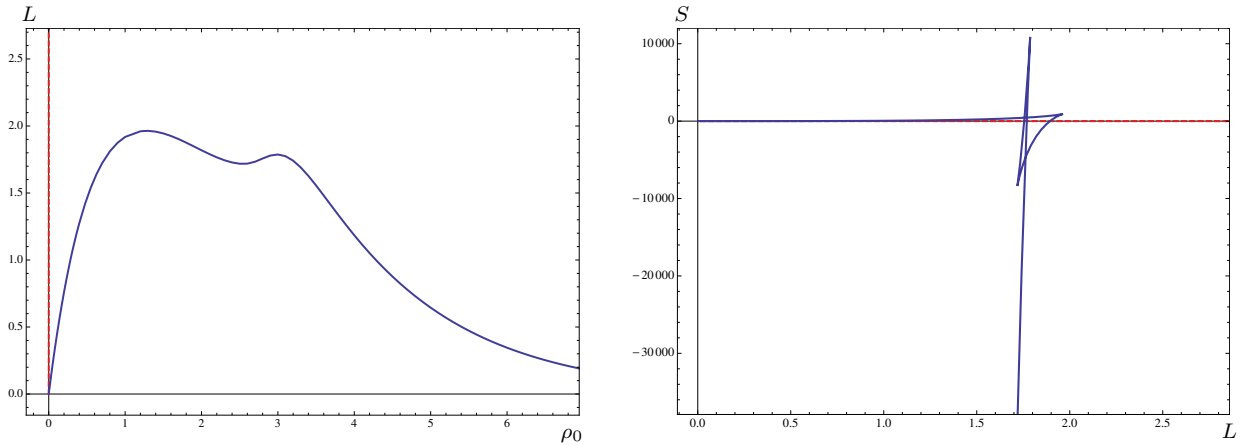
The Figure \ref{Fig:twophase},
illustrate our results for the entanglement entropy,
$L(\r_0)$
and $S(L)$, when
calculated in the mesonic branch of the KS field
theory in the presence of a localised bunch of matter represented by
a D3-D5 bound state.

We wish to close this section by emphasizing that our findings are
clearly making the point of KKM \cite{Klebanov:2007ws}: a phase transition
in the EE is a sign of a confining QFT. But one should be careful about
the UV behavior of this field theory. If non-local,
either cutoff effects or a UV completion will recover the phase transition.
We can also think our findings for the EE
as a diagnostic to decide if a QFT is showing or not the
high energy behaviour
expected from a  four-dimensional (or lower) QFT, free of non-localities.

\section{Conclusions and Future Directions}
Let us first summarise the contents and ideas explained in this paper.

We started by observing the obvious analogy between the
holographic calculation of
Wilson loops and Entanglement Entropy. Indeed,
being both minimisation problems for two- and eight-surfaces,
they display very similar general solutions.
In spite of the analogous formulas derived when minimising surfaces,
the two observables are such that when evaluated
on particular backgrounds the results are quite different.
For example, in confining backgrounds (the topic that mostly
occupied us in this work)
the Wilson loop holographic calculation gives a linear dependence
between the quark-antiquark's Energy and separation
$E_{QQ}\sim \sigma L_{QQ}$ (for large $L_{QQ}$),
while for the Entanglement Entropy,
we observe a very different behavior, including a first
order phase transition. Aside from these observations,
either obvious or made previously in the bibliography,
the first contribution of this paper was to develop a
simple and operative criteria to test under which conditions
the Entanglement Entropy would display a phase transition.
This was done in Section \ref{relationsWEE}, where aside
from the criteria for phase transitions, some
examples were given and an important
question was posed: `Why the relation between Confinement and
the existence of a phase transition in the Entanglement
Entropy  breaks down, for models based on D$(p>4)$ branes ?'

Section \ref{newmaterial} answered this question,
after a detour into non-local QFTs, QFT with a cutoff and the
realisation that, when calculating with string duals to non-local QFTs,
we are potentially missing a  set of configurations that are
very important in the calculation of the Entanglement Entropy.
These configurations that become apparent when considering
the non-local QFT with a cutoff, solve the Physics problem of
having only a disconnected  and
an unstable solution. This is the material discussed in
Section \ref{newmaterial}, together with the explicit
solution for a simple confining model --- D5's on $S^1$ --- where the
problem originally appears.

It may be unpleasant to some physicists that
we need a UV-cutoff to resolve the problem of
stability and to regain the phase transition argued to
be present in confining models in \cite{Klebanov:2007ws}. Nevertheless,
we want to point out that this UV cutoff is actually capturing
the behavior that the QFT, once UV-completed, will display.
Indeed, this is the point made in Section
\ref{sectionabsence}. We studied a trademark model of  confinement
in four dimensions, the string background corresponding to
D5 branes wrapping a two-cycle of the resolved conifold. We
explained  how this model would not display a
phase transition if taken at face value. But upon the
introduction of the cutoff at high energies, we observed
the phase transition and the whole behavior of
a four dimensional confining field theory with
a Hagedorn density of glueball-states.
The effect of this cutoff is the same as the one found in the UV-completed QFT on the D5 branes.
Indeed, an inverse Higgs mechanism takes place completing the non-local
QFT
into the Klebanov-Strassler field theory (in
a  generic point of its Baryonic Branch). This point
is made clear, with calculations and plots of the
$S(L)$ phase transition in Section \ref{recoverbb}.
Further to this, in Section \ref{recoverbb}, we also pointed out
that some backgrounds describing the Mesonic
Branch of the Klebanov-Strassler field theory
(these backgrounds include a large number of D5-D3 sources)
are also afflicted by a non-locality, unless the sources
are introduced with a particular profile proposed
(using completely different arguments) in \cite{Conde:2011aa}.

This completes a very pleasant picture advocated
in Section \ref{getback}, linking confining,
non-local QFTs and their local UV-completed counterparts.
Further a link between the Entanglement Entropy and
its phase transition, that act as a measure of both locality and confinement.
Numerous appendices complement the presentation, and study
a wide variety of other examples, to further illustrate our ideas above.

Let us describe a couple of ideas that this work
suggests as possible extensions of what we have learnt here.

First of all, since we found that confining models
typically imply a phase transition for the EE,
it would be interesting to ask what happens to the
EE when one considers a confining model that
presents also a  phase transition for the Wilson loop.
Indeed, models with various scales have shown
this behavior --- see for example \cite{Bigazzi:2008gd},
 \cite{Elander:2011mh}. It may be, as
we argued in  Section \ref{getback}, a
multi-phase transition is present for the EE in
these cases.
It is also worth analysing if the confining behavior
implies that $H(\rho_\Lambda)=0$, since the argument
we gave involving the central charge of the field theory may be evaded.
In this same line, the criteria for phase
transitions in the EE, discussed in Section
\ref{relationsWEE}, may lead to interesting extensions.
One could try to find out if confinement is actually {\it needed} for
the EE to present a phase transition.
Studying the invariances of the EE under different dualities
seems like another small and nice project. We already know
that S-duality and non-Abelian
T-duality \cite{Itsios:2013wd}, \cite{Caceres:2014uoa},
are invariances of the EE.

On more general grounds, an observation
that this paper suggests is the following:
we know that black holes for D$p$ branes (with $p>4$)
turn out to have negative specific heat.
We also know that the holographic renormalisation
program can be successfully applied to
backgrounds based on D$p$ branes with $p<5$
\cite{Kanitscheider:2008kd}. We found
that the connection between non-locality of the QFT
and the absence of the phase transition in the EE,
is there for solutions based on D$p$ branes
with $p>4$, and this signals that probably, as it happens with the
EE, one may find a way to `fix' the density of states of
the finite Temperature QFT and also, with
a UV-cutoff, or better with a suitable UV-completion,
one may be able to implement the program of
holographic renormalisation. Sorting out
which observables turn out to behave similarly
with the cutoff, or the UV-completion, seems another interesting problem.

\section*{Acknowledgments}

Discussions with various colleagues helped to improve
the contents and  presentation of this paper.
We wish to thank: Ofer Aharony, Adi Armoni, Carlos Hoyos, Prem Kumar, David Kutasov, Yolanda Lozano, Niall T. Macpherson, Patrick Meessen, Maurizio Piai, Alessandro Pini, Diego Rodriguez-Gomez and Johannes Schmude.

This work started while Carlos Nunez was
a Feinberg Foundation Visiting Faculty Program Fellow,
he thanks the hospitality extended at Weizmann Institute
and The Academic Study Group for the Isaiah Berlin Travel award.

\appendix
\renewcommand{\thesection}{\Alph{section}}
\renewcommand{\theequation}{\Alph{section}.\arabic{equation}}
 \addtocontents{toc}{\protect\setcounter{tocdepth}{1}}

\section{Appendix: Wilson Loop-Entanglement Entropy Relation - An Exercise}\label{exercisezz}
The similarities between the EE and the Wilson loop, summarised
in equations \eqref{ell}, \eqref{S}
suggest an interesting small exercise. We can ask what are the conditions
on a given background, so that the EE and the Wilson loop have the same $L$ dependence.
In order to solve this exercise,
we will consider situations in which the EE behaves, for large separations,
as $S\sim L^{-p}$, being $p$ some positive number.
For example, in the case of conformal field theories in $d+1$ dimensions, one finds that $p=d-1$. This is referred to as
an `Area law' for the EE. In some other examples--characteristically in non-local (d+1)-dimensional
theories --- one finds $p=d$, in which case the name of `Volume law' is used.
We will first study, based on the similarities alluded to above and summarised by equations \eqref{ell}, \eqref{S}  the characteristics that a background must have such that
the Wilson loop behaves like the EE, namely $E_{WL}\sim L_{WL}^{-p}$.
Let us start with an IR analysis.
\subsection{Small Radius Expansion of the Wilson Loop}
The functions $\alpha(\rho),g(\rho)$ characterising the Wilson loop computations,
can be expanded around $\rho=\rhol$ as follows
\begin{eqnarray}
  \alpha(\rho) &=& \alpha(\rhol) + a_k \rho^k + \mathcal{O}(\rho^{k+1}), \nonumber\\
  g(\rho) &=& b_j \rho^j + \mathcal{O}(\rho^{j+1})\nonumber .
\end{eqnarray}
If $\alpha(\rhol)\neq0$ the Wilson loop exhibits linear confinement
with $\alpha(\rhol)$ the string tension. We are not interested in a linear
law $E_{WL}\sim f(\rhol) L_{WL}$ and therefore assume $\alpha(\rhol)=0$.
Further assuming $k>0$, $j>-1$ and few other reasonable assumptions on
the functions $\alpha(\rho),g(\rho)$, the authors of \cite{Kinar:1998vq} found that
\begin{equation}\label{IRWL}
  E \sim L ^{- \frac{j+1}{k-j-1}},\nonumber
\end{equation}
for large $L$ and for the case $k>j+1$ (the case $k\leq j+1$ will not result in a negative power of $\ell$).
\subsection{Large Radius Expansion of the Wilson Loop}
On the other hand, the functions $\alpha(\rho),g(\rho)$ can be expanded around the boundary $\rho=\infty$ as follows
\begin{eqnarray}
  \alpha(\rho) &=&  c_n \rho^n + \mathcal{O}(\rho^{n-1}), \nonumber\\
  g(\rho) &=& g(\infty)+   d_m \rho^{-m} + \mathcal{O}(\rho^{-m+1}) ,\nonumber
\end{eqnarray}
with $  n,m>0.$
In the region close to the boundary, i.e. small distances, the Wilson loop then behaves as \cite{Kol:2010fq}
\begin{equation}
\label{UVWL}
  E \sim L ^{-\frac{1}{n-1}}.
\end{equation}
The UV-behavior of the Wilson loop will take the same functional form as the EE
(for the case of a $d+1$ QFT with Area law) $S\sim L^{1-d}$, when
\begin{equation}
  n= \frac{d}{d-1}.\nonumber
\end{equation}

For the D$p$ brane we then have $n=\frac{7-p}{2}$.
The UV-behavior of the Wilson loop in the $p$-dimensional QFT, coincides with the functional dependence of the
EE, calculated for a strip of lenght $L$ in $d$-space dimensions and with Area law, when
\begin{equation}\label{formula}
  p= \frac{5d-7}{d-1}.
\end{equation}

\section{Appendix: A Taxonomy of Behaviours for Systems with Sources}
\label{app:d5s2cases}
\setcounter{equation}{0}

Adding sources to the D$5$ wrapped on a two-cycle as discussed in Sections \ref{losing} and \ref{getback}, creates a rich and complex family of backgrounds, with many free parameters which influence the behaviour of the Entanglement Entropy of the dual QFT.  In this section we will systematically categorize the different cases the parameter space allows.\footnote{It should be noted that we will focus here on the connected solution. In all cases, we chose a renormalization scheme such that the disconnected case will be a horizontal line at $S=0$ in the $S(L)$ plots} The behaviour of the Entanglement Entropy is most severely influenced by the following choices:

{\it{Rotation}} - The U-duality/rotation of the $SU(3)$-structure described in Section \ref{recoverbb} and \ref{losing}, that we will refer to as ``rotation" for brevity, will only be applicable to cases where the solution to the `master' equation 
exhibits exponential behavior in the UV---see eq.(\ref{zzxxaa}).  
However, even after the addition of sources, solutions that are linear in the UV can still be found. In what follows we will make a clear distinction between the linear (and thus unrotated) and rotated exponential behaviour of $P(\rho)$ . Note that here we shall not study unrotated exponentially growing solutions, representing QFTs coupled to gravity and string modes, as they are known to require a non-trivial UV completion and in these cases the Entanglement Entropy $S(L)$ will always diverge as in Figure \ref{Fig:MNunrot}.

{\it{Profile}} - The second important choice is the 
type of profile to be used. In what follows we will focus on two 
types of profile, those profiles in which $\lim_{\rho\to\infty}\mathcal{S}(\rho)=1$, or those that instead have  $\lim_{\rho\to\infty}\mathcal{S}(\rho)=0$.  The first 
type are given by eq.(\ref{sigmoid}) and we will refer to profiles of this type as ``sigmoid" profiles.  The profiles we will adopt for the second case are given by eq.\eqref{bump} and will be referred to as ``bump-like" profiles.

Thus we will divide this section into 4 subsections, each discussing one possible combination of the above choices. The analysis in each subsection involves the study of the interplay of the three relevant scales in the background:
\begin{itemize}
\item{$\rho_*$ - The scale at which the source profile $\mathcal{S}(\rho)$ becomes non-zero}
\item{$\bar{\rho}$ - The scale at which $P$ changes from having linear behavior (when $\r<\bar{\r}$) to having exponential asymptotic behavior (when $\r>\bar{\r}$)}\footnote{Note this only applies in cases with $P\sim e^{4\r/3}$ but not in cases with $P \sim \rho$}
\item{$x=\frac{N_f}{N_c}$ - the ratio of source branes to color branes present}
\end{itemize}

In each section, the discussion will usually start with the case where $\rho_*=0$, to see what effect the addition of sources has. Once this case is understood, predicting the behaviour of the system with non-zero $\rho_*$ becomes trivial. For $\r<\r_*$, the system will behave like the corresponding solution on the Baryonic Branch, and then at $\rho_*$ will smoothly switch to the associated behaviour of the sourced system.

Further, note that the scale $\bar{\rho}$ is only finite in the rotated case. Thus, in a similar fashion to the discussion in the last paragraph, we know that for $\r<\bar{\r}$ the system will behave like its corresponding linear case (with or without the addition of sources). Thus it makes sense to study the linear $P$ solutions with added sources first.

Finally, although we do not always discuss them here directly, the `short configurations' play an important role in the examples below that exhibit non-locality in the corresponding QFT, in a similar way to the cases presented in the main body of the paper.

\subsection{Linear $P$ Sourced Systems with Sigmoid Profiles}

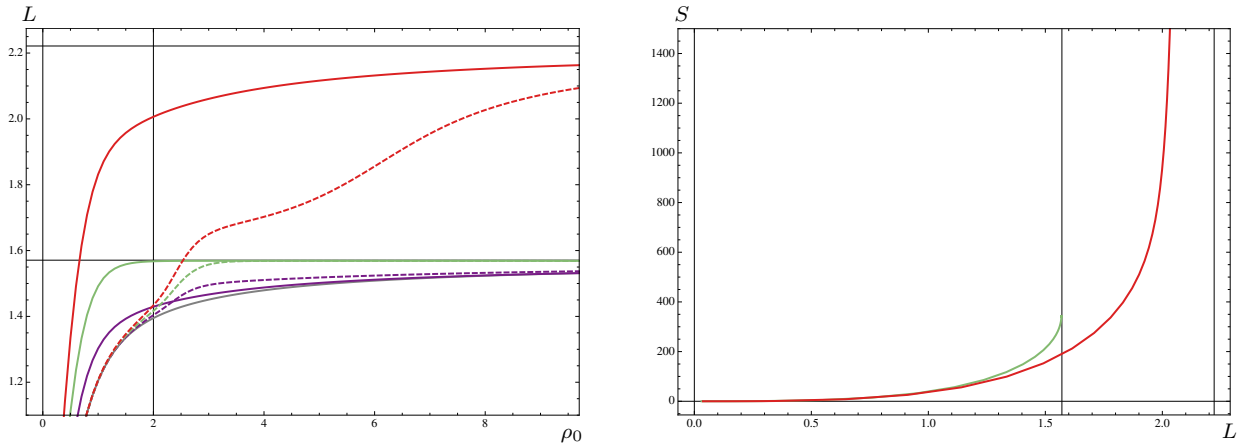
\begin{figure}[h]
\begin{center}
\begin{picture}(220,160)
\put(-124,0){\includegraphics[height=5.35cm]{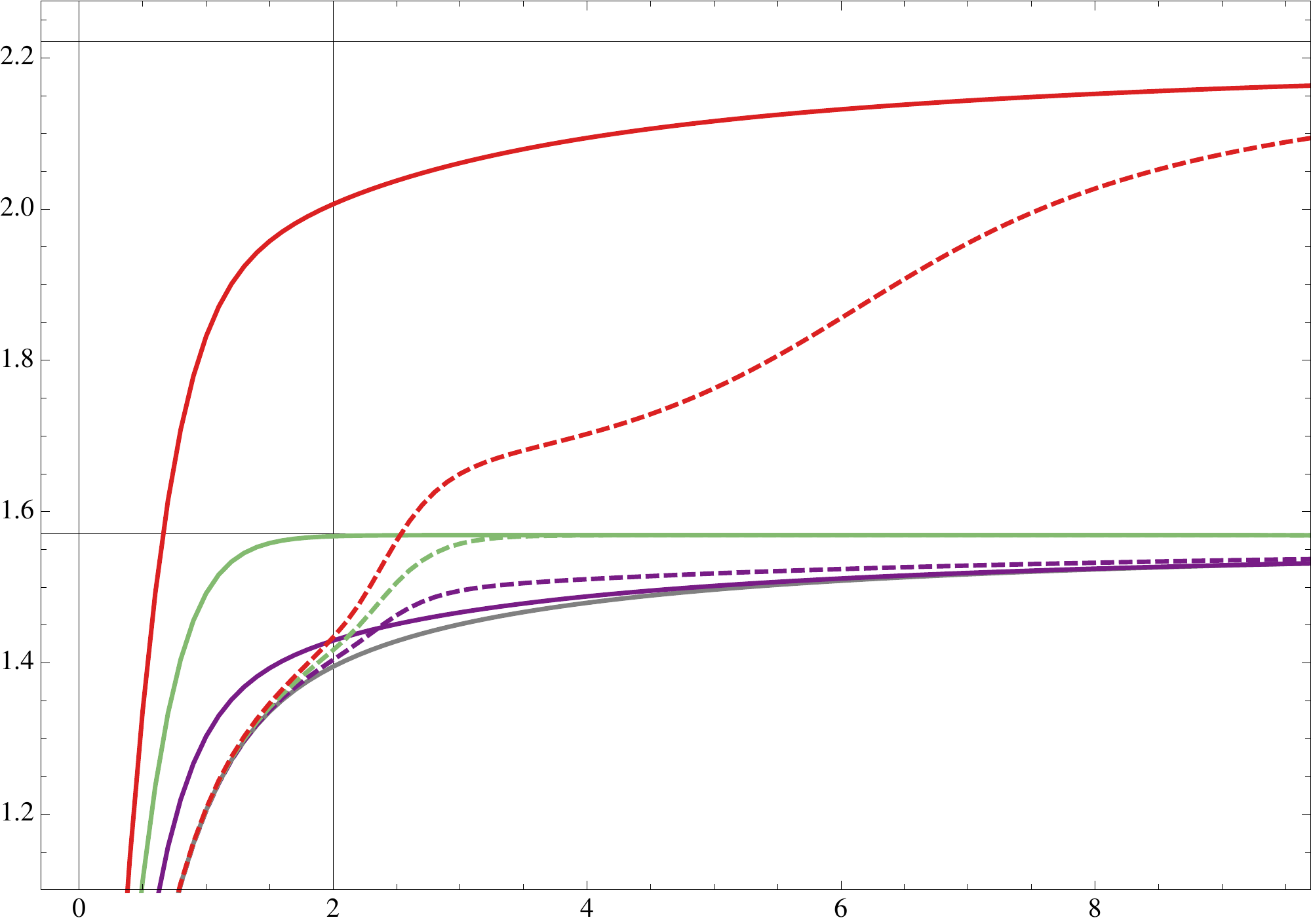}}
\put(-119,155){{\scriptsize{$L$}}}
\put(85,-3){{\scriptsize{$\r_{0}$}}}
\put(119,0){\includegraphics[height=5.35cm]{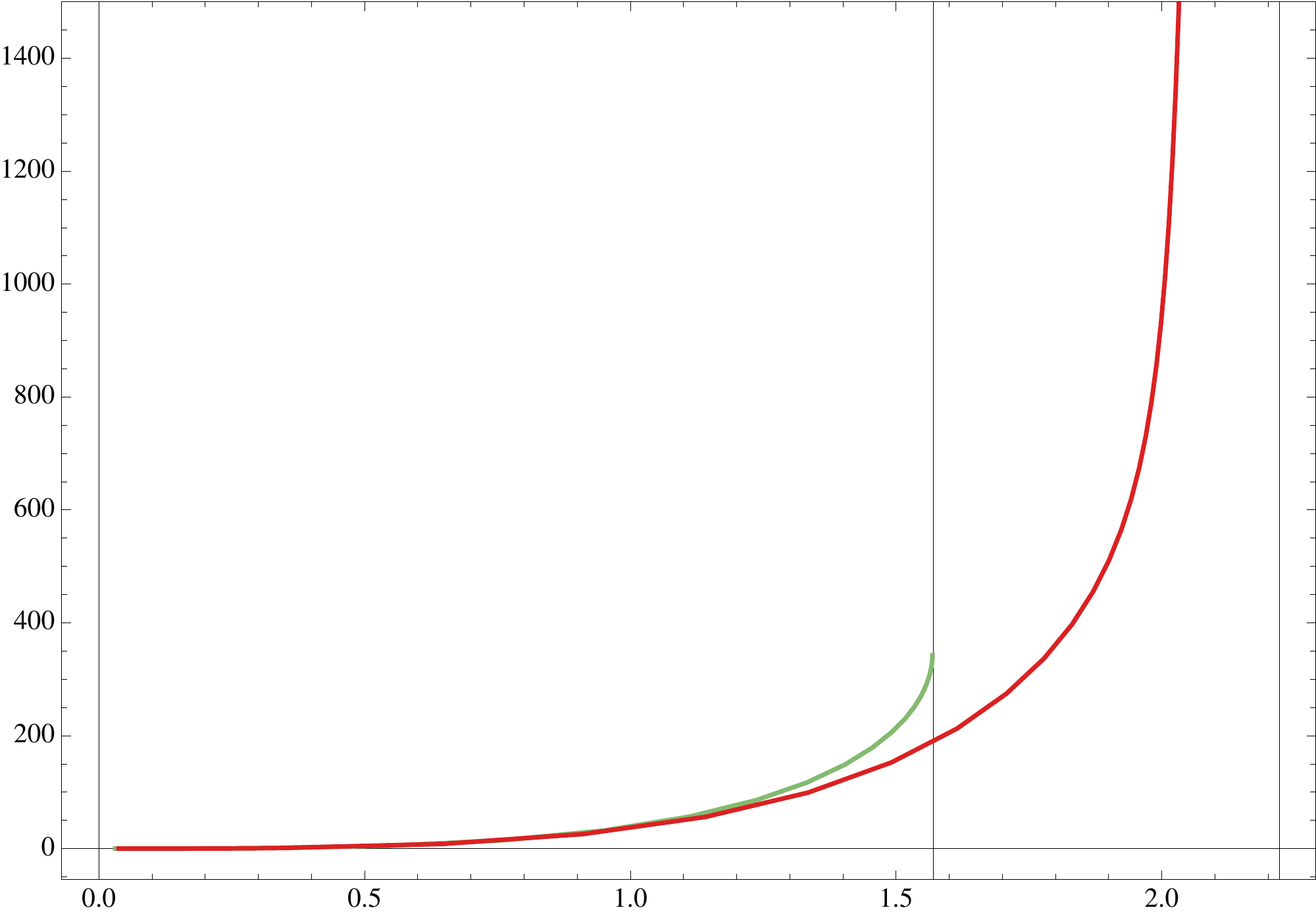}}
\put(128,155){{\scriptsize{$S$}}}
\put(335,-3){{\scriptsize{$L$}}}
\end{picture}
\caption{Various Plots of $L(\rho_0)$ and $S(L)$ in linear backgrounds with sigmoid source profiles. Continuous lines represent solutions for $\r_*=0$, while dashed lines represent $\r_*=2$. The coloring is such that the solutions with $x=\{1,2,3\}$ are given by $\{$purple, green, red$\}$ respectively. The grey line represents the sourceless $P=2N_c\r$ solution.}
\label{Fig:Unrot-Sig}
\end{center}
\end{figure}

All the solutions of this form are such that $L\to \text{const} \neq 0$ from below with no phase transition in $S(L)$ as we move toward the UV. That said, there exists an interesting dependence on $x=\frac{N_f}{N_{c}}$, with $x=2$ playing an important role. For $x<2$ and 
setting $\alpha'=g_s=1$ we have,
\begin{equation}
L(\r_0 \to \infty) = \frac{\pi}{2}  \sqrt{N_c}\left(1-\frac{1}{4 \r_0}\right)+\mathcal{O}\left(\frac{1}{\r_0^{2}}\right),
\end{equation}
which is while for $x>2$ we get
\begin{equation}
L(\r_0 \to \infty) =\frac{\pi}{2} \sqrt{N_f-N_c}\left(1-\frac{1}{4 \r_0}\right)+\mathcal{O}\left(\frac{1}{\r_0^{2}}\right).
\end{equation}
For $x=2$ we get precisely
\begin{equation}
L(\r_0 \to \infty) =\frac{\pi  }{2}\sqrt{N_c}.
\end{equation}
We see that for $x<2$, $L$ approaches $\frac{\pi}{2}\sqrt{N_c}$ (as in the sourceless case), while for $x>2$, $L$ approaches $\frac{\pi}{2} \sqrt{N_f-N_c}$. In the case of $x=2$, the UV expansion is exact, indicating that $L$ reaches its bound at a finite value of $\r$. Taking into account the analysis of this class 
of solutions found in \cite{Barranco:2011vt}, this behaviour can be explained through a Seiberg Duality picture, which involves taking the function $Q\rightarrow-Q$ and $N_c\rightarrow N_f-N_c$, relating solutions below and above $x=2$. For $x=2$, we have an invariance under the aforementioned duality transformation, and this effectively freezes the entanglement entropy in place (away from the IR), causing the $S(L)$ plot of this solution (see Figure
\ref{Fig:Unrot-Sig}) to stop at a finite value and not grow without bound as $S(L)$ for all other solutions of this class.

If we now consider profiles with non-zero $\rho_*$, for $\r_0<\r_{*}$, the solution will follow the sourceless case, whose limit is $L\to\frac{\pi }{2} \sqrt{N_c}$, and thus smaller than or equal to the limit of the equivalent solution with sources. This, and the fact that adding sources leads to the solution approaching its UV limit faster than the sourceless solution, guarantees that adding sources at $\r_*$ will always cause an increase in $L$ (as shown in Figure \ref{Fig:Unrot-Sig}). Hence, it is not possible to produce a phase transition, due to the addition of sources, for these kinds of solutions.

\subsection{Linear $P$ Sourced Systems with Bump-like Profiles} \label{Unrot-Bump}

The most important difference between this and the previous subsection is that the source profile is suppressed in the UV. Thus we expect the deviation from the sourceless case to diminish for large $\r$. Nevertheless, the profile still introduces an interesting dynamic into the system, especially when we look at non-zero $\r_*$.  Although so far we have described this as the scale at which the sources become `active' (forcing the system to deviate from the sourceless case), it now further dictates the point at which the sources effectively become `inactive', thus bringing the system back in line with the corresponding sourceless case.

A simple check agrees with the above analysis: in all cases we have $L\to\frac{\pi}{2}\sqrt{N_c}$ in the UV, as expected. As in the previous subsection, the behaviour has an interesting dependence on $x$. For low $x$, we see no phase transition, only a small deformation on $L$ around $\r_*$ is noticeable. But for large enough $x$, this deformation becomes sharp enough such that a local maximum forms in $L$, and thus a we have a phase transition in $S(L)$. The critical value $x_c$ at which a phase transition forms is a monotonically decreasing positive-definite function of $\r_*$. This is logical as a deformation in $L$ means a change in $L'$. As we can see from Figure \ref{Fig:MN}, $L'$ is a monotonically decreasing positive-definite function, so at larger $\rho$ any effect in $L$ is more pronounced.  It is possible to estimate the value at which a phase transition is introduced. For $\r_*=0$, the critical value is $x_c\approx5.3$, while for $\r_*=2$, $x_c\approx2.3$ already.

Further, it should be noted that while we are able to form a phase transition via the addition of sources, the fact that $L$ in the UV approaches a finite limit from below (which we term $L_{UV}$), guarantees that for every first order phase transition, from the disconnected to the connected solution, would be accompanied by a discontinuous jump, from the continuous solution back to the discontinuous solution. Thus, here we again require the introduction of the branch of solutions living close to the boundary (that we termed `short configurations'), which will remove the problem, as it is the preferred branch for $L<L_{UV}$.
\begin{figure}[h]
\begin{center}
\begin{picture}(220,160)
\put(-124,0){\includegraphics[height=5.35cm]{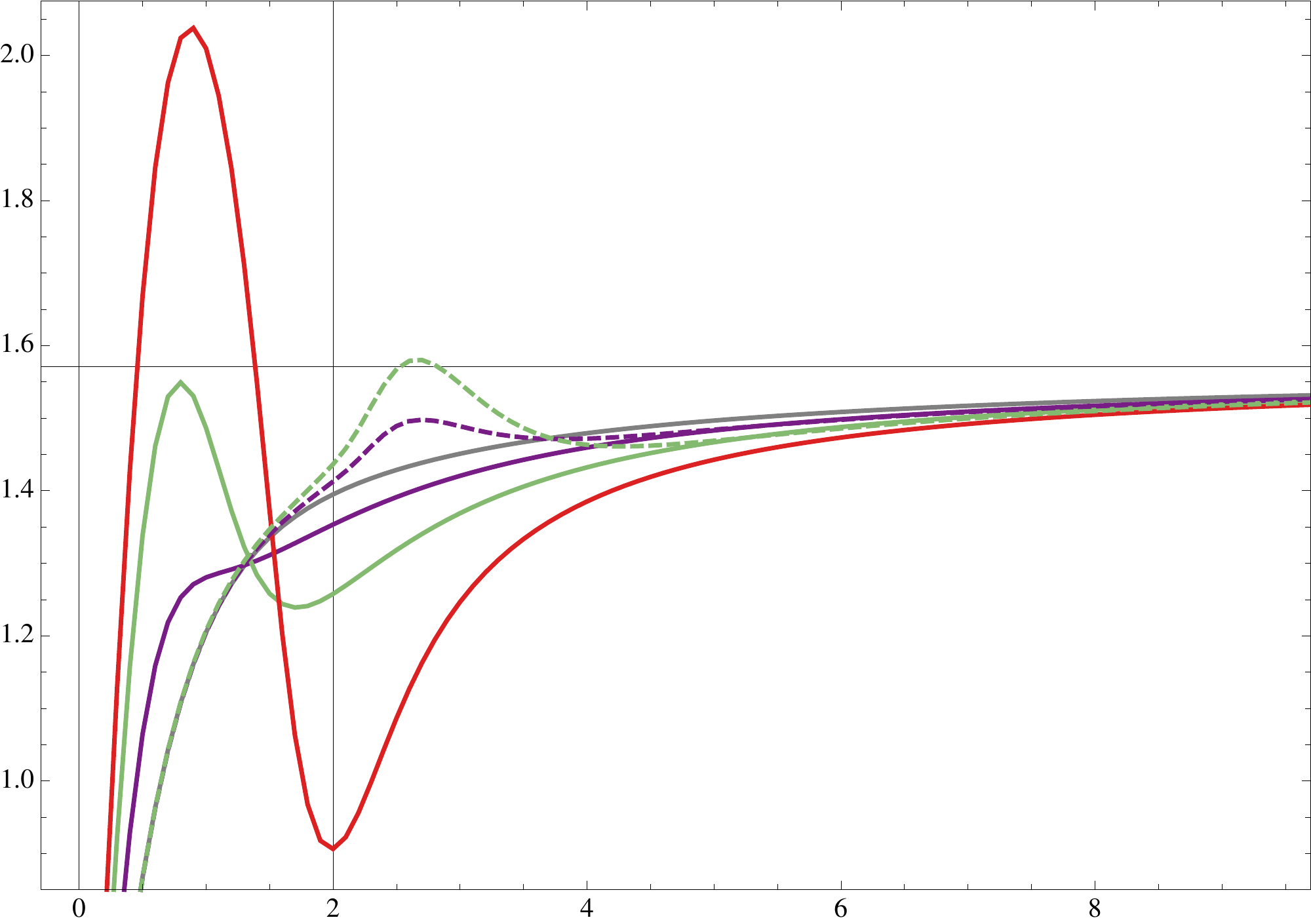}}
\put(-119,155){{\scriptsize{$L$}}}
\put(85,-3){{\scriptsize{$\r_{0}$}}}
\put(116,0){\includegraphics[height=5.35cm]{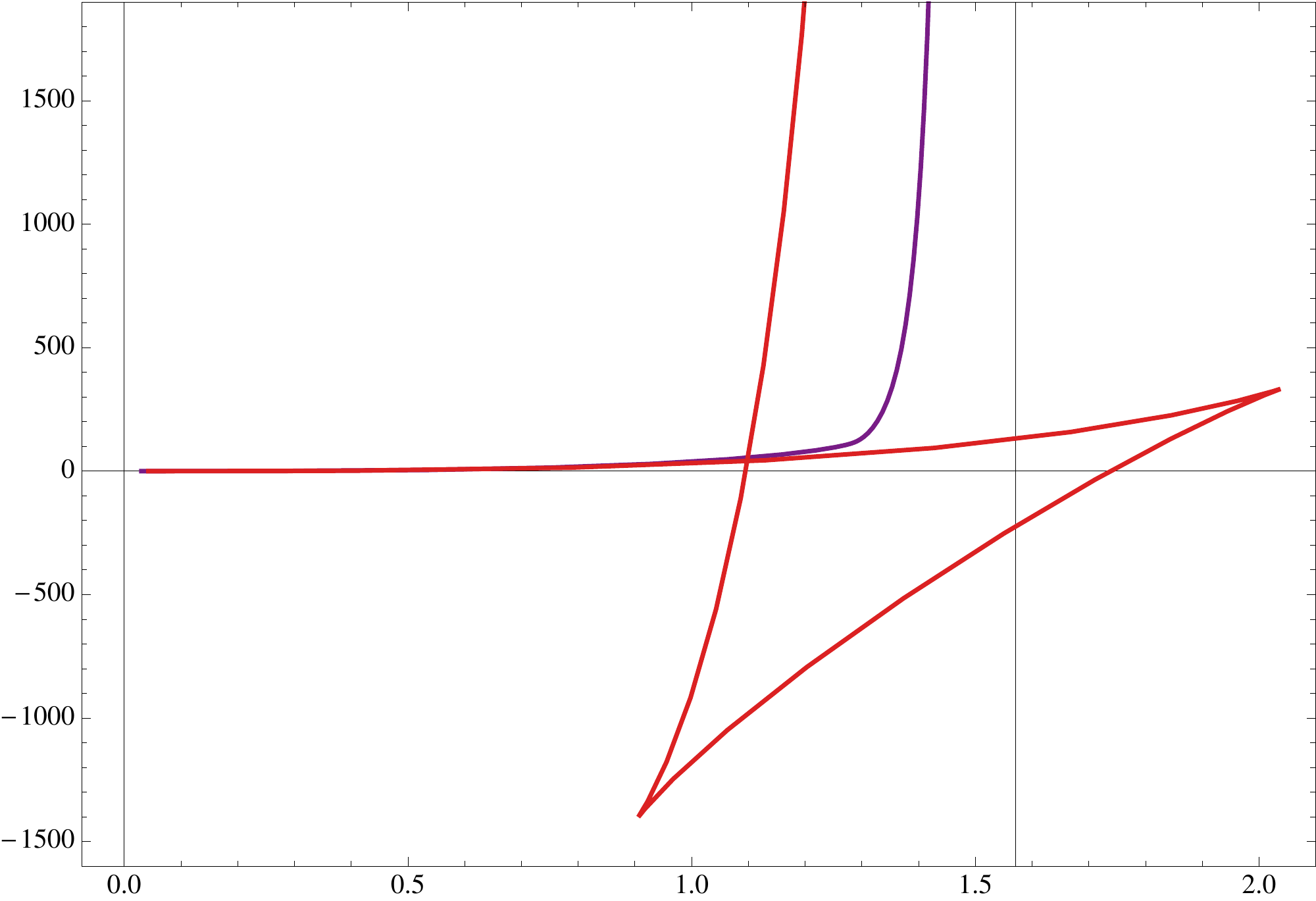}}
\put(128,155){{\scriptsize{$S$}}}
\put(335,-3){{\scriptsize{$L$}}}
\end{picture}
\caption{Various Plots of $L(\rho_0)$ and $S(L)$ in linear $P$ backgrounds with bump-like source profiles. Continuous lines represent solutions for $\r_*=0$, while dashed lines represent $\r_*=2$. The coloring is such that the solutions with $x=\{5,10,15\}$ is given by $\{$purple, green, red$\}$ respectively. The grey line represents the sourceless $P=2N_c\r$ solution.}
\label{Fig:Unrot-Bump}
\end{center}
\end{figure}
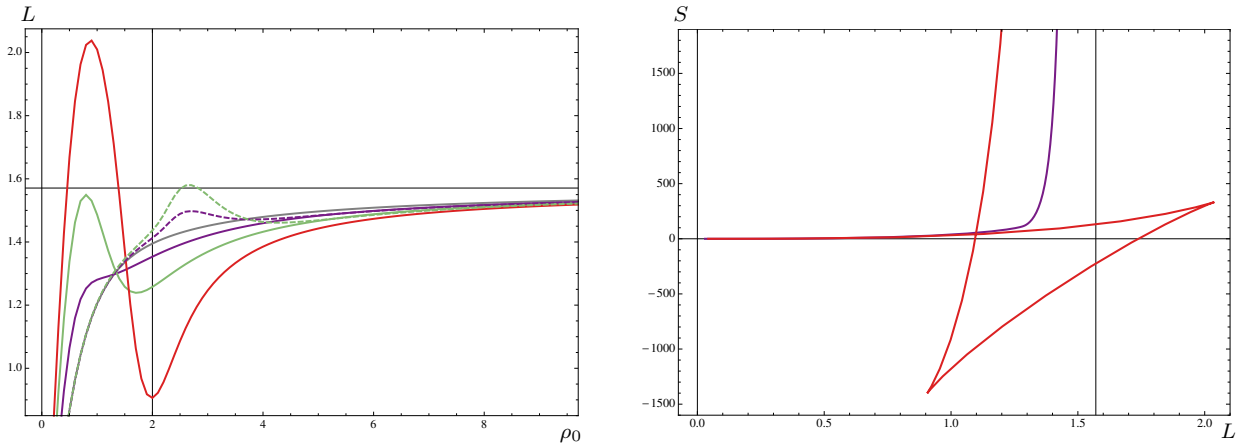

\subsection{Rotated Sourced Systems with Sigmoid Profiles}

All solutions of this class share the following UV asymptotics for $L$:
\begin{equation}
L(\r_0\to\infty)=\frac{3 \pi}{8}\sqrt{N_f}\left(1-\frac{(N_f-2 N_c)^2}{4 c N_f}\r_0\, e^{-4 \r_0/3}\right)+\mathcal{O}\left(e^{-4 \r_0/3}\right)
\end{equation}
From this we can immediately see that $L$ will always eventually approach $\frac{3\pi}{8}\sqrt{N_f}$ from below. This means there are only the two possibilities (depicted in Figure \ref{Fig:Unrot-Bump}): either we will have a phase transition (connected - disconnected) coupled with a discontinuity, or there will be no phase transition at all (disconnected always preferred).  Again, bear in mind, that we would employ the branch of solutions near the boundary to resolve any 
issues with the stability of the configuration and the 
above discontinuity, or lack of phase transition.

Let us study in detail which case occurs for different values in the parameter space.\footnote{Keep in mind that as mentioned, for $\r<\r_*$ the solution will follow its corresponding (rotated) sourceless case, and for $\r<\bar{\r}$ the solution will behave exactly as the solutions discussed so far in this appendix.} As expected from the van der Waals gas analogy, we will never find a phase transition when the two scales $\r_*$ and $\bar{\r}$ are too close together. Thus we will assume that there always is a large enough separation between the two scales in the following.

First let us assume $0\leq\r_*<\bar{\r}$. We know from our previous analysis that we  cannot have a phase transition before $\r<\bar{\r}$. We also know that $\lim_{\r\to\infty} L = \{\frac{\pi  \sqrt{N_c}}{2},\frac{\pi  \sqrt{N_f-N_c}}{2}\}$ for $\{x\leq 2, x>2\}$ respectively and beyond $\bar{\r}$, $L$ tends to $\frac{3\pi}{8}\sqrt{N_f}$. Thus we can see that $L$ will (for almost all values of $x$) first approach a larger value than its UV limit, guaranteeing a phase transition. The exception is the range $\frac{16}{9}\leq x \leq \frac{16}{7}$, for which a phase transition does not occur, no matter the separation between $\r_*$ and $\bar{\r}$, making the region around $x=2$ quite special.

Let us study the case $0\leq\bar{\r}<\r_*$. We know that before $\r_*$ the solution will behave like its corresponding (rotated) sourceless solution. Thus the Entanglement Entropy will behave as depicted in Figure
\ref{Fig:MNrot2} in this region. Thus, as long as we push $\r_*$ sufficiently far away to make sure we pick up the maximum in $L$, we will always produce a phase transition.
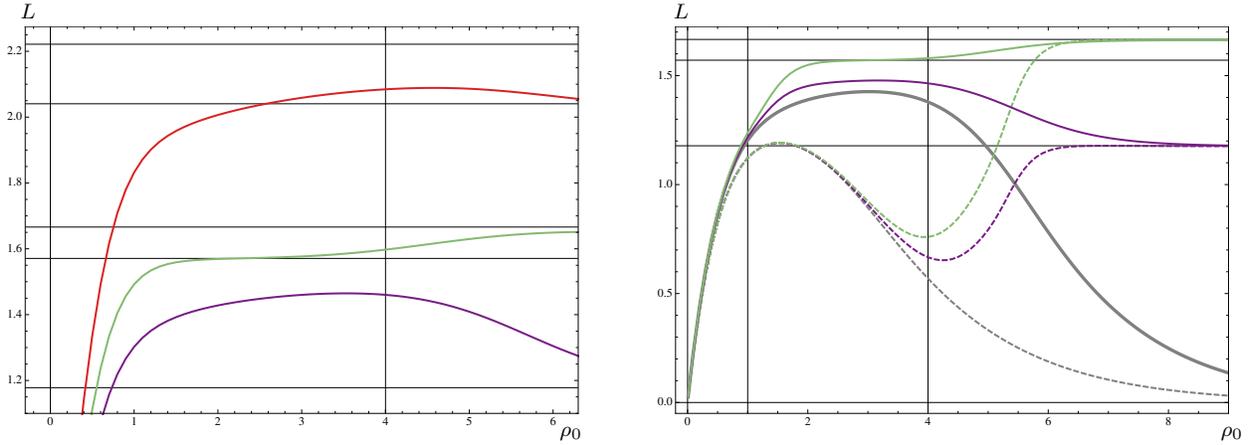
\begin{figure}[h]
\begin{center}
\begin{picture}(220,160)
\put(-124,0){\includegraphics[height=5.35cm]{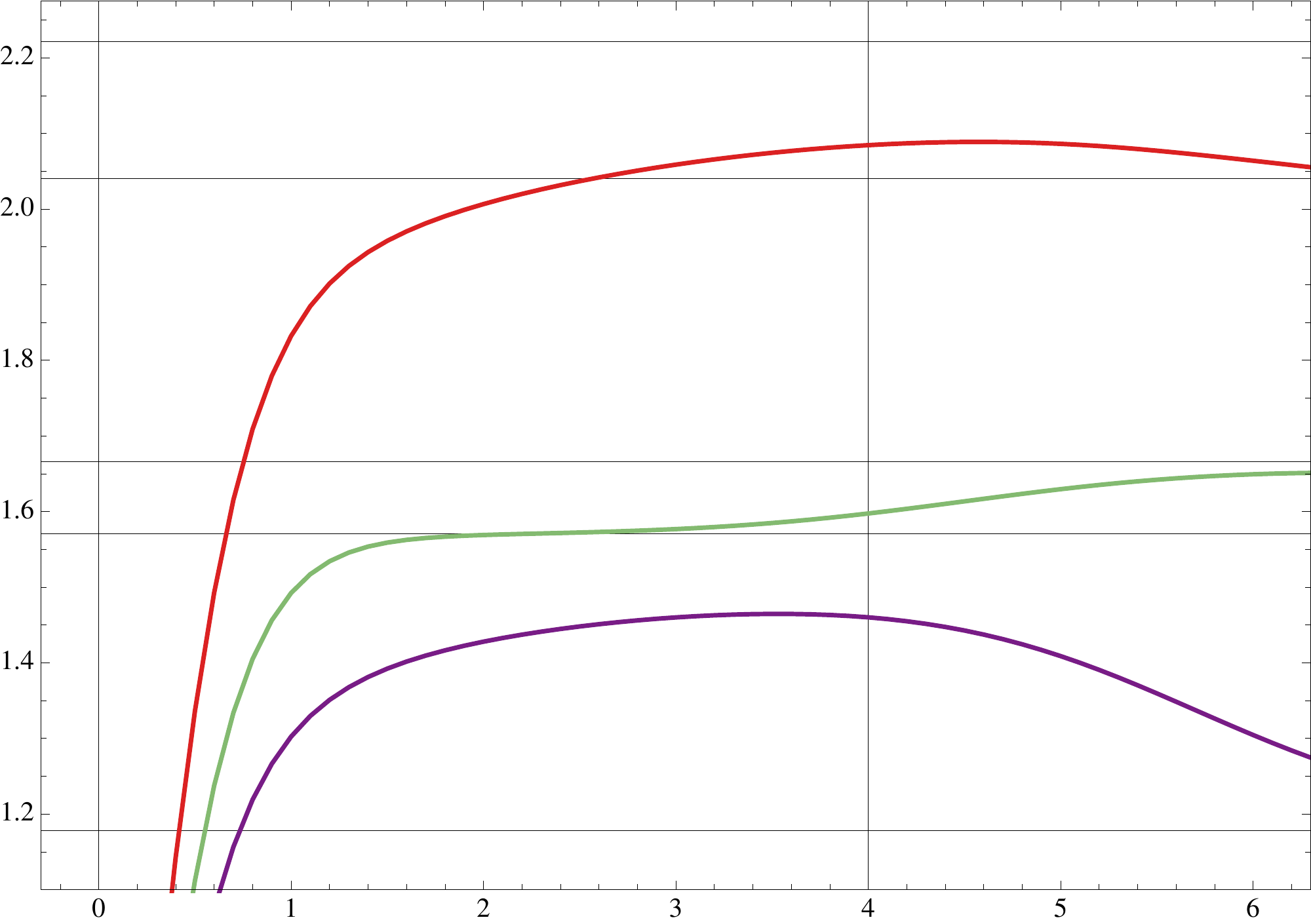}}
\put(-119,155){{\scriptsize{$L$}}}
\put(85,-3){{\scriptsize{$\r_{0}$}}}
\put(122,0){\includegraphics[height=5.35cm]{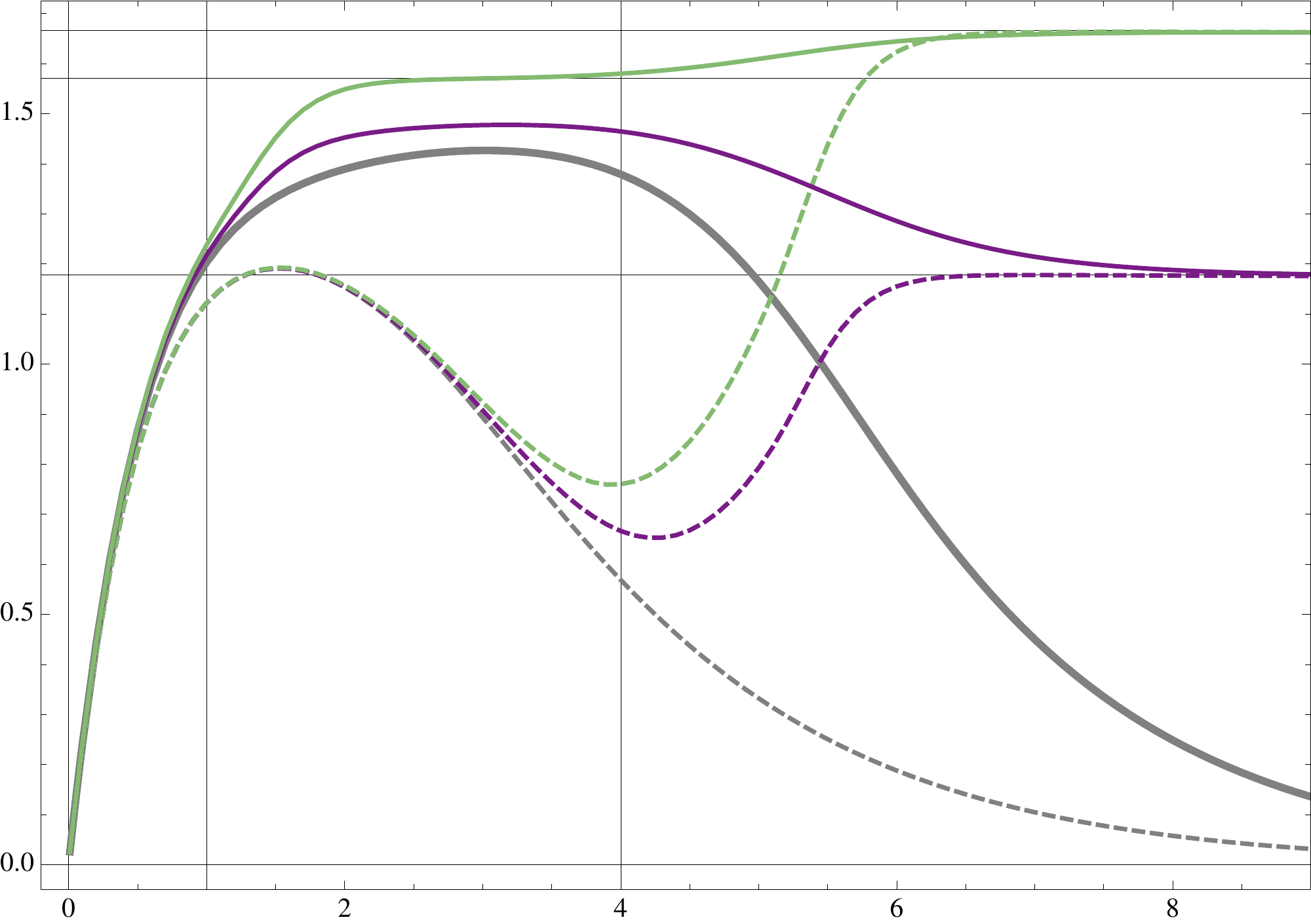}}
\put(128,155){{\scriptsize{$L$}}}
\put(335,-3){{\scriptsize{$\r_0$}}}
\end{picture}
\caption{Various Plots of $L(\rho_0)$ in rotated backgrounds with sigmoid source profiles. The left-hand panel shows solutions for $\r_*=0$, $\bar{\r}=4$. In the right-hand panel, continuous lines are for solutions with $\r_* =4$, $\bar{\r}=1$, while dashed lines represent $\r_*=1$, $\bar{\r}=4$.  The coloring is such that the solutions with $x=\{1,2,3\}$ is given by $\{$purple, green, red$\}$ respectively. The grey graphs represent the corresponding rotated sourceless  solutions.}
\label{Fig:Rot-Sig}
\end{center}
\end{figure}

\subsection{Rotated Sourced Systems with Bump-like Profiles}

For all rotated systems with bump-like profile we have $L\to0$ in the UV. This guarantees that we have at least one phase transition in every case.  It turns out, one can construct solutions with two phase transitions, one between the disconnected solution and the connected, and one between two different branches of the connected solution.

As we know for $\r<\bar{\r}$ the solution will behave exactly as in Section \ref{Unrot-Bump}. Thus we know, that we can pick up a phase transition for large $x$, that is produced through the interplay of the scales where the sources become relevant and are suppressed. Further, now that we have a finite $\bar{\r}$, we effectively introduce a third scale, and it is exactly this scale that gives us the second phase transition. The phase transition due to $\bar{\r}$ is always present.

\begin{figure}
\begin{center}
\begin{picture}(220,450)
\put(-115,300){\includegraphics[height=5cm]{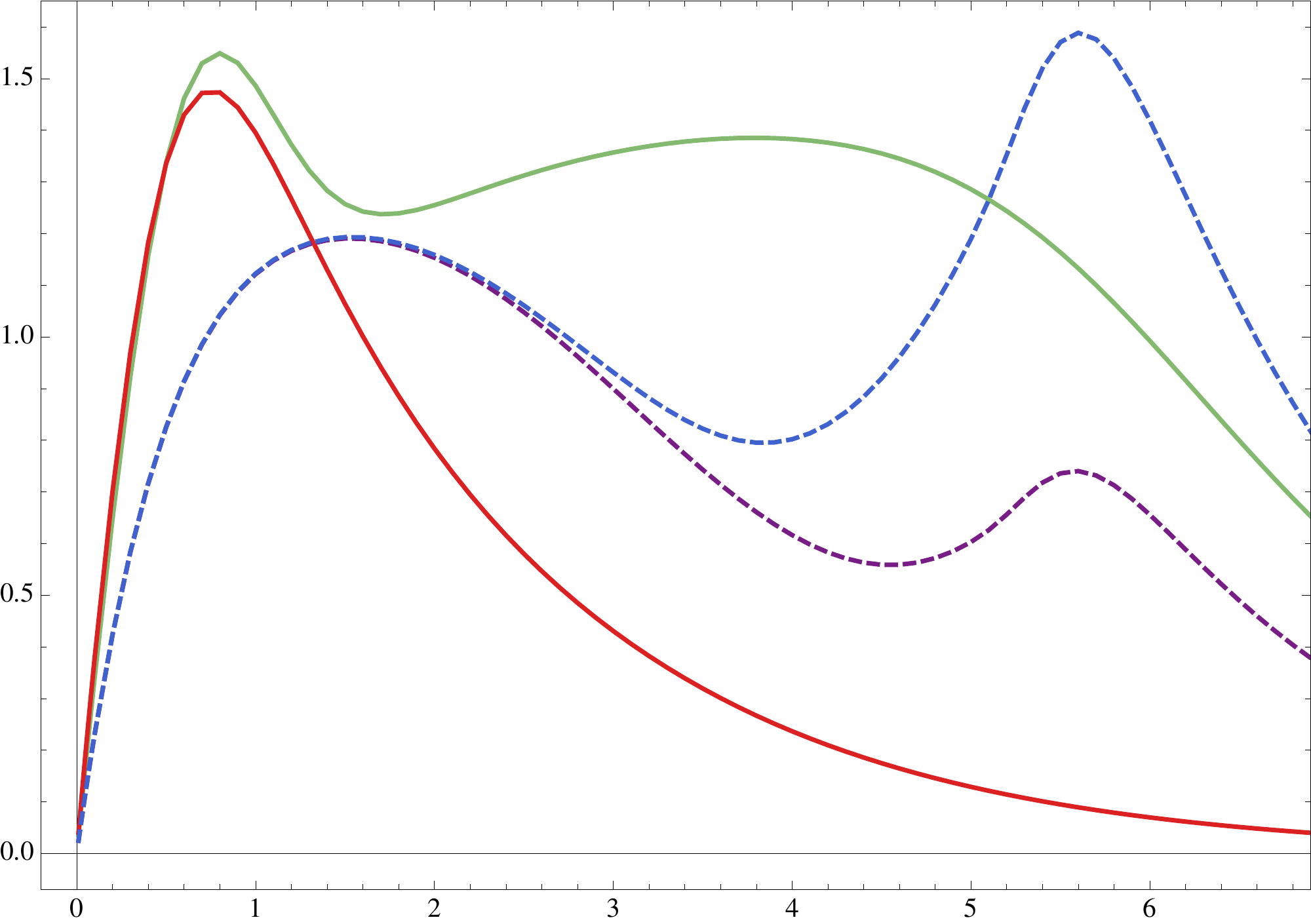}}
\put(-117,438){{\scriptsize{$L$}}}
\put(89,303){{\scriptsize{$\r_{0}$}}}
\put(123,300){\includegraphics[height=5cm]{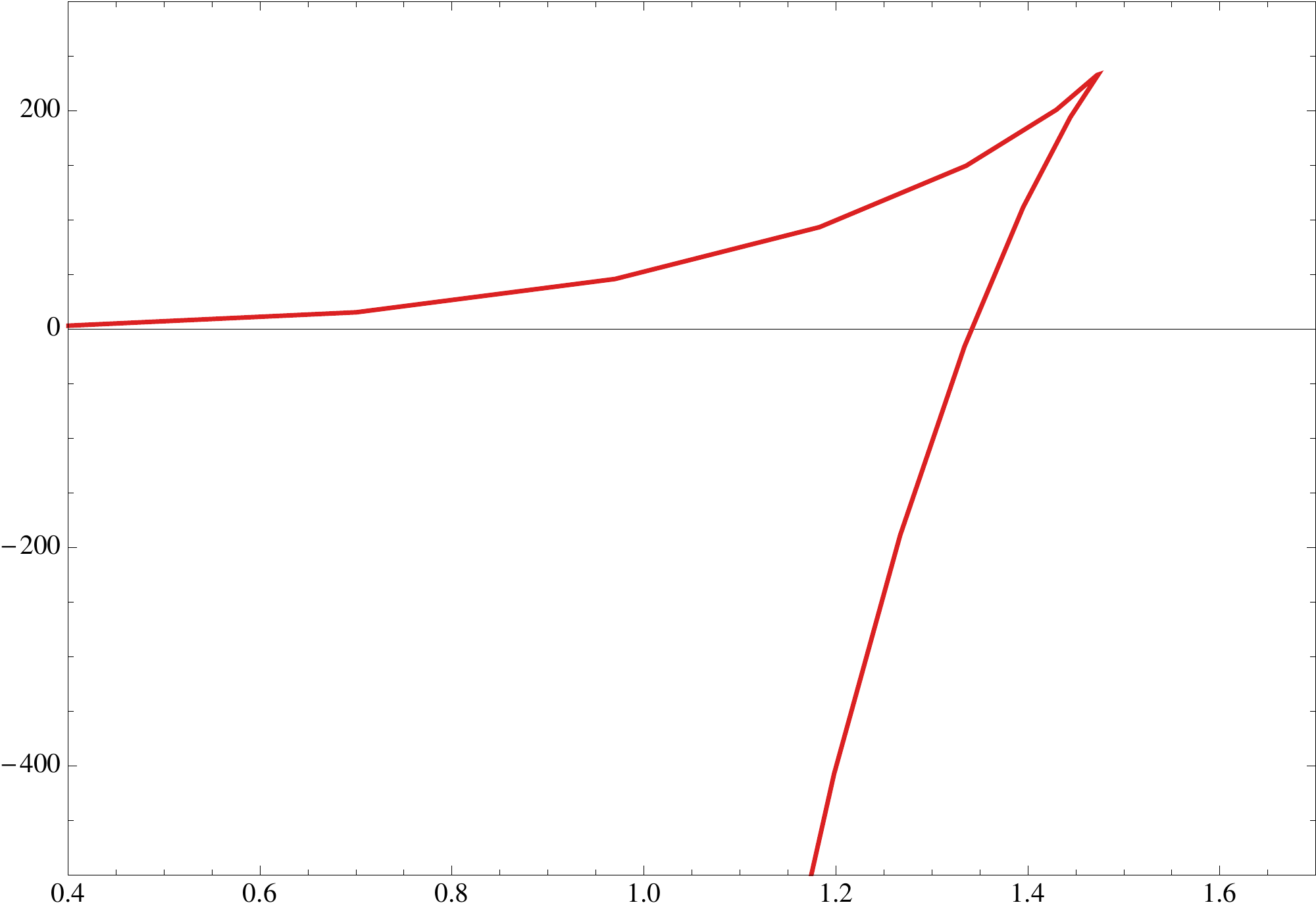}}
\put(126,438){{\scriptsize{$S$}}}
\put(332,303){{\scriptsize{$L$}}}
\put(-120,150){\includegraphics[height=5cm]{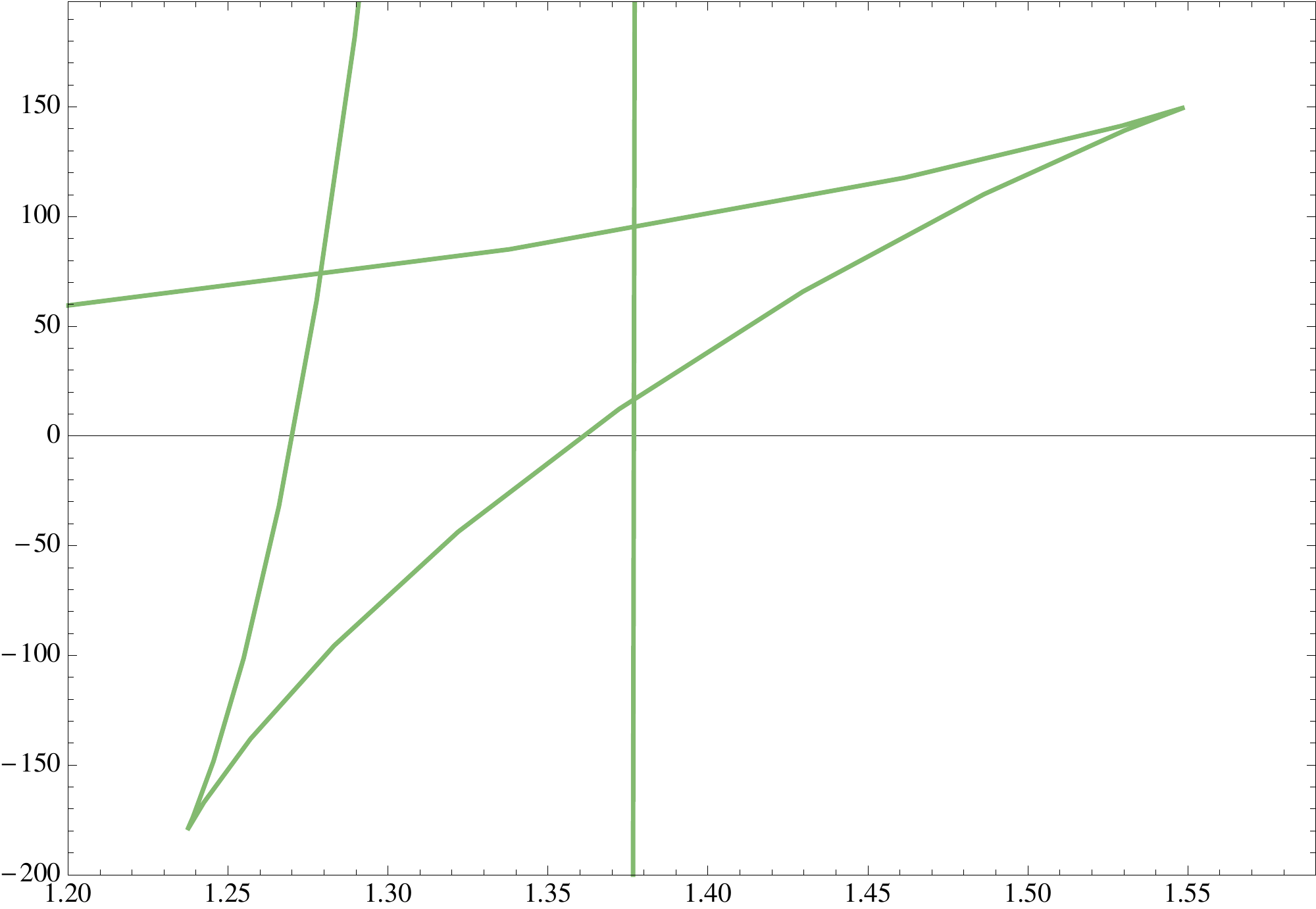}}
\put(-117,287){{\scriptsize{$S$}}}
\put(89,153){{\scriptsize{$L$}}}
\put(116,150){\includegraphics[height=5cm]{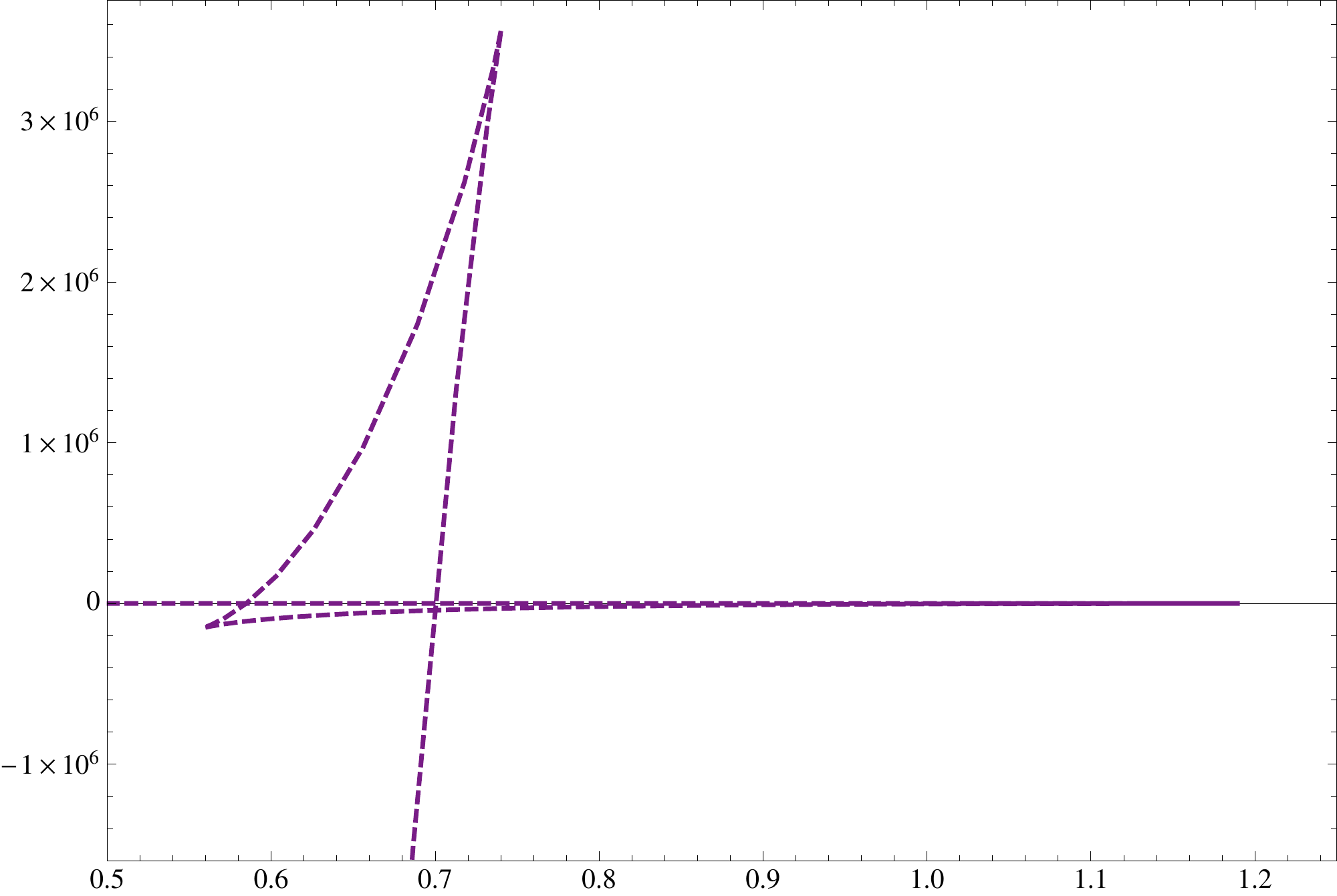}}
\put(126,287){{\scriptsize{$S$}}}
\put(332,153){{\scriptsize{$L$}}}
\put(-128,0){\includegraphics[height=5cm]{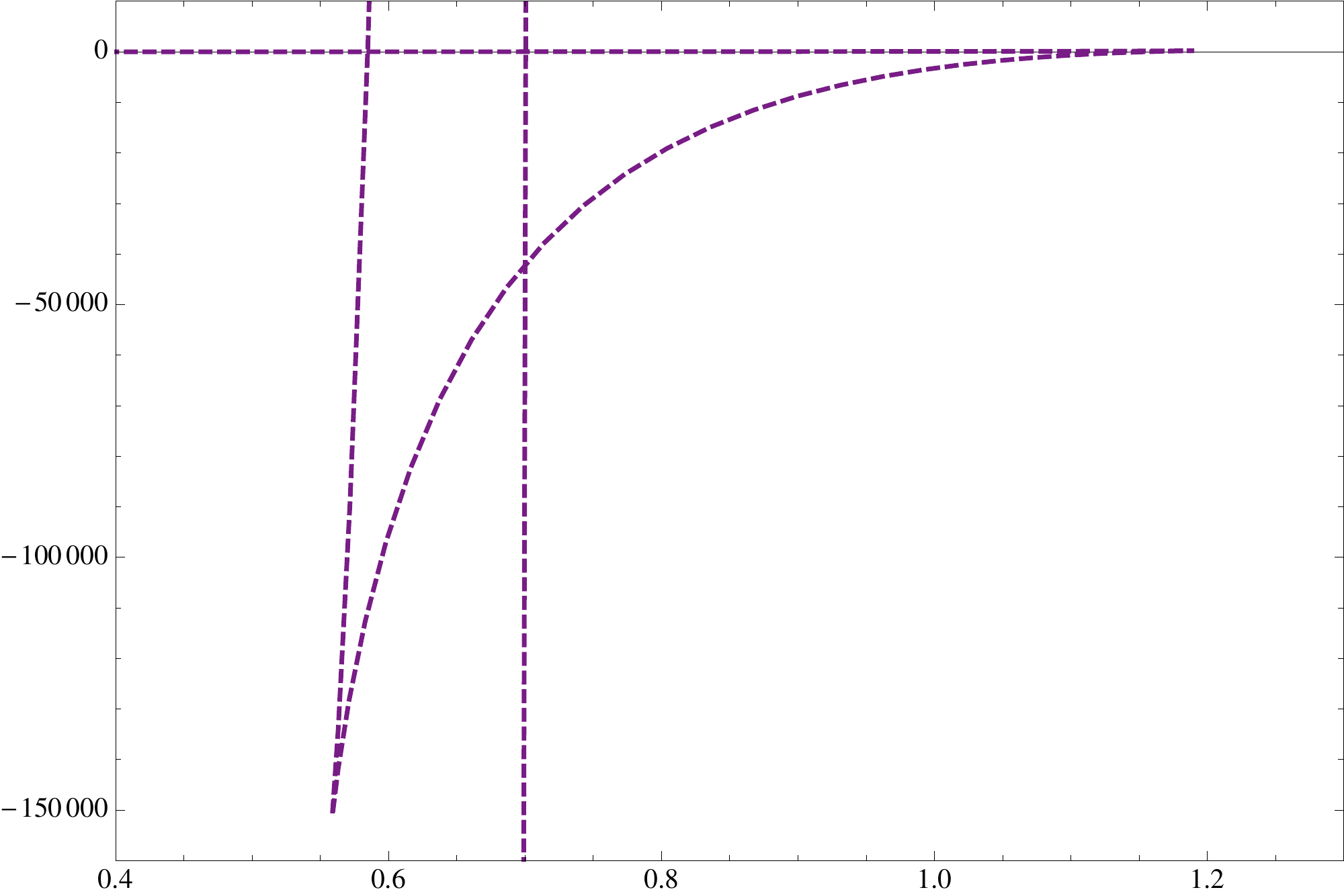}}
\put(-117,137){{\scriptsize{$S$}}}
\put(89,3){{\scriptsize{$L$}}}
\put(122,0){\includegraphics[height=5cm]{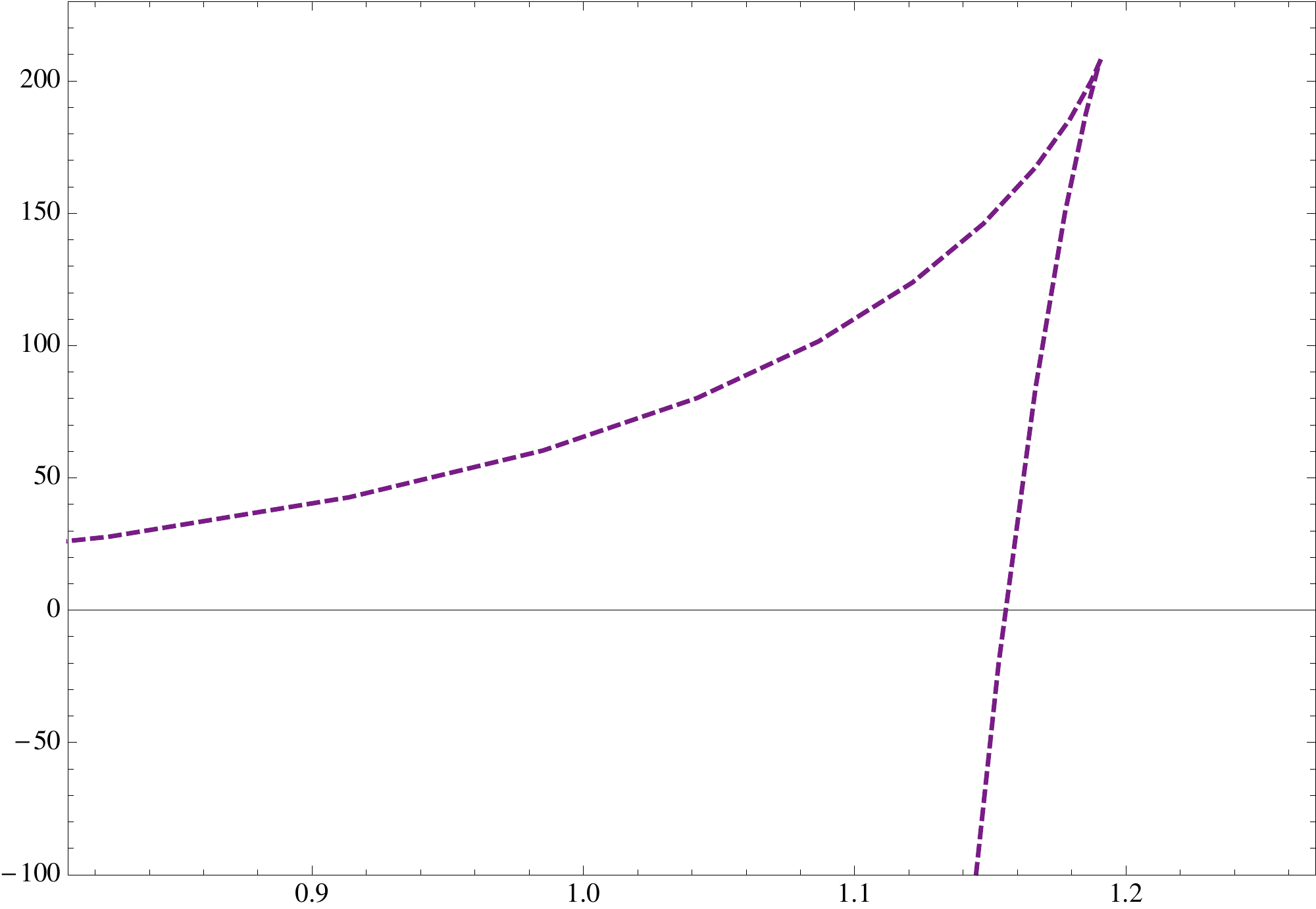}}
\put(126,137){{\scriptsize{$S$}}}
\put(332,3){{\scriptsize{$L$}}}
\end{picture}
\caption{In all figures above: continuous graphs have $\r_*=0$ and $x=10$ and dashed ones have $\r_*=5$ and $\bar{\r}=2$. For the continuous lines, the red line has $\bar{\r}=0$, while the green line has $\bar{\r}=4$. For the dashed lines, blue indicates $x=10$ and purple $x=2$.  In all of the panels involving $S(L)$, the disconnected solution lies at $S=0$, but is not shown. As we can see in the top-right panel, the continuous red solution has a first order phase transition.  The left-hand panel in the middle row is an example of overshooting - the branch that is relevant in the dashed purple case between the transitions, is above the disconnected solution and is thus unphysical. The right-hand figure in the middle row represents a solution that has been tuned to have 2 physical phase transitions. The graph has been enhanced in the two panels in the bottom row to clearly show each of the phase transitions.}
\label{Fig:Rot-Bump}
\end{center}
\end{figure}
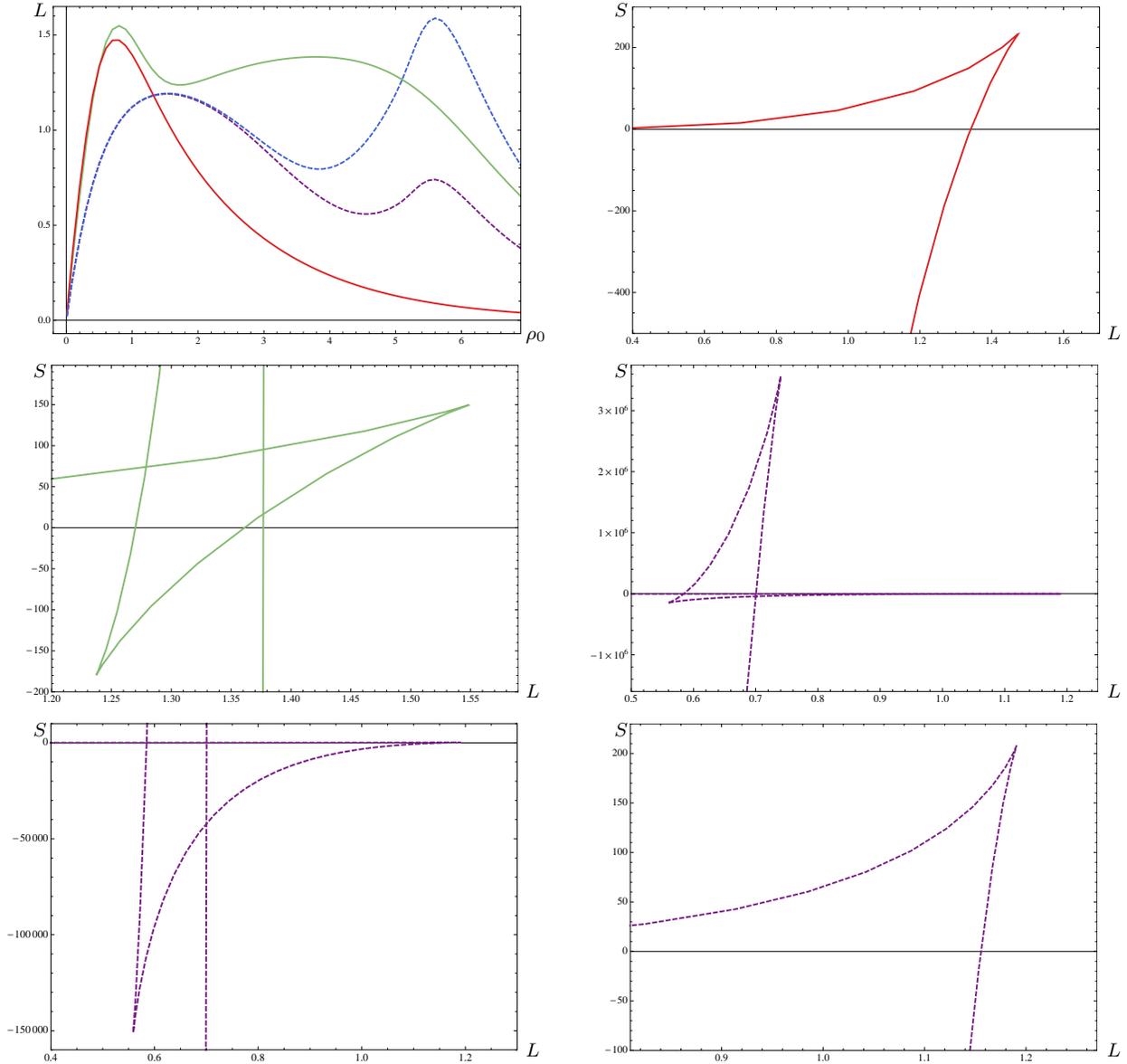

Thus we know that we will always have two phase transitions for $x>x_c$.  If we choose a non-zero $\r_*$, we additionally need the condition $|\bar{\r}-\r_*| \gg 0$. However, there is a third, more subtle condition. It follows from the fact, that while the above conditions are sufficient to produce two phase transitions, it is possible for the system to ``overshoot", leading to one of the phase transitions being in the unphysical branches of the Entanglement Entropy. A case like this is presented in Figure
\ref{Fig:Rot-Bump} in the middle-left panel.  It is easy to see, that the physical branch of the Entanglement Entropy will contain two phase transitions if the second maximum of $L$ is noticeably lower than the first maximum. Armed with this knowledge we deduce that the third condition is:
\begin{itemize}
\item{$0\leq\rho_*<\bar{\r}$ - $x$ must be very large. The critical value $x_{cc}$ above which we have two phase transition is always larger than $x_c$. Thus, here, this condition is always strictly stronger than the first one.}
\item{$\bar{\rho}<\r_*$ - $x$ increases the size of the second bump, so we know that we will have two physical phase transitions only for a bounded range of values of $x$, namely $x_c<x<x_{cc}$.}
\end{itemize}
Last, but not least it should be noted, that we can increase the number of phase transitions by introducing additional scales. For example, we could use source profiles such as \cite{Warschawski:2012sx}:
\begin{equation}\label{Double-Bump}
  \hat{\mathcal{S}}(\rho) = \begin{cases}
  \tanh^4(2(\r-\r_{*1})) e^\frac{-4(\r-\r_{*1})}{3}+\tanh^4(2(\r-\r_{*2})) e^\frac{-4(\r-\r_{*2})}{3} & \text{if } \rho \geq \r_{*2} \\
 \tanh^4(2(\r-\r_{*1})) e^\frac{-4(\r-\r_{*1})}{3}          & \text{otherwise}
  \end{cases}
   \end{equation}
Each time we let the sources become relevant again, we will introduce additional phase transitions that can be tuned to lie on the physical branch.

Notice that in this paper, we have used the word `source' branes instead of 
`flavor' branes. We made this distinction
because in many cases studied here, the fluctuations on the branes 
(the `mesons') are non-normalisable, and hence non-dynamical. 
The connection with studies and ideas
presented by the authors in \cite{Chang:2013mca} 
is worth pursuing and is left
for future work.
\section{Appendix: Other Models without Phase Transitions}
\label{app:ZZ}
\setcounter{equation}{0}
Here we will summarise the situation in other known confining 
models that do not produce a phase transition in 
the EE, unless of course, a cutoff is introduced as explained in 
the main body of this paper. 

We will start off by discussing the cases of D5 branes wrapped 
on a three-cycle and see how it is similar to that of the 
D5 branes wrapped on two-cycle.  
We shall then go on to look at the backgrounds discussed in KKM, 
but modified by changing the warp factor in such a way that can 
be thought of as adding a relevant operator
that changes the UV, but leaves the IR unaffected.  We then finally go on to discuss D6 branes wrapped on three-cycle, discuss the similarities, and make a comment about hard cut-offs.

We will notice, that making a more natural choice
of the radial coordinate implies, that when 
$\lim_{\r_{0}\to\infty} L(\r_{0})=\infty$, 
$L$ diverges linearly. 
This is true for many of the `divergent' cases we have studied in 
this paper. For example, 
in systems generated by D5 branes, a closer analysis reveals 
that a coordinate transformation $e^{\frac{2\r}{3}}= r $ is 
needed to properly compare those cases to the ones presented below. 
It is easy to check that this will recover the 
linear behaviour for $L(\r_0)$.

\subsection{D$5$ on Wrapped on a Three-Cycle}

Here we look at the backgrounds presented in \cite{MN:2001}, and generalised in \cite{CMR:2008},\cite{M:2013}.  We start by defining the ansatz: there are two sets of $SU(2)$ left-invariant 1-forms, $\s^i$ and $\tilde{w}^i$ ($i=1,2,3$), which obey
\begin{align}
d\s^i = -\frac{1}{2} \eps_{ijk} \s^j \wedge \s^k, \qquad d\tilde{w}^i = -\frac{1}{2} \eps_{ijk} \tilde{w}^j \wedge \tilde{w}^k.
\label{LI1Fd}
\end{align}
Each parametrises a three-sphere, and can be represented 
by three angles, $(\q,\varphi,\psi)$,
\begin{align}
	\s_1	&=	\cos\psi d\q + \sin\psi\sin\q d\varphi, \quad \s_2 = -\sin\psi d\q + \cos\psi\sin\q d \varphi, \quad \s_3	=	d\psi + \cos \q d\varphi
\label{LI1FS}
\end{align}
and similarly, three angles $(\tilde{\q},\tilde{\phi},\tilde{\psi})$ for $\tilde{w}$, which take a similar explicit form\footnote{The range of the angles here is $0\leq\q,\tilde{\q}<\p$, $0\leq \varphi,\tilde{\varphi} < 2\p$ and $0 \leq \psi,\tilde{\psi} < 4\p$}. Our spheres will also be fibered with a one-form $A^i$. The $A^i$ take the form
\begin{align}
A^{i}=\frac{1}{2}\left(1+w\right)\s^{i}
\label{As}
\end{align}
where $w$ is a function of the radial coordinate.  
We can then write down our Type IIB metric ansatz (in Einstein Frame), 
in terms of the following vielbeins,
\begin{align}
&E^{x_i}	 =	e^{f}	dx_i					,\quad
E^{\r}	 =	e^{f+g}d\r					,\quad
E^{\q}	 =  	\frac{e^{f+h}}{2}\s^1  			,\quad
E^{\varphi}	 =  	\frac{e^{f+h}}{2}\s^2 			,\quad
E^{\psi}	 =  	\frac{e^{f+h}}{2}\s^3 			,         \nonumber\\
&E^1		 =  	\frac{e^{f+g}}{2}(\tilde{w}^1-A^1)		,\quad
E^2		 =  	\frac{e^{f+g}}{2}(\tilde{w}^2-A^2)		,\quad
E^3	  	 =  	\frac{e^{f+g}}{2}(\tilde{w}^3-A^3)
\label{vielbeinUnrot}
\end{align}
where $x_i$ represents the Minkowski metric in $2+1$ dimensions, $\r$ is the radial coordinate, and $\{f, g, h\}$ are only functions of $\r$. This means we can write the metric compactly as
\begin{align}
	ds_E^2 =	\sum_{i} (E^i)^2	
\label{Unrotmetric}
\end{align}
The theory also contain a non-trivial dilaton $\Phi$. There is also a RR three-form $F_{3}$ but we shall not require its expression 
here.  From the usual SUSY requirements we find $\Phi=4f$.

There is a solution generating procedure \cite{GM:2011} (similar to the one discussed in the case of D5 branes wrapped on a two-cycle in the main text)  which takes us from this solution to one of Type IIA (with extra fluxes).  Here we write the relevant parts:  we can write our metric in the string frame using $ds^2_{s}=e^{\Phi/2} ds^2$, then use an S-duality.  The S-duality takes
\begin{align}
ds_s^2 \,  \rightarrow \, e^{-\Phi} ds^2_{s}=ds^2_{str}, \qquad F_{3} \, \rightarrow \,  H_{3}, \qquad \Phi \, \rightarrow \, -\Phi,
\label{SDual}
\end{align}
leaving us in the common Type II NS-sector.  Then after applying the dualities we generate the following Type IIA solution
\begin{align}
d{\hat{s}}^2_{str}&=\hat{h}^{-1/2}dx^{2}_{i}+\hat{h}^{1/2}ds^2_7,	\qquad	 e^{{2\hat{\Phi}}}=\hat{h}^{1/2}e^{-2\Phi},	\qquad \hat{h}=\frac{1}{\cosh^2 \b}\left(1-\tanh^{2}\b \, e^{2(\Phi-\Phi_{\infty})}\right),
\label{rotMetric}
\end{align}
where hatted quantities denote the new rotated 
solution and the unhatted are the original Type IIB functions.  
Again, we shall not require the explicit forms of $F_{4}$ and $H_{3}$ 
for what follows.  
We can recover the original string frame metric by 
taking $\hat{h}\rightarrow 1$.

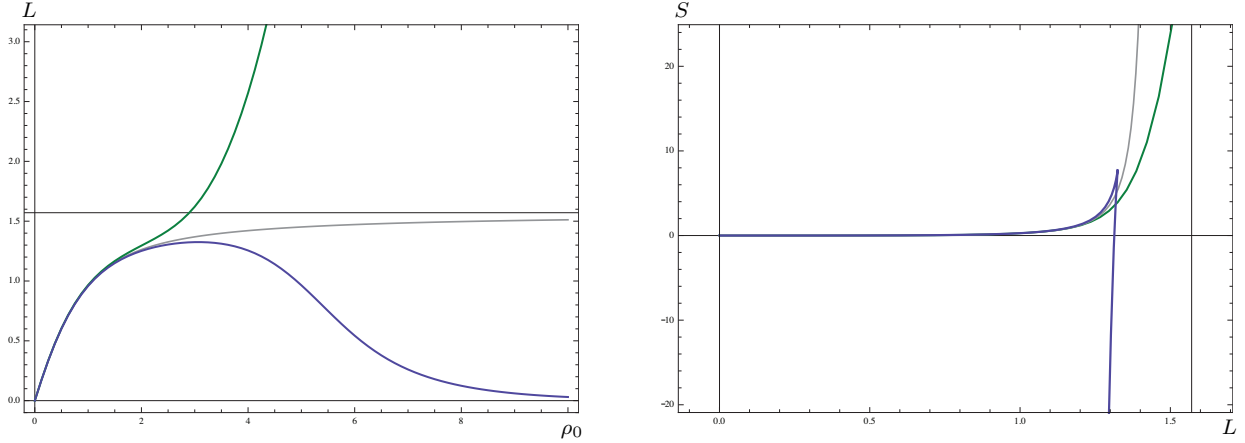
\begin{figure}[h]
\begin{center}
\begin{picture}(220,160)
\put(-124,0){\includegraphics[height=5.35cm]{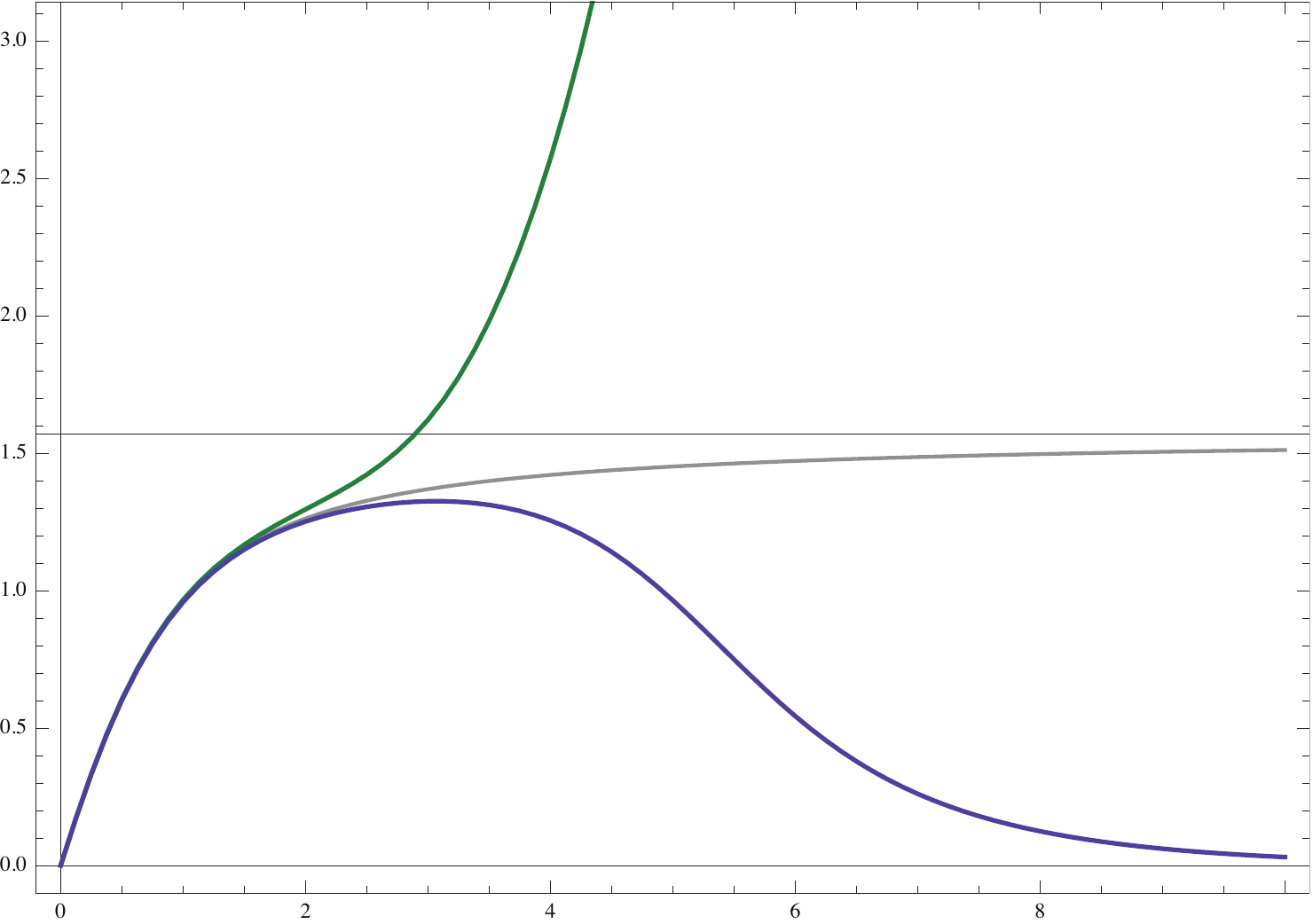}}
\put(-119,155){{\scriptsize{$L$}}}
\put(85,-3){{\scriptsize{$\r_{0}$}}}
\put(122,0){\includegraphics[height=5.35cm]{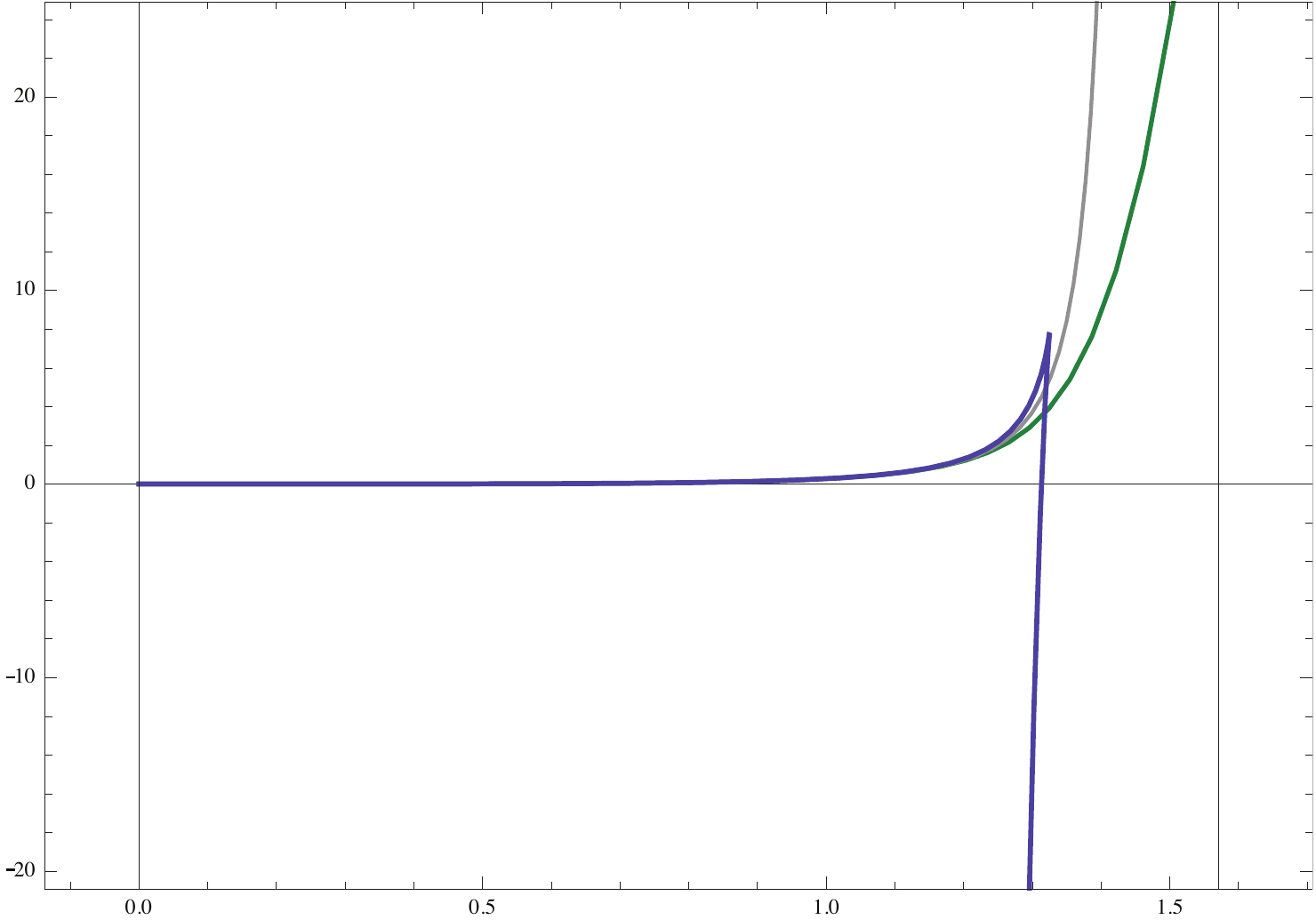}}
\put(128,155){{\scriptsize{$S$}}}
\put(335,-3){{\scriptsize{$L$}}}
\end{picture}
\caption{Here we have plots of the separation $L(\r_0)$ on the left and the entanglement entropy $S(L)$ on the right.  The MNa background is in grey ($g_{0}=1)$, the unrotated in green and rotated in blue (both have $g_{0}=1+10^{-2}$).}
\label{Fig:MNa}
\end{center}
\end{figure}

We again read off the relevant quantities to calculate the EE as
\begin{align}
V_{\mathrm{int}}=4\p^{4}\times  \hat{h}^{3/2} e^{3g+3h},\qquad H(\r)=e^{-4\hat{\Phi}}V_{\mathrm{int}}^2 \alpha^{2},\qquad \alpha(\r)=\hat{h}^{-1/2},\qquad \beta(\r)=\hat{h} e^{2g}
\label{EEQuant}
\end{align}
and making the appropriate substitutions we find that
\begin{align}
\sqrt{H(\r)}=2\p^2\times\hat{h}^{1/2} e^{3g+3h+2\Phi},\qquad \sqrt{\beta}=\hat{h}^{1/2}e^{g}.
\label{EEQuant2}
\end{align}
We can now discuss the behavior of the EE in each case.  For the 
Maldacena-Nastase (MNa) case, we find the same as in the D5 wrapped on a two-cycle (with linear $P=2N_c\r$), such that the separation $L$ grows with $\r_{0}$, and finds a maximum at $L_{\infty}=\p/2$.  The `unrotated' case initially follows the MNa behavior in the IR, but then blows up, whereas the `rotated' result (which again follows the MNa result up to 
around the same scale as the `unrotated'), goes to zero for larger $\r_{0}$. This can all be seen in Figure \ref{Fig:MNa}.

This means that for the EE, 
in both the MNa and unrotated case, 
the disconnected branch is always the lower than the connected 
branch and thus we require the `short configurations'.  
This is not true in the rotated case, where we find a behavior 
like that of a first order phase transition (thanks to the presence of
the function $\hat{h}$), 
akin to what happened with the D5 branes wrapped on a two-cycle, after
completing the system into the Baryonic Branch solution.

\subsection{Deformed D$4$ on a $S^{1}$, D$3$ on a $S^{1}$ and $AdS_5\times S^5$}
\begin{figure}[h]
\begin{center}
\begin{picture}(220,160)
\put(-124,0){\includegraphics[height=5.35cm]{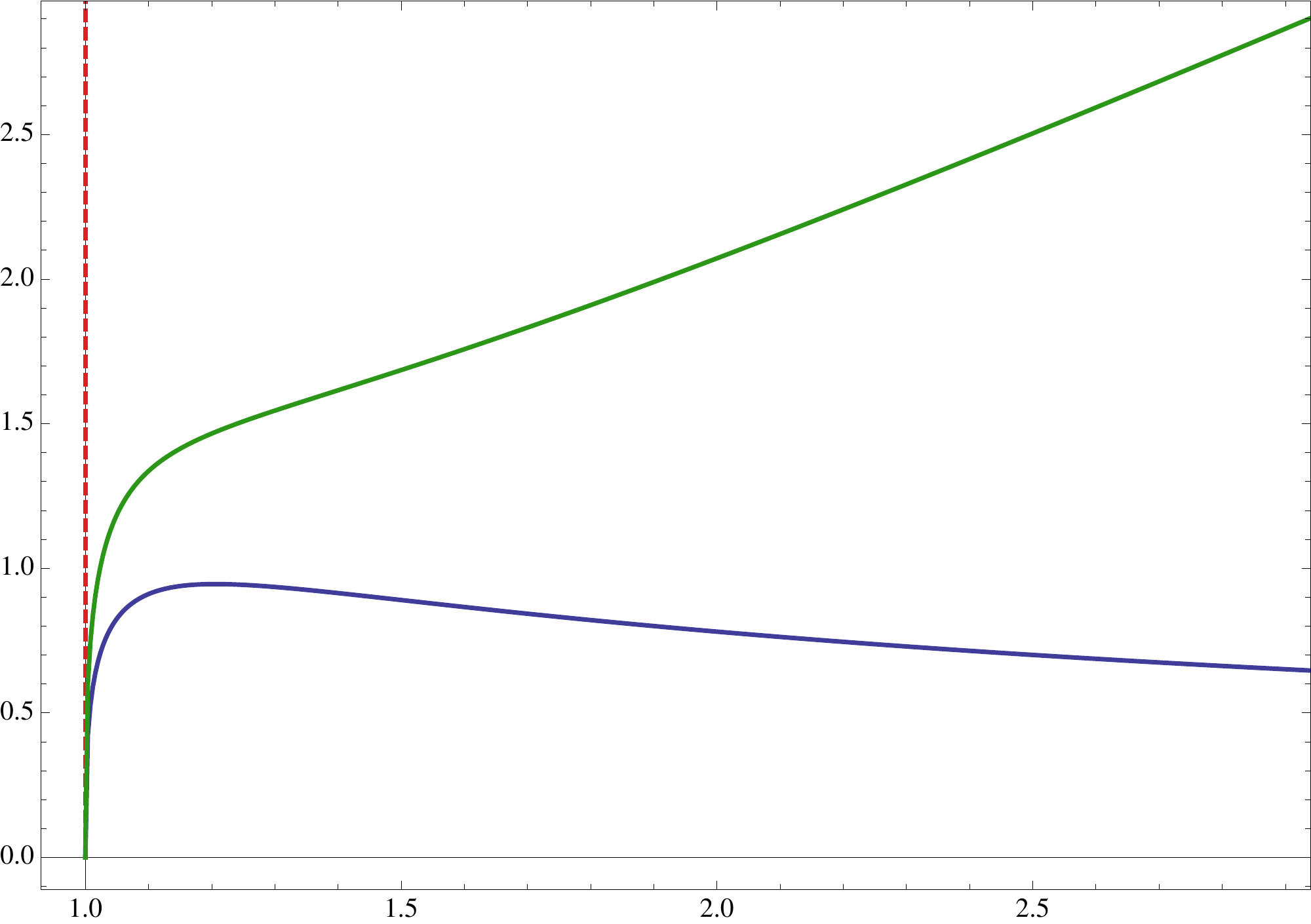}}
\put(-119,155){{\scriptsize{$L$}}}
\put(85,-3){{\scriptsize{$\r_{0}$}}}
\put(122,0){\includegraphics[height=5.35cm]{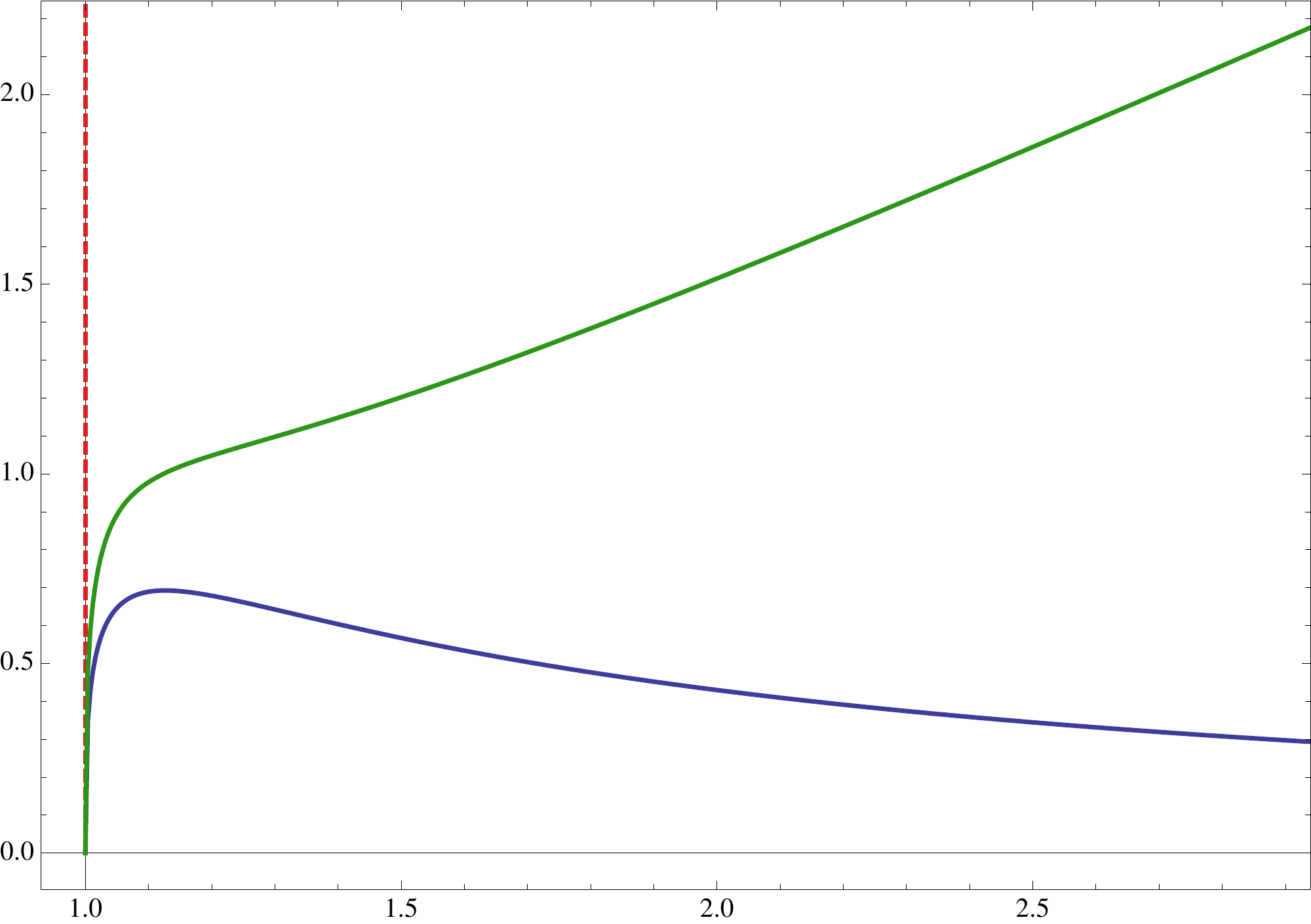}}
\put(128,155){{\scriptsize{$L$}}}
\put(335,-3){{\scriptsize{$\r_0$}}}
\end{picture}
\caption{In the left panel we have $L(\r_0)$ for the deformed D4 on $S^{1}$ and in the right panel we have $L(\r_0)$ for the deformed D3 on $S^{1}$.  The blue lines represented the undeformed solutions compared with those with the deformation in green.}
\label{Fig:KKM-D4-L}
\end{center}
\end{figure}
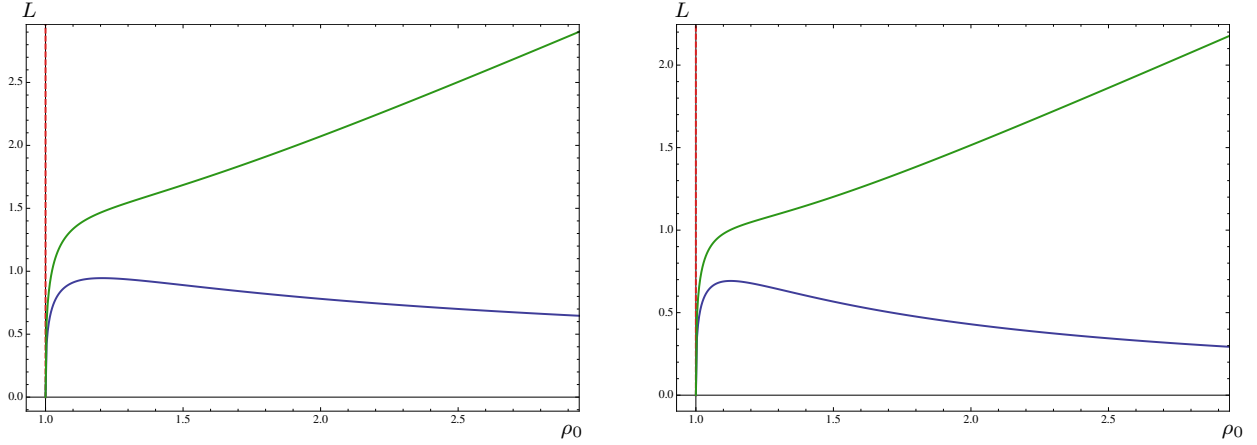

Initially, we start with the deformation of the D4 branes on $S^{1}$, where we write the metric as 
\begin{equation}
ds^2 =  \hat{h}(\r)^{-\frac{1}{2}} \left[dx^\mu
dx_\mu +f(\r)(dx_4)^2 \right] + \hat{h}(\r)^\frac{1}{2}\left[\dfrac{d\r^2}{f(\r)}+\r^2d\Omega_4^2\right],\quad
e^{-4\phi}=\hat{h}(\r)
\end{equation}
 We then get as usual
\begin{equation} \label{KKM-a}
\a(\r)= \hat{h}(\r)^{-1/2},\qquad \b(\r)= \dfrac{\hat{h}(\r)}{f(\r)},\qquad V_{int}(\r)=\dfrac{32\p^3 R^\frac{3}{2}}{9\sqrt{\rhol}}\hat{h}(\r)^{\frac{3}{4}}\r^4\sqrt{f(\r)}.
\end{equation}
It is easy to check that if we instead choose the warp factor as (this
choice emphasizes the 'non-locality' discussed in the 
main body of the paper),
\begin{equation}
\hat{h}=1+\left(\dfrac{R}{\r}\right)^3
\end{equation}
we find
\begin{equation}
\b(\r)=\frac{\r^{3}+R^3}{\r^{3}-\rhol^{3}} , \qquad H(\r)=\frac{1024 \pi ^6 \rho ^2 R^3 \left(\rho ^3+R^3\right) \left(\rho ^3-\rhol^3\right)}{81 \rhol},
\end{equation}
This leads to the behavior presented in the left-hand panel of Figure \ref{Fig:KKM-D4-L}. Note that now, $L$ grows linearly in the UV. This is easy to see if we look at $L$ in the UV as
\begin{equation}
L_{dD4}(\r_0\rightarrow\infty)=\frac{2 \pi  \rho_0  \left(\rho_0 ^3+R^3\right)^{3/2} \sqrt{\rho_0 ^3-\rhol^3}}{8 \rho_0 ^6+5 \rho_0 ^3 \left(R^3-\rhol^3\right)-2 R^3 \rhol^3}\sim\rho_0.
\end{equation}
Similarly for the deformed D$3$, with
\begin{equation}
ds^2 =  \hat{h}(\r)^{-\frac{1}{2}} \left[dx^\mu
dx_\mu +f(\r)(dx_3)^2 \right] + \hat{h}(\r)^\frac{1}{2}\left[\dfrac{d\r^2}{f(\r)}+\r^2d\o_5^2\right],\quad
e^{-4\phi}= \textrm{const},
\end{equation}
and $\a(\r)$ and $\b(\r)$ as in \eqref{KKM-a}, but with the warp factor modified as
\begin{equation}
\hat{h}(\r)=1+\left(\dfrac{R}{\r}\right)^4
\end{equation}
\begin{equation}
V_{int}(\r)=\frac{\p^4R^2}{\rhol}\hat{h}(\r)\r^5\sqrt{f(\r)},
\end{equation}
and thus
\begin{equation}
\b(\r)=\frac{\r^{4}+R^4}{\r^{4}-\rhol^{4}} , \qquad H(\r)=\frac{\pi ^8 \rho ^4 R^8 \sqrt{R^{4}+\r^{4}} \left(\rho ^4-\rhol^3\right)}{\rhol^2}.
\end{equation}
For deformed $AdS_5\times S^5$, we simply neglected the standard periodic association of the $x_3$ coordinate (be aware that this leaves 
a conical singularity at $\rhol$). 
As $x_3$ is not compact, it no longer is a part of $V_{int}$ but of $dx^\mu dx_\mu$. Note that the metric is no longer of the form of \eqref{background}. But this can trivially be dealt with by modifying the equation of $H(\r)$ to
\begin{equation}
H(\r) = e^{-4 \phi} V_{\rm int}^2  \alpha^d f(\r)~.
\end{equation}
We get
\begin{equation}
V_{int}(\r)= \p^3\hat{h}(\r)^{\frac{5}{4}}\r^5, \qquad
H(\r)=\pi ^6 \rho ^2 \left(\rho ^4+R^4\right) \left(\rho ^4-\rhol^4\right)
\end{equation}
The results are qualitatively equal to what we found for the deformed D4 on $S^{1}$. Choosing the warp factor as we did, causes $L$ to grow linearly in the $UV$, preventing 
the phase transition.  The UV expansion for $L$ yield the following
\begin{equation}
L_{dD3}(\r_0\rightarrow\infty)=\frac{\pi  \rho_0   \sqrt{\left(\rho_0 ^4+R^4\right)^{3}\rho_0 ^4-\rhol^4}}{5 \rho_0 ^8-2 R^4 \left(\rhol^4-2 \rho_0 ^4\right)-3 \rho_0 ^4 \rhol^4}\sim\rho_0,
\end{equation}
\begin{equation}
L_{dAdS}(\r_0\rightarrow\infty)=\frac{\pi  \r_0 \sqrt{ \left(\rho_0 ^4+R^4\right)^3 \left(\rho_0 ^4-\rhol^4\right)}}{5 \rho_0 ^8+3 \rho_0 ^4 \left(R^4-\rhol^4\right)-  R^4 \rhol^4}\sim\rho_0.
\end{equation}
The right-hand panel of Figure \ref{Fig:KKM-D4-L} shows the deformed D3 result and the deformed $AdS_5\times S^5$ behaviour is very similar.

\subsection{D$6$ Wrapped on a Three-Cycle}
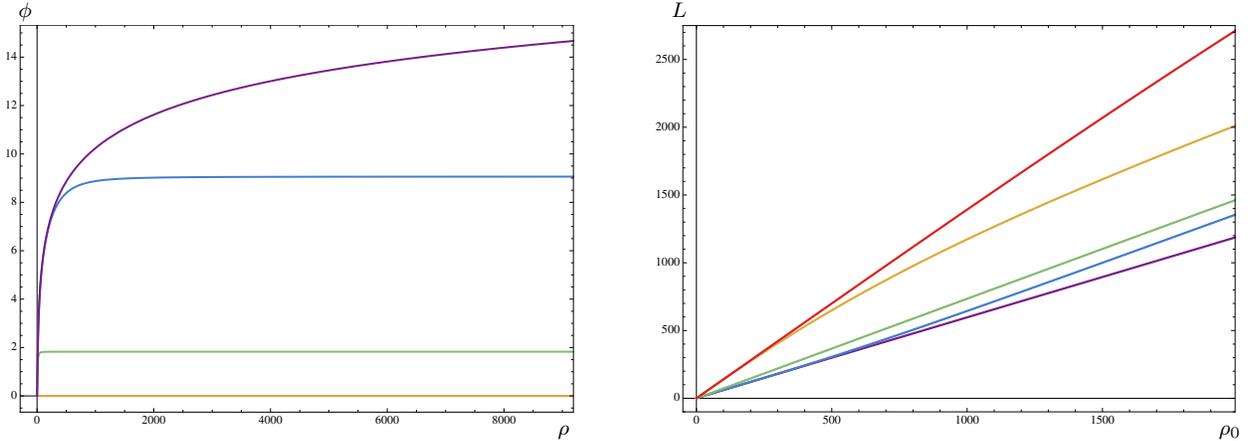
\begin{figure}[h]
\begin{center}
\begin{picture}(220,160)
\put(-124,0){\includegraphics[height=5.35cm]{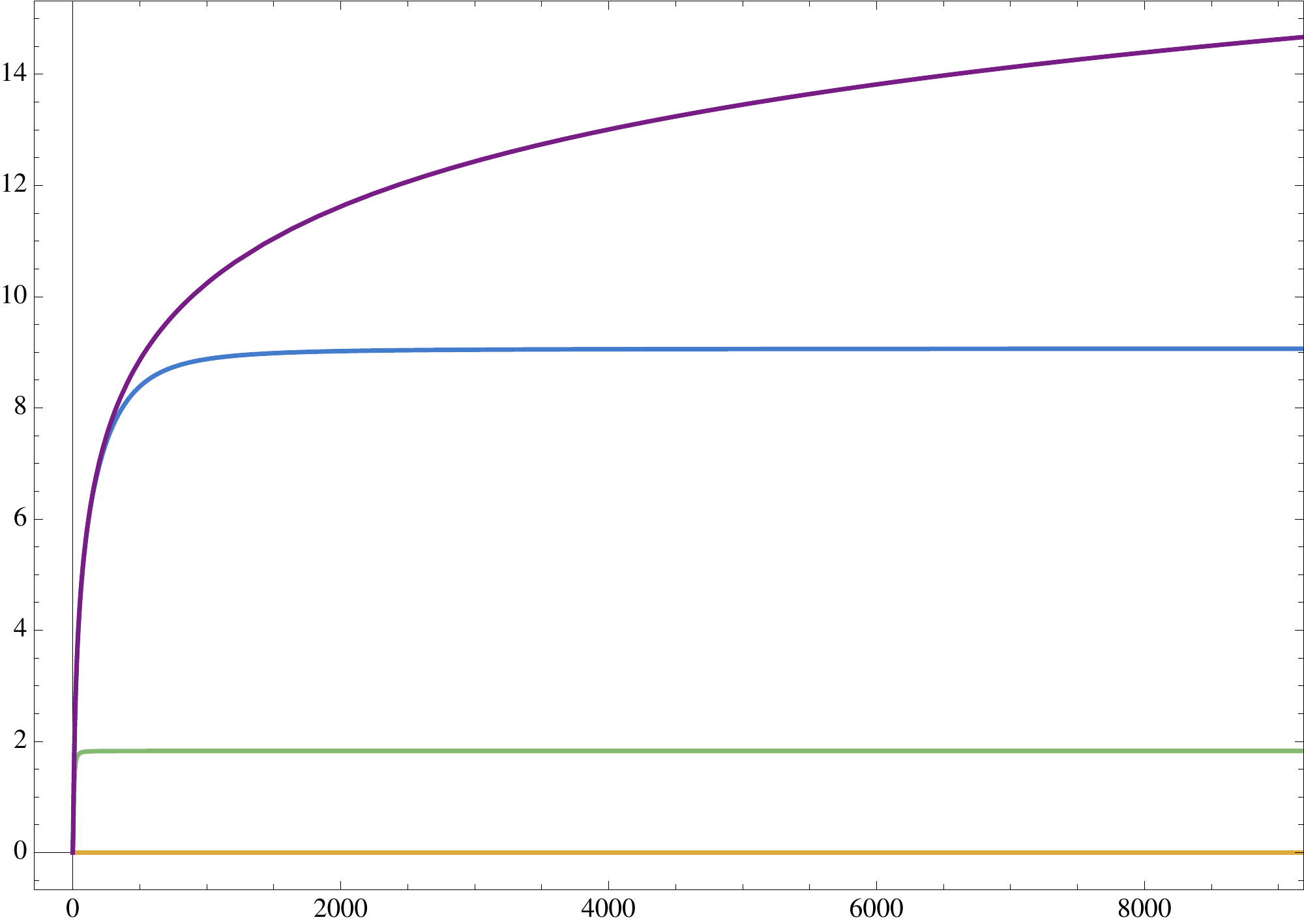}}
\put(-119,155){{\scriptsize{$\phi$}}}
\put(85,-3){{\scriptsize{$\r$}}}
\put(122,0){\includegraphics[height=5.35cm]{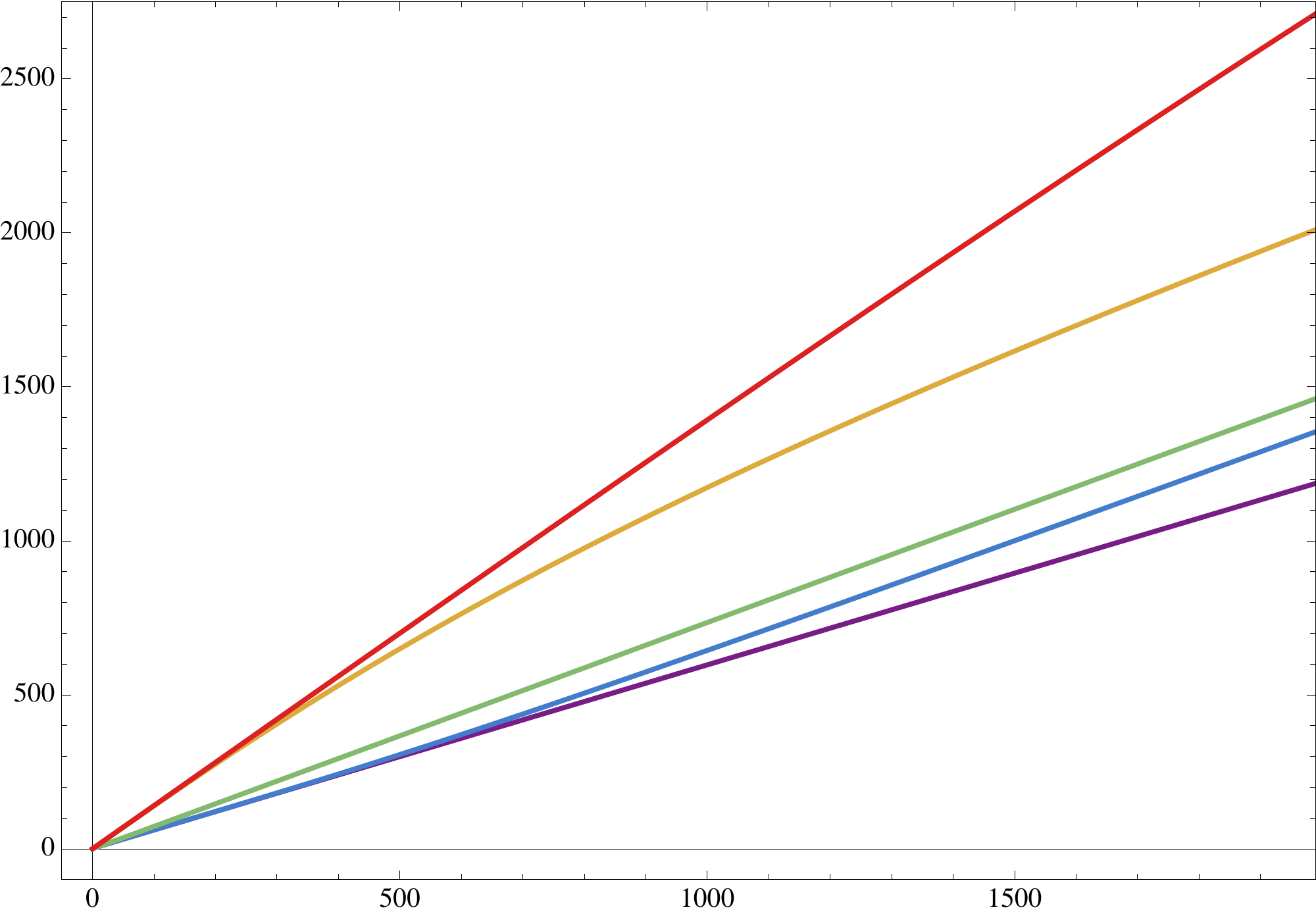}}
\put(128,155){{\scriptsize{$L$}}}
\put(335,-3){{\scriptsize{$\r_0$}}}
\end{picture}
\caption{Here we have plots of the dilaton in the left panel and the separation $L(\r_0)$ in the right panel in the backgrounds based on the D6 branes wrapped on a three-cycle. The range of $c$ is given in the text, with $c=\frac{3}{2}$ in purple and larger values $c$ in the range of colours to up to red.}
\label{Fig:D6L}
\end{center}
\end{figure}

Here we look at the behaviour of the 
solutions recently discussed 
in \cite{Caceres:2014uoa}. The full metric and dilaton are given by
\cite{Brandhuber:2001kq},
\begin{equation}
\begin{array}{ll}
\vspace{3 mm}
ds_{IIA,st}^2&=N e^{2\phi/3}\Bigg[\frac{dx_{1,3}^2}{N} + dr^2 + b^2(d\theta^2 + \sin^2\theta d\varphi^2)+\\
\vspace{3 mm}
&~~~~~~~a^2(\tilde{\omega}^1+gd\theta)^2+a^2(\tilde{\omega}^2+g\sin\theta d\varphi)^2+h^2(\tilde{\omega}^3-\cos\theta d\varphi)^2\Bigg]\\
\vspace{3 mm}
h^2&=\frac{c^2f^2}{f^2+c^2(1+g_3)^2}\\
e^{4/3\phi}&=\frac{c^2 f^2}{4 N h^2}
\label{metricxxx}\end{array}
\end{equation}
The background functions $a$, $b$, $c$, $f$, $g$ and $g_{3}$ are determined through the BPS equations
\begin{eqnarray}\label{D6BPS}
\dot{a} = -\frac{c}{2a} + \frac{a^5 f^2}{8 b^4 c^3}, & \;\; &
\dot{b} = -\frac{c}{2b} - \frac{a^2 (a^2-3c^2)f^2}{8b^3c^3}, \nn \\
\dot{c} = -1+\frac{c^2}{2 a^2}+\frac{c^2}{2 b^2}-\frac{3 a^2
f^2}{8b^4}, & \;\; &
\dot{f} = -\frac{a^4 f^3}{4 b^4 c^3},
\end{eqnarray}
as well as the relations
\begin{equation}
g(r) = \frac{-a(r)f(r)}{2b(r)c(r)} , \;\; g_3(r) = -1 + 2g(r)^2 .
\end{equation}
The forms of the functions required for the calculation of the Entanglement Entropy are given by
\begin{align}
&V_{int}= \int d^{8-d}y \sqrt{\det[g_{ij}]}=
(4\pi)^3 b^2 a^2 h (\alpha' g_s N)^{5/2} e^{5\phi/3},\nonumber\\
&\alpha = \mu e^{2\phi/3},\;\;\;\; \beta= \frac{\alpha' g_s N}{\mu},
\;\;\;, d=3,\\
&H= e^{-4\phi}V_{int}^2\alpha^d= (4\pi)^6 \mu^3 b^4 a^4 h^2 (\alpha' g_s
N)^{5}
e^{4\phi/3},\nonumber\\
& ds_5^2= \kappa [dx_{1,3}^2 + dr^2],\;\;\;\; \kappa^3=H\nonumber
\end{align}
The solutions in which the dilaton stabilises 
are interesting because the associated
backgrounds do not need an M-theory completion, 
so we will focus on them. We will re-express the expansion 
parameters used in \cite{Caceres:2014uoa}
as follows,
\begin{equation}
q_0=\dfrac{2}{\frac{1}{2}+c},\qquad R_0=\dfrac{1}{2}+c.
\end{equation}
So the parameter space is then defined through $c$ thus we choose to explore the following values for c:
\begin{equation}
c= \frac{3}{2}, \frac{3}{2} + 10^{-5}, \frac{3}{2}+10^{-1}, \frac{3}{2}+10^3, \frac{3}{2}+10^5.
\end{equation}
It appears that $c=\frac{3}{2}$ is a limiting case and the dilaton diverges, while for $c>\frac{3}{2}$ it eventually stabilises.
The results are presented in Figure \ref{Fig:D6L}.
Note that in all the cases, $L$ grows without bound for large $\r_0$. This agrees with the UV expansion of $L$ which is given by
\begin{equation}
L(\r_0\rightarrow\infty)= \dfrac{\pi }{5}\r_0-\dfrac{3}{5}(\pi q_1 R_1)+ \mathcal{O}\left(\frac{1}{\r_0}\right).
\end{equation}
In these solutions we find initially that $L$ will shoot off at different gradients depending on $c$, but eventually curves down to approach a line with the same gradient, but shifted by a constant.

\subsubsection{Illustrating the Dependence on the UV Cutoff}
The equations \eqref{D6BPS} also have a known analytical solution
\cite{Edelstein:2001pu}, that reads
\begin{align}\label{eq:exactsol}
	a(r)&=-\frac{r}{3}\left(1-\frac{r_{\Lambda}^{3}}{r^{3}}\right),\qquad\qquad
	b(r)= \frac{r}{2\sqrt{3}},\nonumber\\
	c(r)&=-\frac{r}{3}\left(1-\frac{r_{\Lambda}^{3}}{r^{3}}\right), \qquad\qquad
	f(r) =  \frac{r}{2 \sqrt{3}}.
\end{align}
From this form, one can work out the relevant functions and find that
\begin{equation}
L(\r_0\rightarrow\infty)= -\dfrac{2\sqrt{\pi}\Gamma(\frac{5}{12})}{\Gamma(-\frac{1}{12})}\r_0.
\end{equation}
Note that here $L$ also grows linearly into the UV.  In the main body of the paper we discuss the need for additional solutions, especially in backgrounds that have some form of non-locality. We found that these solutions are given by `short configurations' that never go far from the boundary. In Section \ref{StudyD5S1}, we explicitly mention that one can find these solution by studying the behaviour of the system close to the UV cutoff.

\begin{figure}
\begin{center}
\begin{picture}(220,160)
\put(3,0){\includegraphics[height=5.35cm]{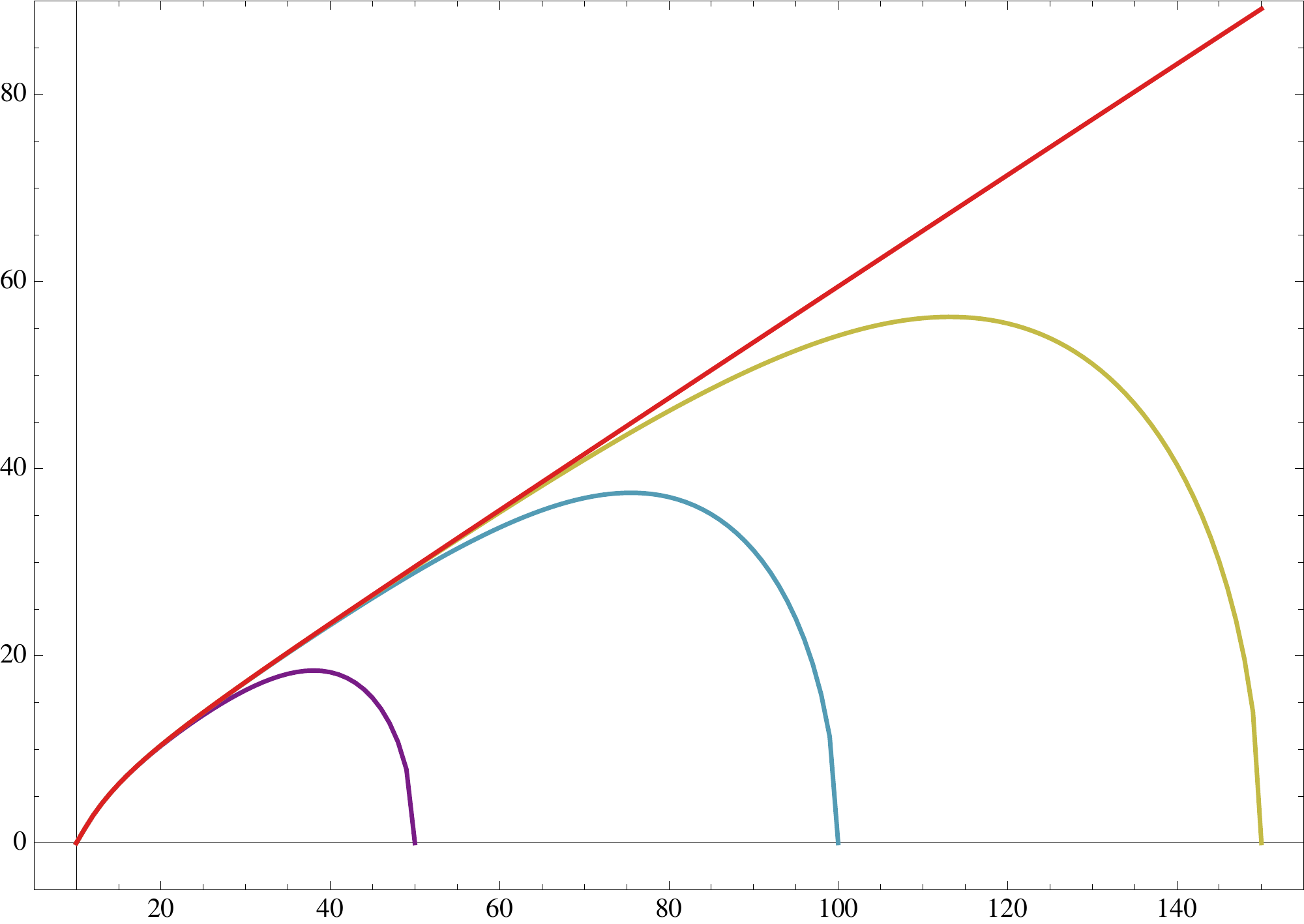}}
\put(2,148){{\scriptsize{$L$}}}
\put(220,3){{\scriptsize{$\r_0$}}}
\end{picture}
\caption{These are various plots of $L(\r_0)$ in the D6 wrapped on a three-cycle solution with $r_{\Lambda}=10$. The linear graph is the solution without cutoff, while the other three are numerical solutions with cutoff $50,100,150$ respectively.}
\label{Fig:cutoff}
\end{center}
\end{figure}

In the cases discussed where $L$ diverges, if we look at the `short configurations', we find that they change as we vary the UV cutoff.  Thus, we may want to think of this as viewing a system with strong IR/UV mixing, which causes the divergence, as discussed in \cite{Fischler:2013gsa},\cite{Karczmarek:2013xxa} (see to Figure \ref{Fig:cutoff}).

One could easily conclude that this solution has a phase transition, 
at a particular separation $L$, for a particular value of the cutoff, 
when in fact it is entirely cutoff dependent.  
Another case where the cutoff has caused similar effects was 
presented in \cite{Kim:2013ysa}.





\section{Appendix : Hints at Invariances of the Entanglement Entropy}
\label{app:YY}
\setcounter{equation}{0}

When calculating the Holographic Entanglement Entropy Density, we 
always seem to be able to reduce the problem to a one-dimensional system 
such that
\begin{equation}
 \frac{S_A}{V_{d-1}}=\int d\r \mathcal{L}(\r,\dot{\r})
 \end{equation}
For backgrounds of the form given in eq.(\eqref{background}) 
this is a trivial observation. However, as we will show in the
two examples examples, 
this remains true even for more complicated backgrounds, with warp
factors depending on the coordinates of the internal space 
and various fibrations.

\subsection{The D$4$-D$8$ System}
Let us look at the system presented by Brandhuber and Oz in 
\cite{Brandhuber:1999np}, the D4-D8 system.  
Here the ten dimensional space is a fibration of $AdS_{6}$ over $S^{4}$.  
Let us study if this causes 
complications with the entanglement entropy calculation.  The metric can be written as
\begin{equation}
ds_{str}^{2}=M(\w)\left[U^{2}dx_{1,4}^{2}+\frac{9Q}{4}\frac{dU^{2}}{U^{2}}+Q\, d\Omega_{4}^{2}\right],
\label{D4D8metric}
\end{equation}
where
\begin{equation}
M(\w)=\a' Q^{-1/2}\left[\frac{3}{4\p}C(8-N_{f})\sin\w\right]^{-1/3}, \qquad d\Omega_{4}^{2}=d\w^{2}+\cos^{2}\w\, d\Omega_{3}^{2},
\label{D4D8Malpha}
\end{equation}
and the dilaton has a profile given by
\begin{equation}
e^{\Phi}=Q^{-1/4} C \left[\frac{3}{4\p}C(8-N_{f})\sin\w\right]^{-5/6}.
\label{D4D8dil}
\end{equation}
Choosing the appropriate 8-dimensional surface $\Sigma_{8}=\{\w,\q_{1},\q_{2},\q_{3},x_{2},x_{3},x_{4},\s\}$ and allowing the radial coordinate to be $U=U(x_1)$ and for fixed $t$ we find that the induced metric takes the form
\begin{equation}
ds_{\Sigma_8}^{2}=M(\w)\left[U^{2}(dx_{2}^2 + dx_{3}^{2} + dx_{4}^{2})+U^{2} dx_{1}^{2}\left[1+\frac{9Q}{4}\frac{(U')^{2}}{U^{4}}\right]+Q\left(d\w^{2}+\cos^{2}\w\, d\Omega_{3}^{2}\right)\right].
\label{D4D8induced}
\end{equation}
We can then write
\begin{equation}
\sqrt{\det g_{8}}= \sin^{2}\q_{1} \sin\q_{2}\, M(\w)^{4}\, \cos^{3}\w \, U^{4} \left(1+\frac{9Q}{4}\frac{(U')^{2}}{U^{4}}\right)^{1/2}
\label{D4D8det}
\end{equation}
From here it is easy to see that we can perform the integrals in the action for the entanglement entropy and the result will be of the form
\begin{equation}
S=2\p^{2}(\a')^{4}\left[\frac{3}{4\p}C(8-N_{f})\right]^{1/3} \frac{Q^{1/2}}{C^{2}}\times\frac{9}{20}\int\mathrm{d}\s \, U^{4} \sqrt{1+\frac{9Q}{4}\frac{(U')^{2}}{U^4}}
\label{D4D8action}
\end{equation}
where the last part which is now in the usual form (the factor of $\frac{9}{20}$ comes from the $\w$ integral) and is the standard result for $AdS_{6}$.  Thus the D4-D8 system can be solved using the method we have employed previously.

\subsection{A Background with a Cyclic RG Flow}

We now turn our attention to the setup described by 
Balasubramanian in \cite{Balasubramanian:2013ux}.  
In this paper they construct non-singular solutions of 
a six dimensional theory which is a warped product of $AdS_{5}$ and 
a circle. These solutions have very non-trivial warp factors 
which break the symmetries of $AdS_{5}$ to discrete scale invariance 
and also break the translational symmetry along the circle.  Let us study
 if these causes troubles in 
our calculation of the entanglement entropy of a strip.  
The metric takes the form
\begin{equation}
ds_{6}^{2}=e^{2C[\w,\q]}\left[e^{2\w/L}(-dt^{2}+dx_{i}^{2})+d\w^2\right]+e^{2 B[\w,\q]}(d\q+\mathcal{A}[\w,\q]d\w)^2
\label{BalaMetric}
\end{equation}
where the functions $B$, $C$ and $\mathcal{A}$ are non-trivial 
functions of the Jacobi Elliptic functions $\mathfrak{sn}$, 
$\mathfrak{cn}$ and $\mathfrak{dn}$. Their exact form will not 
be important in what follows.  We are interested in whether 
the mixing in the metric due to the fibration 
represented by $\mathcal{A}$ causes any issue in the calculation of the EE.   If we calculate the form of the corresponding pullback of the metric onto the now 4-dimensional surface $\Sigma_{4}=\{x_{2},x_{3},x_{1},\q\}$, and setting the radial coordinate $\w=\w(x_1)$ we find that when we take the determinant it gives
\begin{equation}
\sqrt{\det{g_{4}}}=e^{B[\w,\q]+3 C[\w,\q]+\frac{3\w}{L}}\sqrt{1+e^{-\frac{2\w}{L}}(\w')^2}.
\label{BalaMetricDet}
\end{equation}
From this we see that there are no terms 
involving the function $\mathcal{A}$ and that this again falls 
into the simple form and we can again use the standard procedure.

\end{document}